\documentclass[preprint2]{aastex}

\usepackage{epsfig,amssymb}
\usepackage[dvips]{color}
\usepackage{graphicx}

\shorttitle{Cold Filaments from Colliding Superbubbles}
\shortauthors{Ntormousi et al.}

\begin{document}


\title{Formation of Cold Filamentary Structure from Wind Blown Superbubbles}


\author{Evangelia Ntormousi \altaffilmark{1,2}, Andreas Burkert \altaffilmark{1,2,3}, Katharina Fierlinger \altaffilmark{1,3}, 
and Fabian Heitsch \altaffilmark{4}}


\altaffiltext{1}{University Observatory Munich, Scheinerstr. 1, D-81679 Germany}
\altaffiltext{2}{Max Planck Institute for Extraterrestrial Physics, Giessenbachstr.
85748 Garching Germany}
\altaffiltext{3}{Excellence Cluster Universe, Boltzmannstr. 2
D-85748 Garching Germany}
\altaffiltext{4}{University of North Carolina Chapel Hill
CB 3255, Phillips Hall}


\begin{abstract}

The expansion and collision of two wind-blown superbubbles is investigated numerically.
Our models go beyond previous simulations of molecular cloud formation from converging gas flows by exploring 
this process with realistic flow parameters, sizes and timescales.
The superbubbles are blown by time-dependent winds and supernova explosions, calculated
from population synthesis models.  They expand into a uniform or turbulent diffuse medium.  

We find that dense, cold gas clumps and filaments form naturally in the compressed collision zone of the two superbubbles.
Their shapes resemble the elongated, irregular structure of observed cold, molecular gas filaments and clumps.
At the end of the simulations, between 65 and 80 percent of the total gas mass in our simulation box
is contained in these structures.

The clumps are found in a variety of physical states, ranging from pressure equilibrium with the
surrounding medium to highly under-pressured clumps with large irregular internal motions 
and structures which are rotationally supported.

\end{abstract}




\section{Introduction}

Even though molecular gas amounts to only a small fraction of the volume of the Galactic interstellar medium, it
dominates interstellar mass \citep{Ferriere_2001}.  Moreover, being the site of all star formation, it plays an essential role in galaxy evolution. 

Molecular gas exists in galaxies in highly irregular clouds, with very clumpy structure and large non-thermal line widths, 
an indication of internal supersonic turbulence \citep{Falgarone_1990,Williams_2000}.  
The large-scale structure of these clouds is very filamentary, according to observations in 
different wavelengths \citep{Schneider_Elmegreen_1979, LAL_1999, Hartmann_2002, LAL_2007, Molinari_2010}, 
a fact that can be attributed to their formation process \citep{Myers_2009}.

A natural mechanism for assembling the large amounts of gas necessary for the formation of 
molecular clouds is from large-scale converging atomic flows, such as expanding shells \citep{Elmegreen_1977, McCray_1987},
colliding shells \citep{Nigra_2008}, or 
flows generated by large-scale gravitational instabilities \citep{Yang_2007, KO_2002, KO_2006}.
Gas in these collision environments is subject to a number of fluid instabilities that can cool and condense it 
enough for it to become molecular \citep{Heitsch_2005}. 
This scenario can not only explain the rapid formation of molecular hydrogen
from atomic hydrogen, but also accounts for the turbulent nature of molecular clouds,
attributing it to their formation process \citep{Burkert_2006}.

In the context of local star-forming regions, the converging flow idea was proposed by \citet{Ballesteros-Paredes_1999b}
and \citet{Hartmann_2001} as an explanation of the fact that stellar populations 
in many star-forming clouds have age spreads that are significantly smaller
than the lateral crossing time of the cloud.

\citet{Pringle_2001} proposed an alternative mechanism for molecular cloud formation
by agglomeration of gas which is already molecular, arguing that, in order for the 
resulting clouds to be shielded from the background UV radiation which could evaporate them, the 
material in the converging flows had to be already dense and cold enough to be mostly molecular.  
Although the formation of large clouds from smaller, already molecular clumps
is a possible scenario (\cite{Dobbs_2008}, Dobbs, Pringle \& Burkert, submitted), 
there is still no observational evidence for the existence of a 
large molecular gas reservoir in the ISM outside giant molecular clouds.
Moreover, \citet{Bergin_2004} showed that dust shielding could allow
gas of much smaller densities to form molecules behind shocks and \citet{Heitsch_Hartmann_2008}
showed in colliding flows simulations with gravity
that the collapse along the second and third dimensions can rapidly lead to the high column densities required for molecule formation
and for efficient shielding from UV radiation.

\citet{Heitsch_2006, Heitsch08} studied the formation of cold and dense clumps between two infinite
flows which collide on a perturbed interface. 
Their study showed that cold structure can arise even from initially uniform flows if the
conditions favor certain fluid instabilities, such as the Non-linear Thin shell instability (NTSI, \citet{Vishniac_1994}),
the Thermal Instability \citep{Field_1965} and the Kelvin-Helmholtz instability. 
 
\citet{VS_2006, VS_2007} used colliding cylindrical flows of tens of parsecs length, 
adding random velocity perturbations to the average flow velocity.  
This allowed them to study star formation efficiencies and the fates of individual clouds.

These numerical experiments have investigated the mechanisms which lead to the condensation and cooling of atomic to molecular gas and which cause the complex structure of the resulting clouds, independent of the specific mechanism driving the flows.
In a further improvement of these models, we study the possibility for molecular cloud formation from colliding flows of limited thickness, 
such as thin shells from expanding superbubbles, and without explicit control of the collision parameters. 

We present two-dimensional hydrodynamical simulations of colliding superbubbles, 
both in a uniform and in a turbulent diffuse medium.  
The superbubbles are produced by stellar feedback in the form of time-varying winds and supernova explosions,
calculated from population synthesis models. 
We show that in this environment perturbations in the colliding flows arise naturally through the growth of instabilities 
at the edge of the bubbles, forming cold and dense clumps with a very filamentary structure and high internal velocity dispersions. 
We also find that, although turbulence naturally arises even in an initially quiescent background, 
the presence of background turbulence introduces anisotropies in the shell fragmentation
and more pronounced filaments.    

In section \ref{nm} we explain our numerical method, 
in section \ref{res} we present the results of our simulations 
and in section \ref{sum} we discuss our results and their implications.  


\section{Numerical Method}\label{nm}

\subsection{Numerical Setup}

We perform two-dimensional hydrodynamical simulations using the RAMSES code \citep{Teyssier_2002}, 
which uses a second-order Godunov scheme to solve the equations of ideal hydrodynamics on a Cartesian grid.   
In these calculations we have not made use of the Adaptive Mesh Refinement (AMR) feature of the code.

We simulate a region of physical size equal to 500$^2$ pc$^2$, with a resolution of $4096$ points at each dimension.  
Thus we achieve a spatial resolution of about 0.1 pc.  
We have also performed an $8192^2$ resolution run but, due to its high computational cost, only for limited integrations and for testing some resolution effects.
Although a 0.1 pc resolution is not sufficient to resolve the smallest structures in the ISM  
(see for example, discussion in \citet{Hennebelle_atomic_2007}), 
we can resolve many of the clumps which form in these simulations adequately.  
In our analysis we do not take into account structures that fall near our resolution limit
(see discussion in section \ref{res}).

As an initial condition we use either a uniform or a turbulent diffuse medium, 
of hydrogen density $n_H=1/cm^3$ 
and temperature T=8000 K. 
In the case of the turbulent medium these are just average quantities.  
We assume an ideal monoatomic gas with a ratio of specific heats $\gamma$ equal to 5/3 and mean molecular weight $\mu=1.2m_H$.
The setup of the turbulent medium is discussed in more detail in section \ref{sec:turb}.

In this diffuse medium we set up time-dependent winds that are meant to mimic the combined effects of winds and
supernova explosions in young OB asociations (-- see section \ref{wind} for details).  
We include two wind regions, placed on the edges of the computational domain, 
that is, 500 pc apart, and we assume them to form simultaneously.
We follow their expansion into the diffuse medium until they collide 
and a turbulent region arises at their interaction region.
Reflecting boundaries are used along the x-direction 
and outflow boundaries along the y-direction.

Our runs include a cooling-heating function 
following \citet{Wolfire_1995} and \citet{Dalgarno_1972} 
for the low temperature regime ($T<25000K$) and 
\citet{Sutherland_1993} in the high-temperature 
($25000K<T<10^8K$) regime.  
The metallicity assumed in the simulations is solar.  
Our initial condition is chosen to lie on a stable point of the heating-cooling 
equilibrium curve, so that no cooling instability can occur unless the medium is externally perturbed.  
Of course, this is not exactly true everywhere for the turbulent case, 
but density fluctuations are in any case not large enough 
to lead to a two-phase medium without triggering from the bubbles.

These calculations include no external gravity field or self-gravity of the gas.

Time zero for the simulation is when star formation occurs in the wind areas.
The system is evolved in time until boundary effects start becoming potentially important,
which is 7 Myrs for both simulations.  

\subsection{Wind model}\label{wind}

We simulate the mass and energy input from a young OB association 
to its surroundings according to \citet{Voss_2009}.  The population synthesis model presented there
provides the energy and mass injection from an average star of an OB association with time.  

\begin{figure}
   \plotone{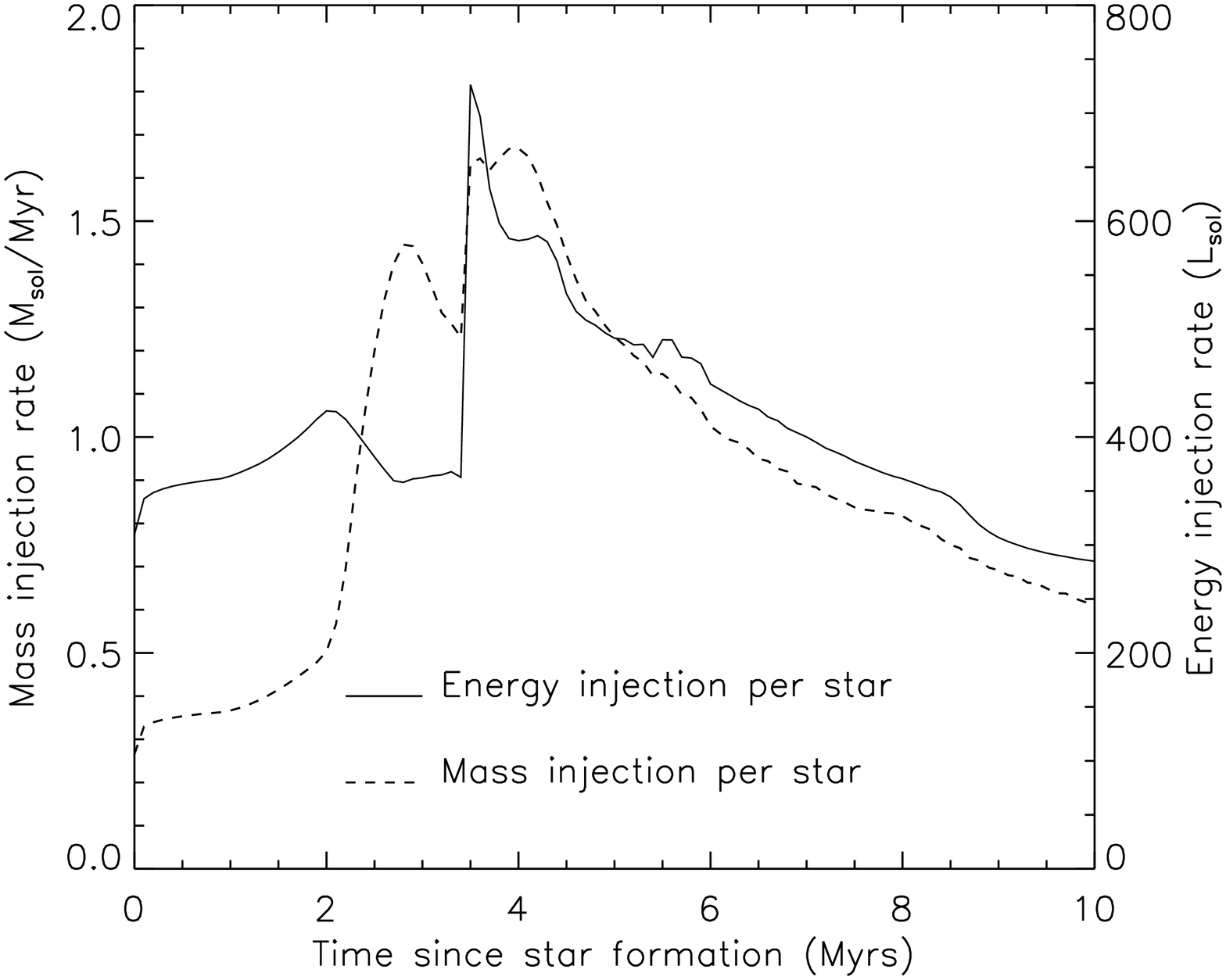}
   \caption{Time dependence of wind properties for an average star.
   The solid line shows the energy injection rate in units of solar luminosity
   and the dashed line shows the mass injection rate in solar masses per Myr.
    The data are from \citet{Voss_2009}}
   \label{wind_time}
\end{figure}

Figure \ref{wind_time} shows the energy and mass injection rate from one such star with time, 
from the combined effects of stellar winds and supernova explosions.
More details on the wind implementation in RAMSES can be found in Fierlinger et al. (2010, in prep.)

Two circular regions are selected, located on the edges of our computational domain 
and 50 of these average stars are placed in each domain.  
These stars insert energy and momentum to the selected region at each timestep, 
according to the model illustrated in figure \ref{wind_time}.  
Although we would be able to resolve the spatial distribution of stars, 
we avoid this complication and treat the whole population as a uniform source,  occupying a region of radius equal to 10 parsecs.
This also minimizes the initial noise in the expanding shell.

\subsection{Turbulence setup}\label{sec:turb}

In order to achieve a turbulent initial condition for the diffuse medium, 
we set up a turbulent velocity field 
according to \citet{MacLow_1998}, that is,
we introduce Gaussian random perturbations in Fourier space, 
in a range of wave numbers from $k=1$ to $k=4$.  

The rms Mach number of the turbulent flow is chosen close to unity, 
so that the turbulent kinetic energy of the gas is equal to its thermal energy.
The energy equipartition assumption between thermal and turbulent kinetic energy is consistent 
with the turbulent velocity dispersions calculated for Galactic HI \citep{Verschuur, Haud}. 

After this initial velocity field is calculated, 
it is introduced as an initial condition to the code, 
using a uniform hydrogen density of $n_H=1/cm^3$ and 
temperature T=8000 K and follow its evolution isothermally 
and with periodic boundary conditions until the density-weighted power spectrum of the 
turbulent velocity field has a power-law slope close to 
\citet{K41} and the density field loses the signature of the initial condition.
During this time the Mach number is kept constant by driving.

The resulting velocity, density and pressure structure is used
as an initial condition for the simulation of bubble expansion, which is 
otherwise identical to the simulations in a uniform medium.

During the calculations we have neglected the driving of turbulence in the diffuse
medium, since the turbulence crossing time for our computational domain, 
$t_{cr}\simeq86$ Myrs is much longer than our entire calculation.
This makes driving unnecessary during the simulation, as  there are no significant energy losses due to dissipation on this time scale.


\section{Results}\label{res}

\subsection{Simulations in a uniform background medium}

\begin{figure*}
    \includegraphics[width=0.5\linewidth]{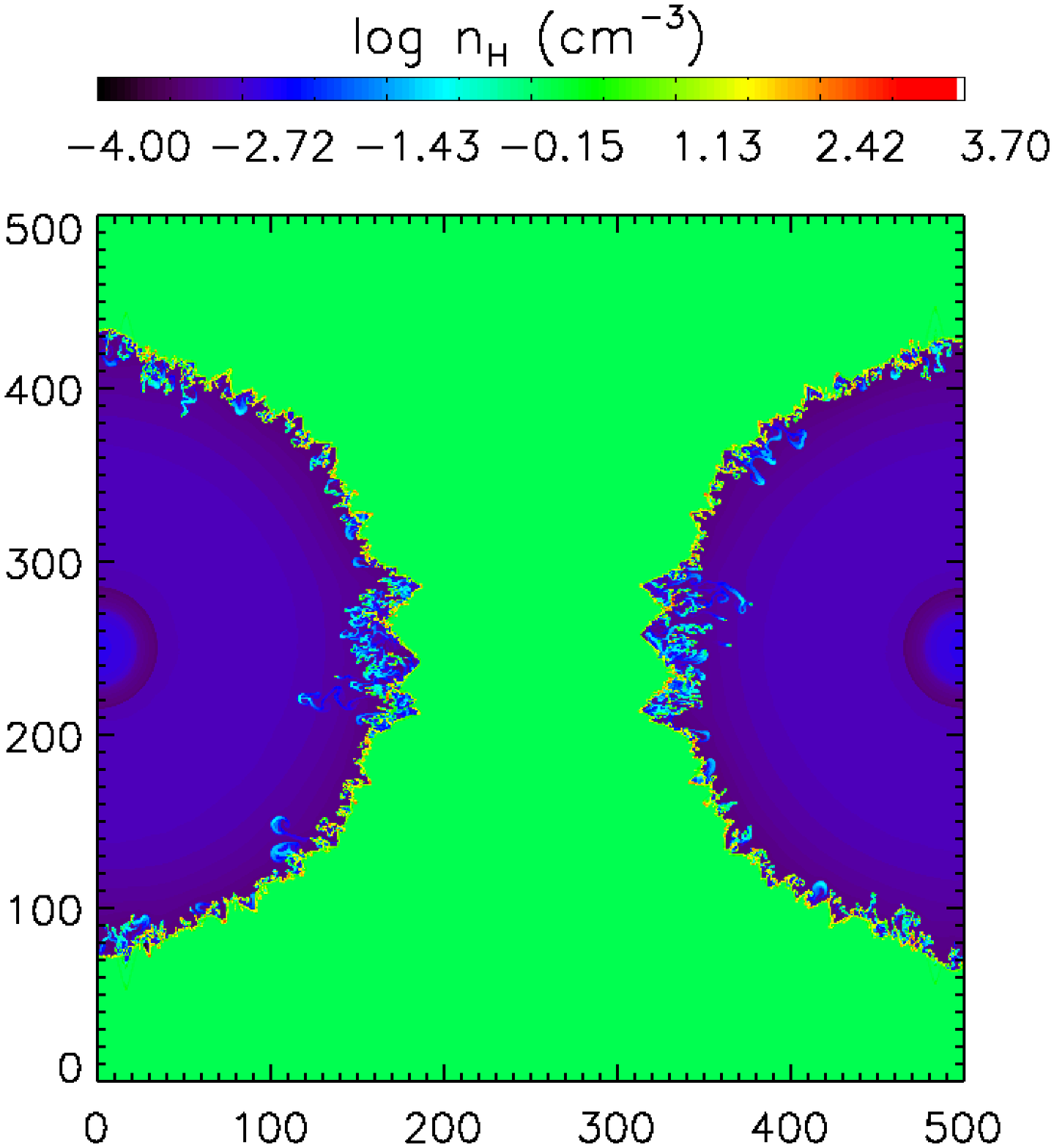} 
    \includegraphics[width=0.5\linewidth]{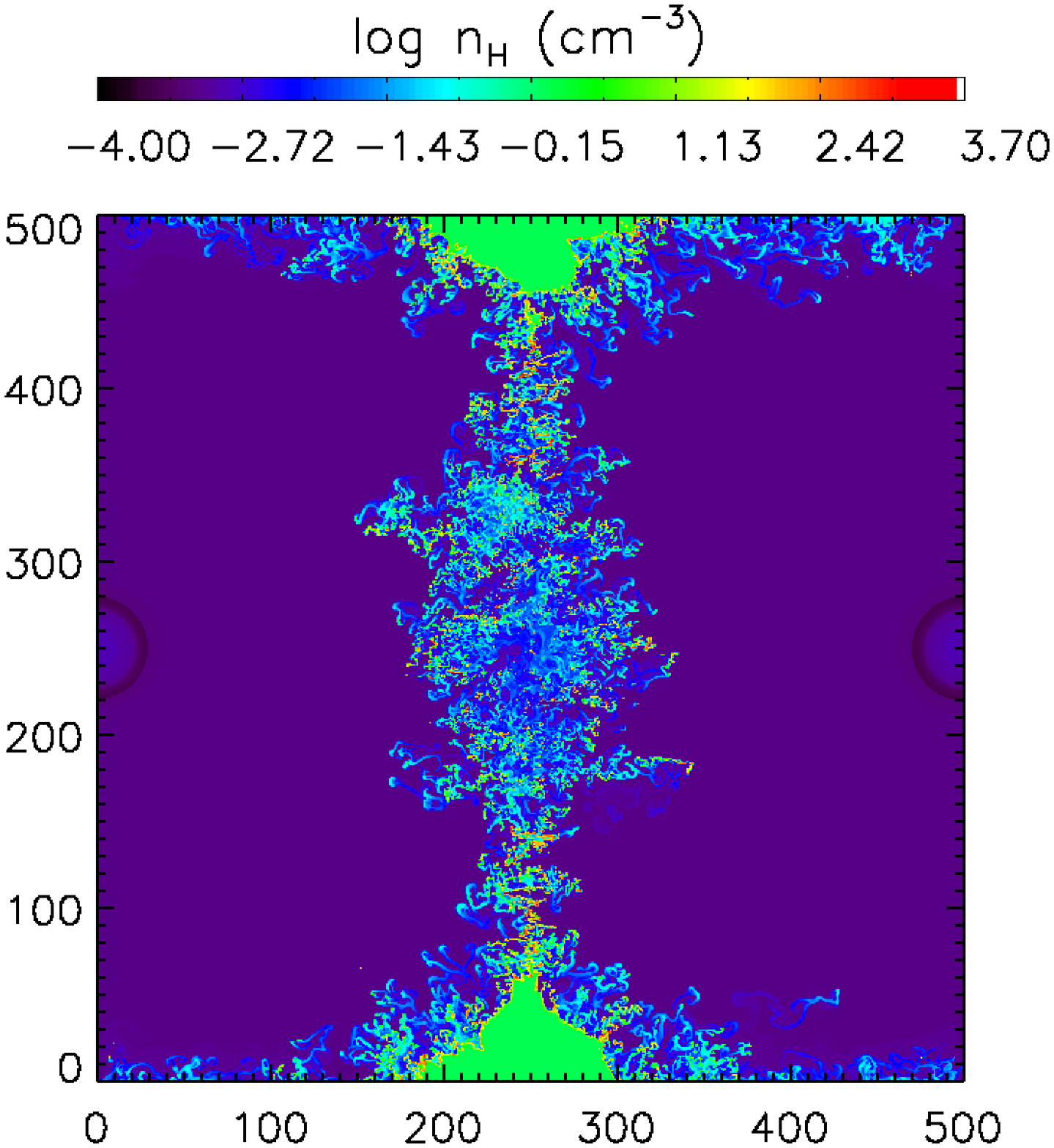}  
    \includegraphics[width=0.5\linewidth]{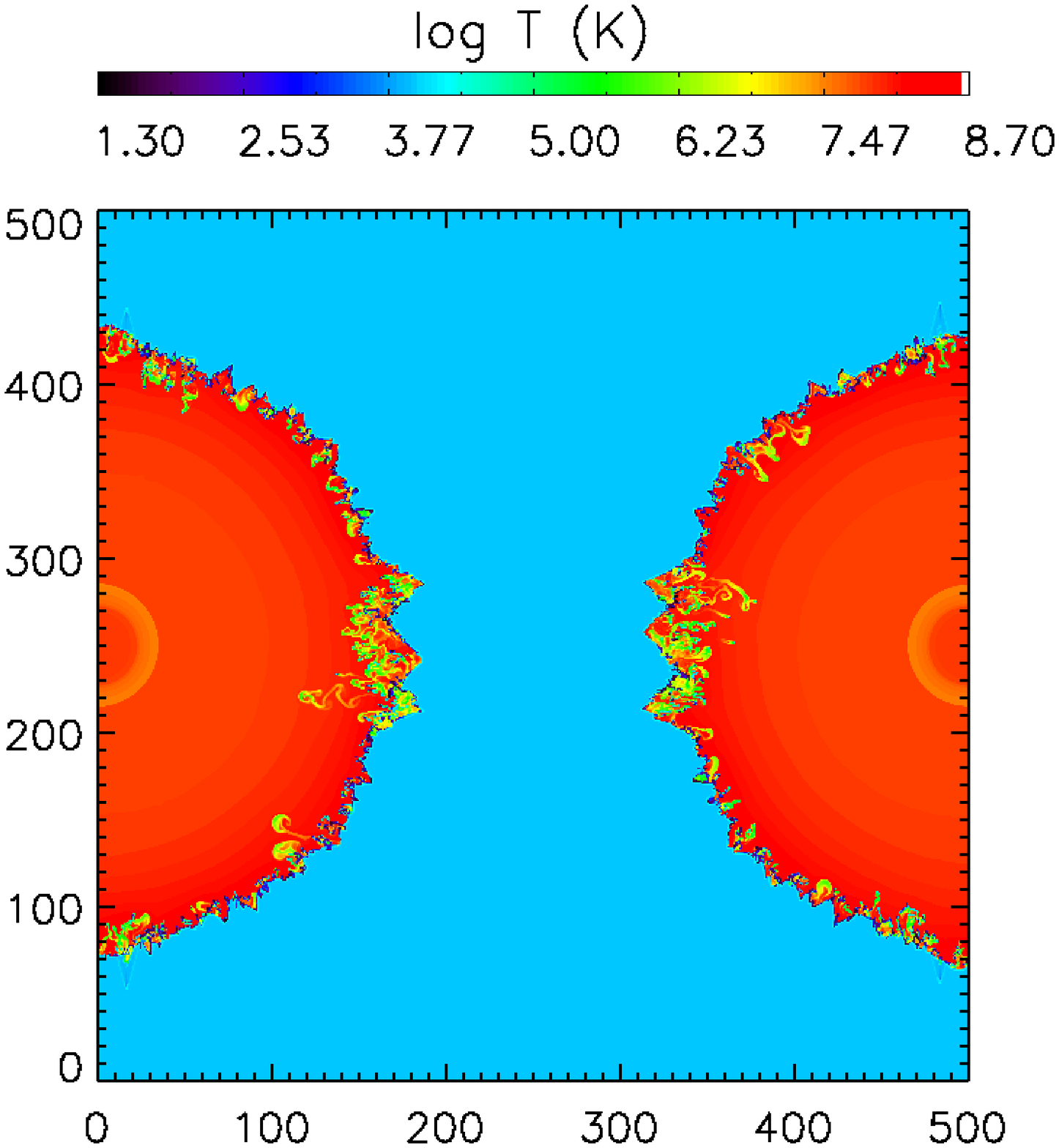} 
    \includegraphics[width=0.5\linewidth]{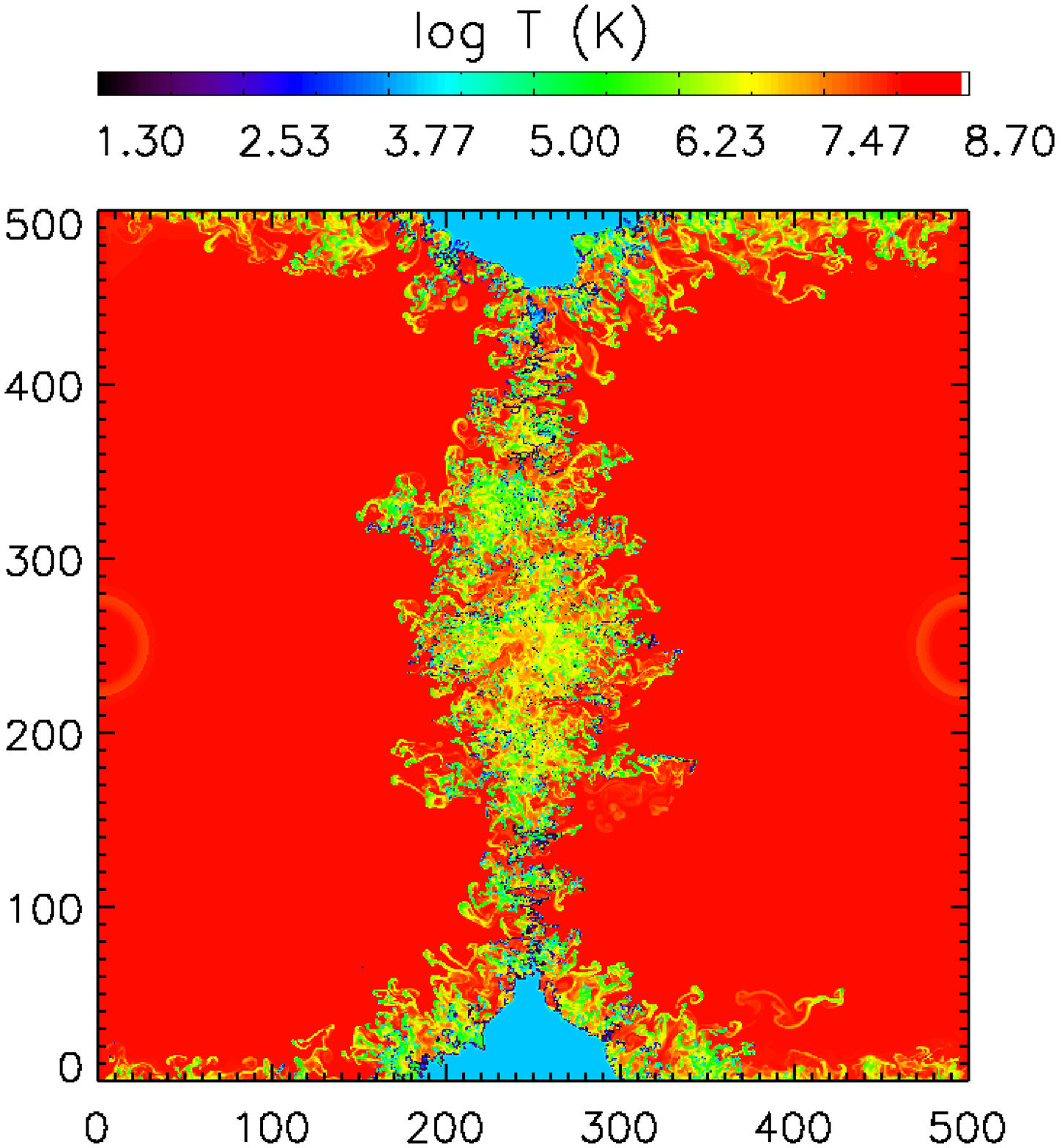}
    \caption{Superbubble collision snapshots in a uniform diffuse medium.
Plotted on the top panels is the logarithm of the hydrogen number density in log(cm$^{-3}$)
and on the bottom panels the logarithm of the gas temperature in log(K).
Left: 3 Myrs after star formation, right: 7 Myrs after star formation}
   \label{noturb}
\end{figure*}

In order to study the expansion of the superbubbles 
and the development of fluid instabilities relevant to it, 
we first simulated the two wind regions in a homogeneous diffuse background medium.  
Figure \ref{noturb} shows two snapshots of the simulation 
in logarithm of temperature and logarithm of density.  

During the expansion of the superbubbles we observe three effects from three different fluid instabilities.
The acceleration of the shock leads to the NTSI, which focuses material on fluctuation peaks.  These condensations
are unstable to Thermal Instability, so they condense and cool further. 
The velocity shear caused by the NTSI at the same time also triggers the Kelvin-Helmholtz instability.
\begin{figure*}
    \includegraphics[width=0.5\linewidth]{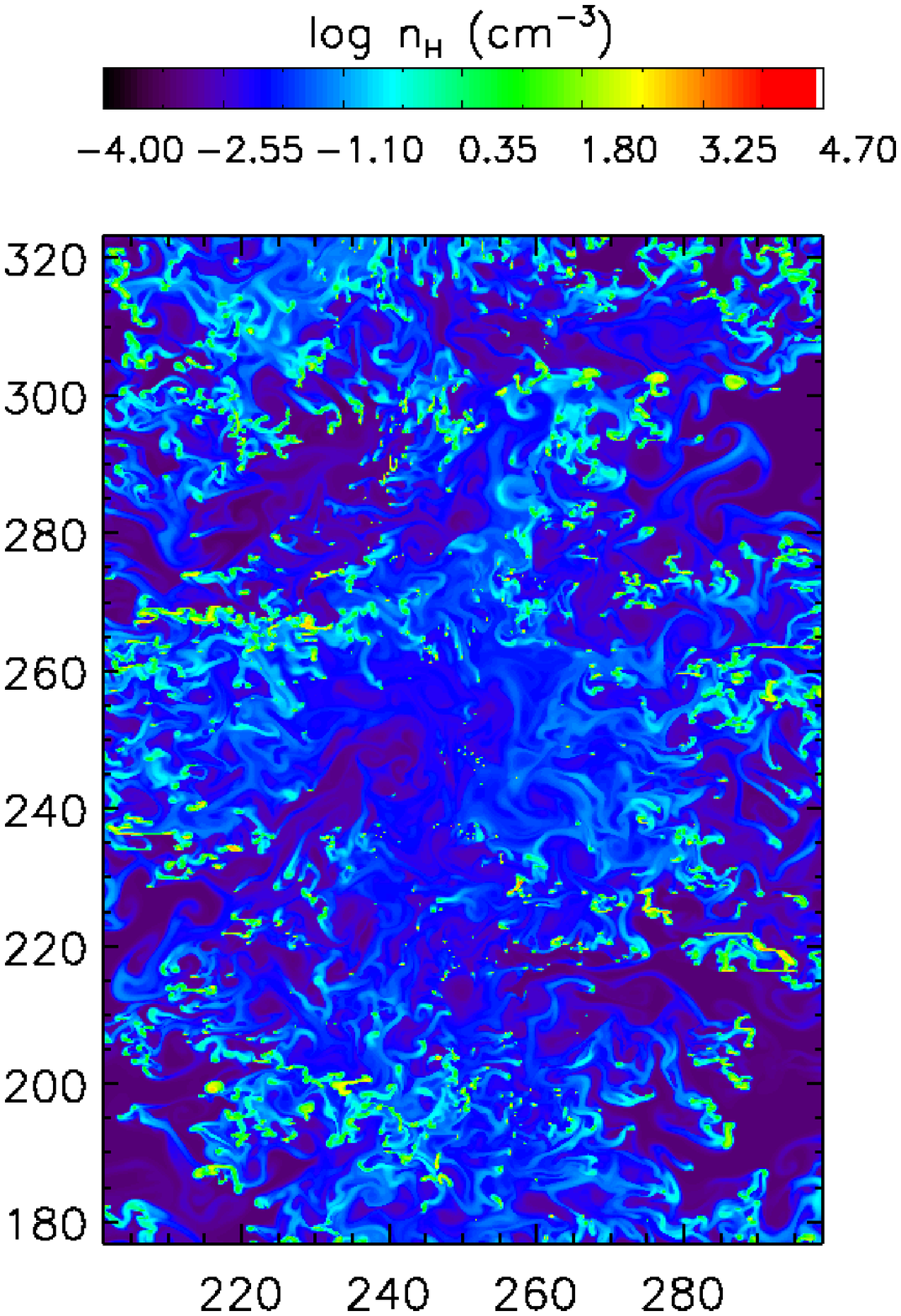} 
    \includegraphics[width=0.5\linewidth]{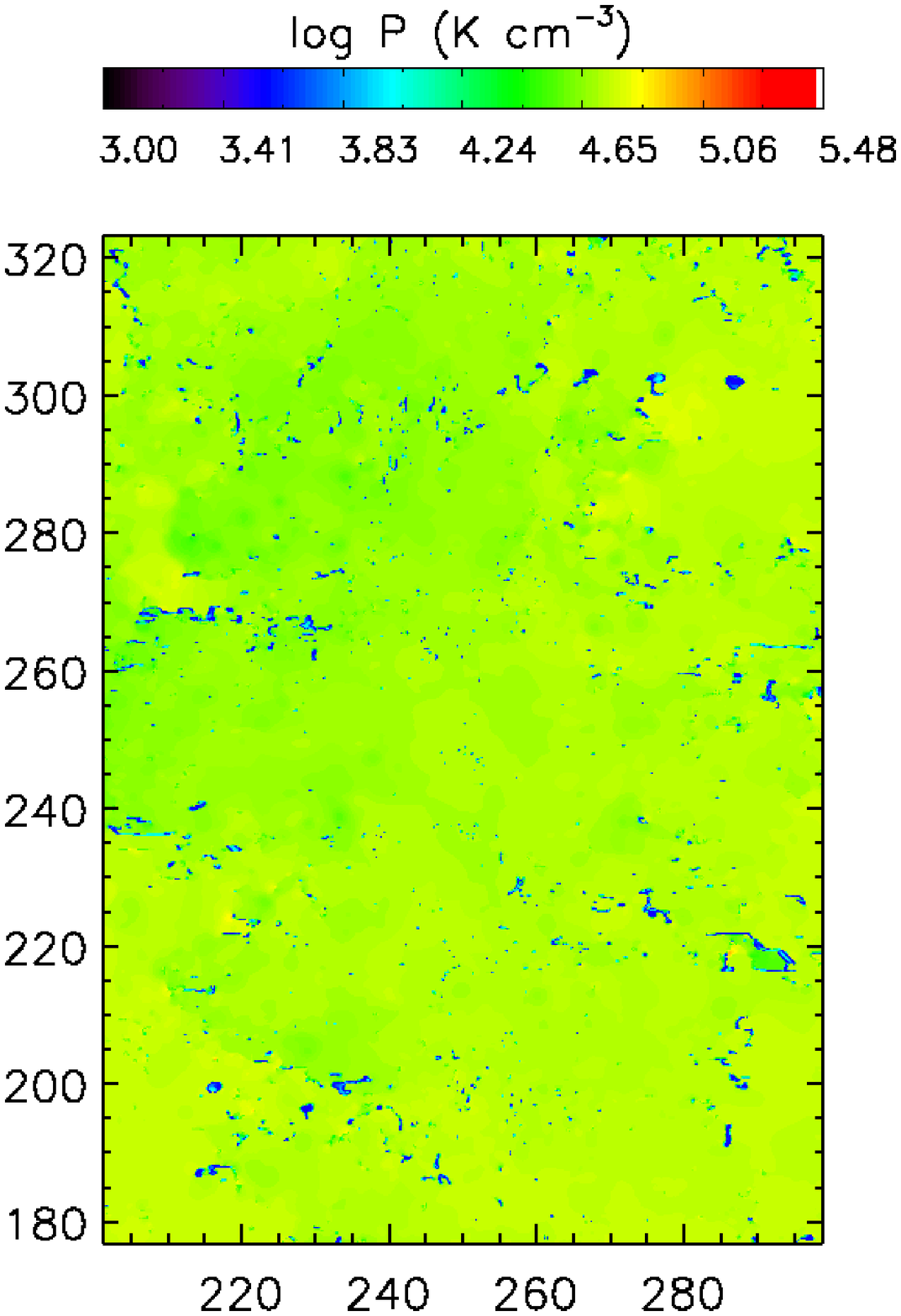}
    \caption{Zoom-in of the last snapshot (7 Myrs after star formation) of the uniform diffuse medium run.
Plotted on the left is the logarithm of the hydrogen number density in log(cm$^{-3}$)
and on the right the logarithm of the gas pressure in log(K cm$^{-3}$).  The axes coordinates are in parsecs.}
   \label{noturb_zoom}
\end{figure*}

As one can see in figure \ref{noturb},  the NTSI develops faster along the x and y axes, where we observe more pronounced "finger-like" structures,
characteristic of this instability.  This is clearly a resolution effect.  \citet{Vishniac_1989} showed that
for an expanding decelerating shock
there is a critical overdensity with respect to the post-shock gas above which the shell is unstable.  
The authors estimated this critical overdensity to be of the order 25 for a wind-blown shock.  
Since in our simulations we do not resolve the smallest cooling length adequately,
gas cannot be condensed into as small a volume as it should be according to its 
cooling rate.  Along the x and y axes though, for a Cartesian grid, the grid cells are closer together than along the diagonals, leading to an effective smaller distance
over which the shock can compress the gas in one timestep.  Thus along the x and y axes the fluid can be compressed slightly faster, then the critical overdensity can be achieved earlier and the instability arises earlier.  As mentioned in previous sections, we  have performed a test run with $8192^2$ resolution for comparison, 
but have not been able to suppress this feature.  However, the faster growth of the instability along 
these two lines does not affect the average clump formation time and the clump properties in any measurable way.  

As mentioned above, the resulting cold and dense structures are a result of the Thermal Instability.
However, our simulations do not include thermal conduction and therefore cannot, by definition, fulfill the "Field criterion".  Simulations not including thermal conduction may be susceptible to artificial phenomena near the resolution limit with no apparent convergence with increasing resolution, since thermal conduction has a stabilizing effect against the Thermal Instability \citep{Koyama_Inutsuka}.

This stabilization is discussed in \citet{Burkert_Lin_2000}, who studied the Thermal Instability both analytically and numerically for a generic, power-law cooling function. They find that density perturbations below a certain wavelength are always damped by thermal conduction.  This wavelength depends on the shape of the cooling function, and in our case would be roughly 4 times the Field length.  For typical warm ISM values, that is, a thermal conductivity of about $10^4$ ergs cm$^{-1}$ K$^{-1}$ sec$^{-1}$, a temperature of T=8000 K and a hydrogen number density of n=1 cm$^{-3}$, and for a cooling rate of $10^{-26}$ ergs sec$^{-1}$, the Field length is about 0.03 pc and four times this length is about our resolution limit.  This means that, even though our resolution is marginal for resolving the smallest unstable fluctuations, higher resolution without thermal conduction would only produce artificially small structure.  Of course, for lower temperatures and higher densities the Field length is much smaller, so unstable density fluctuations due to Thermal Instability within our clumps are not resolved.
    
Figure \ref{noturb_zoom} shows a close-up of the last snapshot of this run, in logarithm of density and pressure.
It is easier to identify the clumps in this figure, so one clearly sees all of them are located in under-pressured regions
with respect to the rest of the gas.

We identify clumps by selecting locations with hydrogen  number density greater than 50 $cm^{-3}$ and temperature smaller than 100 K
and using a friends-of friends algorithm to link such adjacent locations together. The group of cells is then identified as a single clump.
The density and temperature threshold is clearly an approximation, to account for the fact that our simulations 
do not include molecule formation and our cooling function assumes optically thin gas.  
For our analysis we only use structures which contain more than 16 cells, as smaller structures do not contain sufficient information and are considered under-resolved.

\begin{figure}
    \plotone{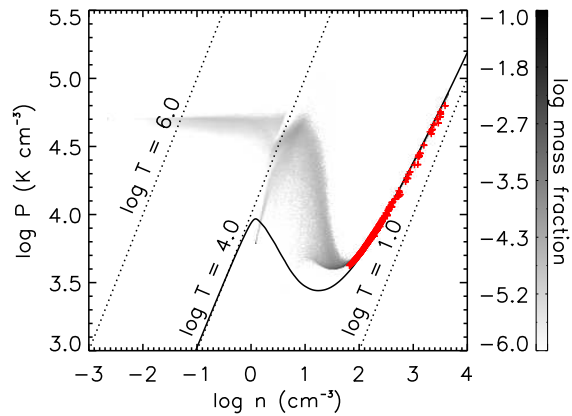}
    \caption{Pressure versus hydrogen density of the fluid in the box for the uniform medium run.
The plot is from the last snapshot, 7 Myrs after star formation.
The gas has been binned and color-coded according to the mass fraction it represents. 
The solid line is the cooling-heating equilibrium curve for the warm and cold gas
and the dashed lines show the locations of three isotherms.
Red crosses are average clump quantities.}                                                     
   \label{phases_noturb}
\end{figure}

\begin{figure}
   \plotone{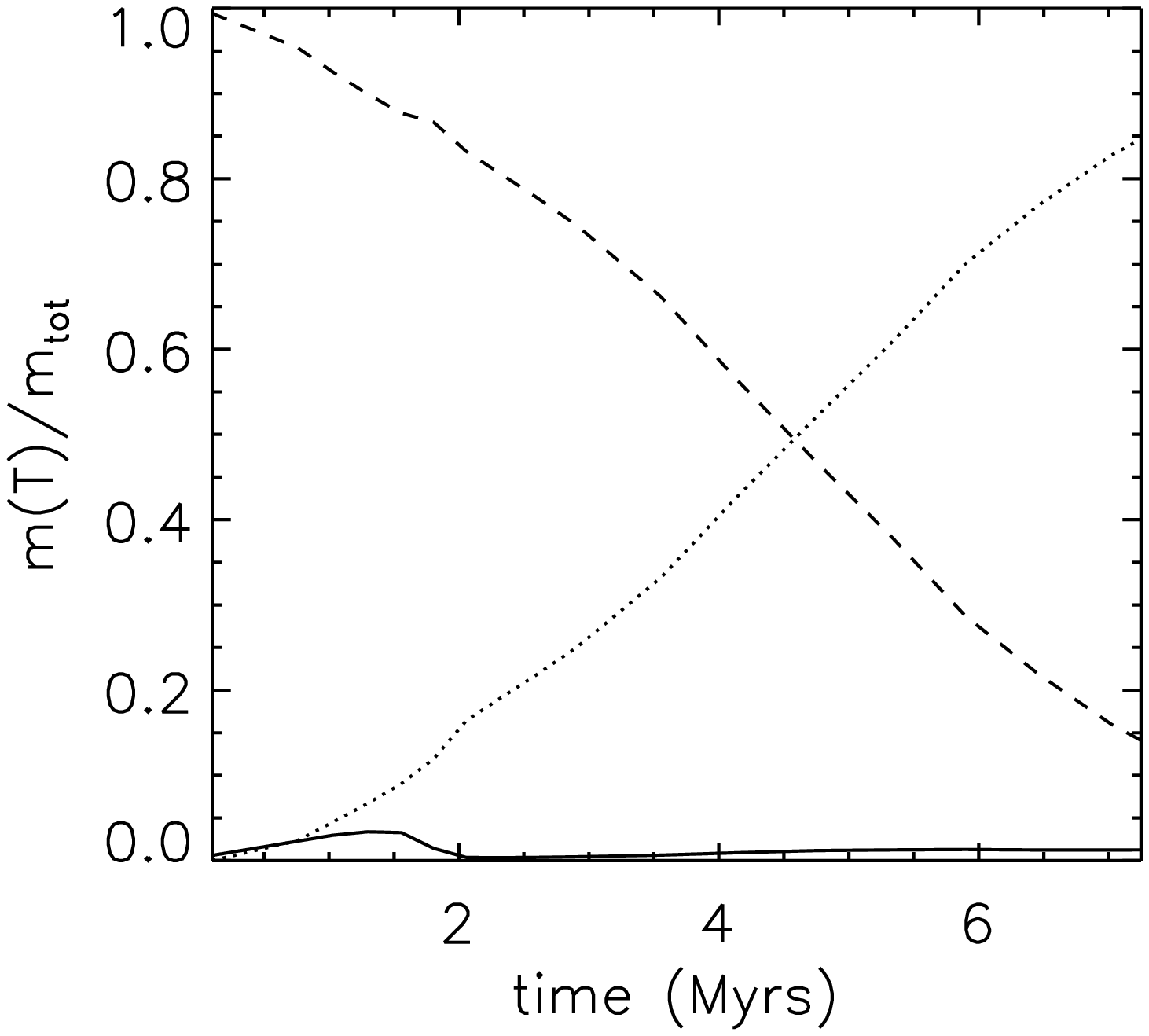}
   \caption{Gas fractions with time for the uniform medium run.  
   The dashed curve shows the mass fraction in the warm gas phase, the 
   solid curve shows the mass fraction in the hot gas phase and
   the dotted curve the mass fraction in the cold phase. 
   (Hot phase: gas with $T\ge25000K$, warm phase: gas with $100K<T<25000K$, cold phase: gas with $T\le100K$)}
   \label{gas_frac_noturb}
\end{figure}

Figure \ref{phases_noturb} shows the gas phase diagram at the end of the simulation. 
The plot shows all the gas in the simulation, color-coded to its corresponding
mass fraction in the simulated box. 
The solid line is the cooling-heating equilibrium curve  for the warm and cold gas and the red crosses are average clump properties.
As an indication for the gas temperatures, three isotherms have been overplotted.
Most of the gas mass seems to be in the cold phase (-- see also figure \ref{gas_frac_noturb}), but there is also about 14\%  of the gas mass in the warm, thermally unstable regime.  This gas is pushed from the stable to the unstable regime by the momentum inserted by the wind. As explained in section \ref{clouds} 
some of this thermally unstable gas is located around the cold clumps, forming a warm corona.  
The clumps lie on the equilibrium curve, as does the coldest gas, as expected from clumps formed by the Thermal Instability.  

Figure \ref{gas_frac_noturb} shows the mass fraction of the computational domain in each phase of the gas with time.
We define cold gas to have temperatures $T < 100K$, warm gas to lie in the temperature regime of $100K < T <  25000K$
and hot gas to have temperatures $T > 25000K$. 
As expected, we start with almost entirely warm gas in the box and as time goes by more and more cold gas is created.
Since the hot gas is very dilute, it only amounts to approximately 1 percent of the mass throughout the simulation.
In the end of the simulation we have approximately 85 per cent of the mass in cold clumps and 14 percent in the warm phase.
The mass injected by the OB association amounts to less than $10^{-5}$ of the total mass in the domain throughout the simulation. 

\begin{figure}
 \plotone{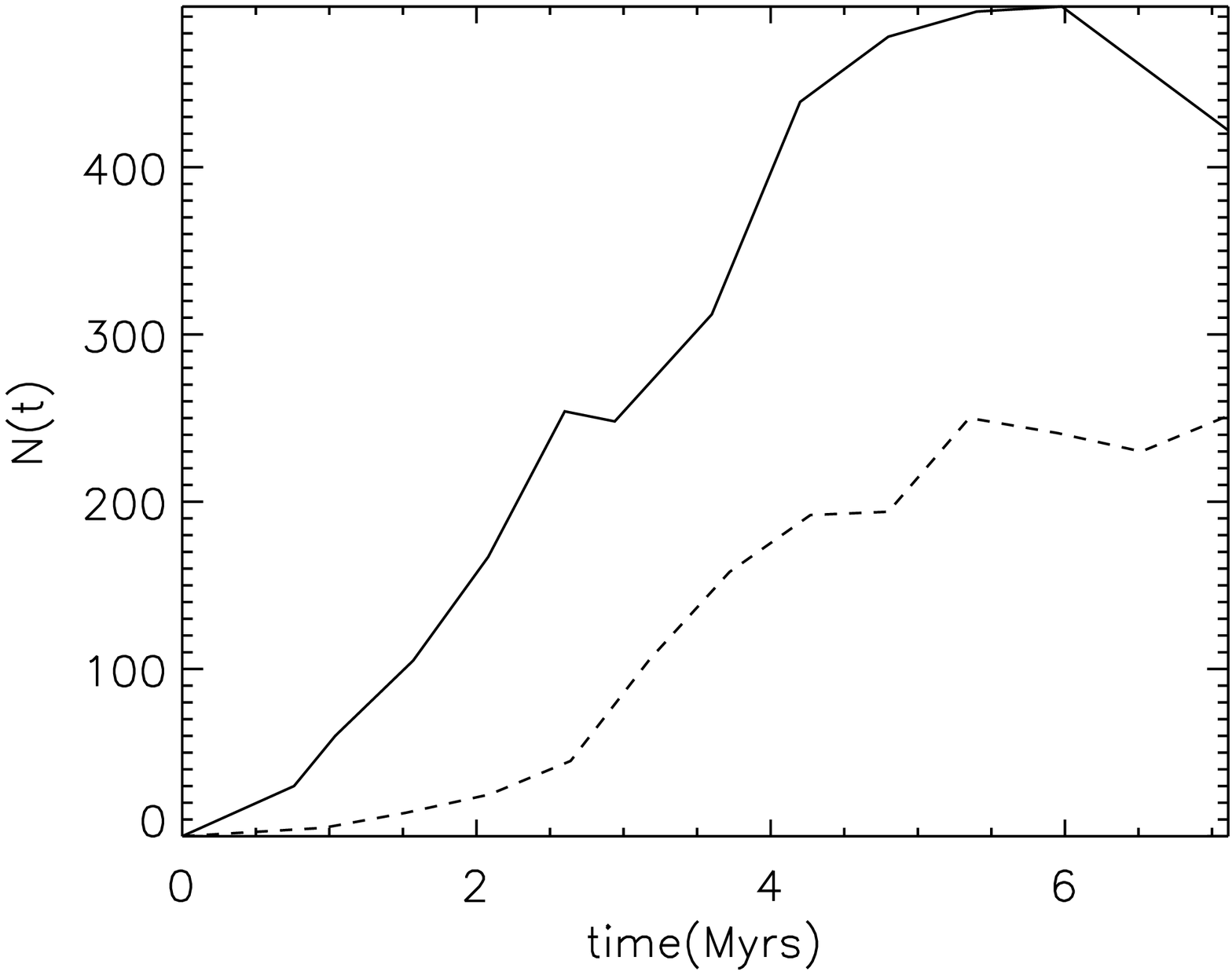}
 \caption{Number of identified clumps with time.  Clumps were only counted if they included more than 16 dense and cold 
         cells.  The solid curve corresponds to the run in a uniform diffuse medium, while the dashed curve to the run
         in a turbulent diffuse medium.}
 \label{number_time} 
\end{figure}

Figure \ref{number_time} shows the total number of identified clumps with time for each simulation.  
At the end of the uniform background simulation, more than 400 clumps have formed.  This is not only caused by the TI creating new clumps, but also by the fragmentation of already existing clumps.


\subsection{Simulations with background turbulence}
\begin{figure*}
    \includegraphics[width=0.5\linewidth]{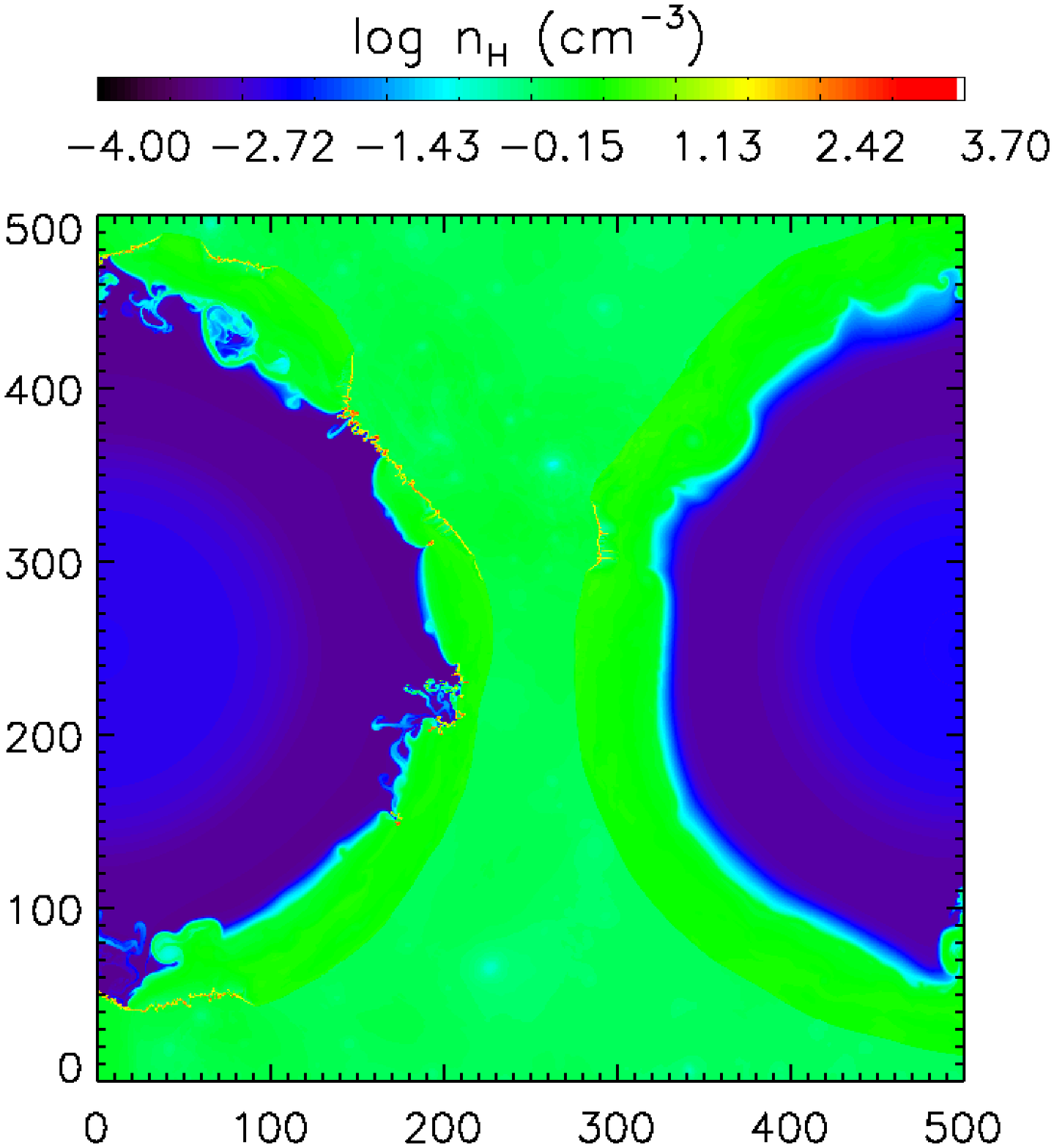} 
    \includegraphics[width=0.5\linewidth]{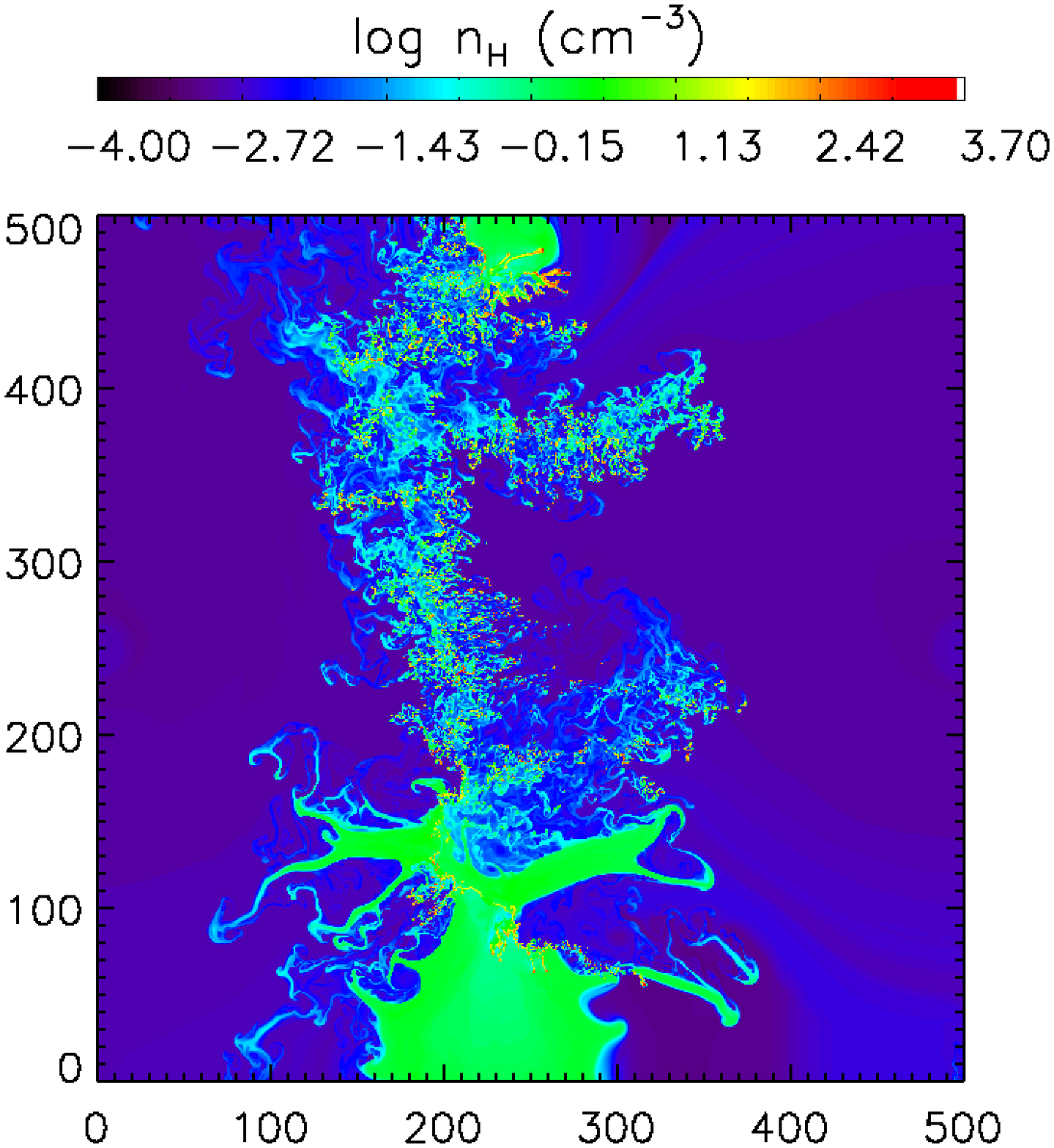} 
    \includegraphics[width=0.5\linewidth]{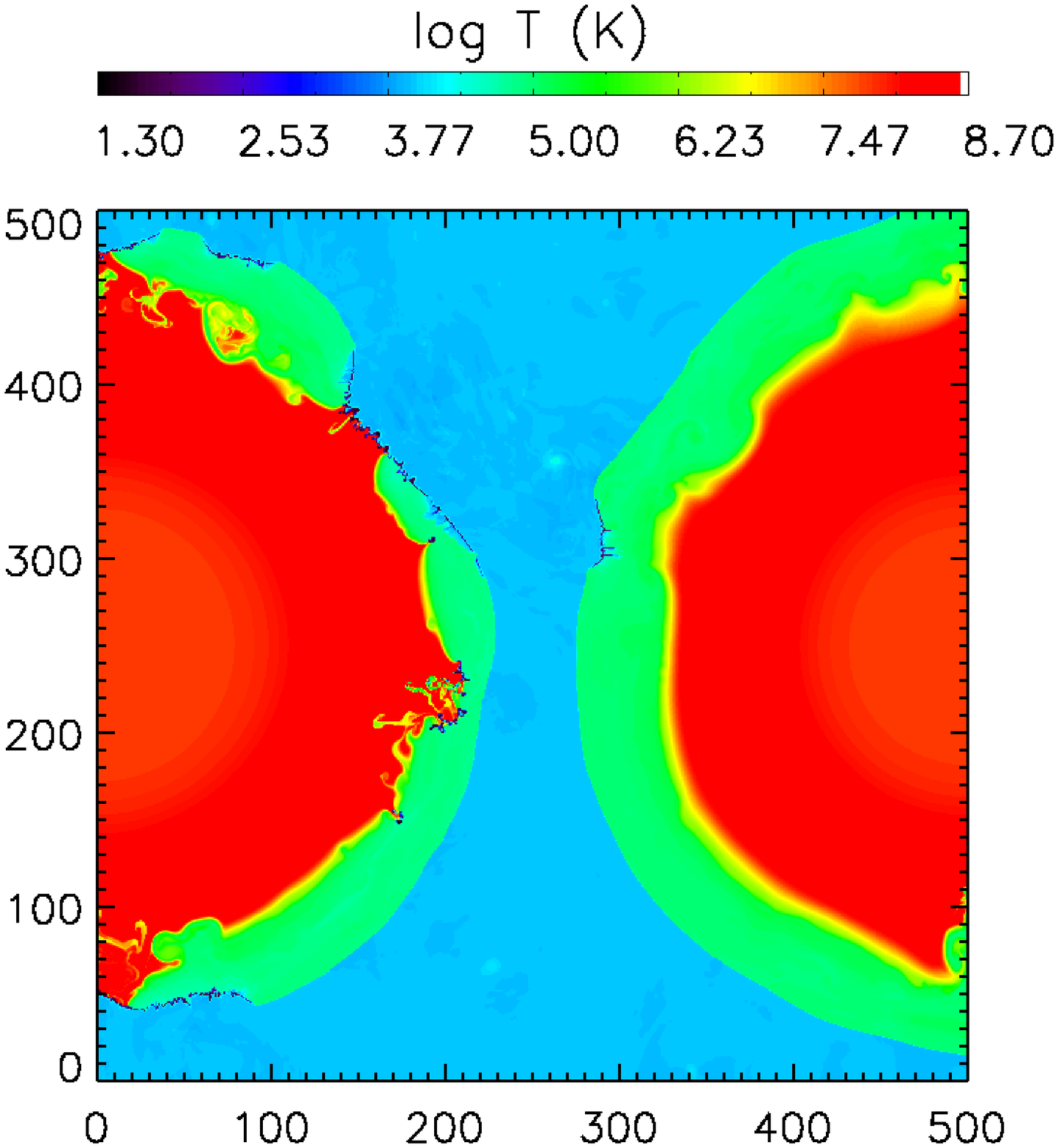} 
    \includegraphics[width=0.5\linewidth]{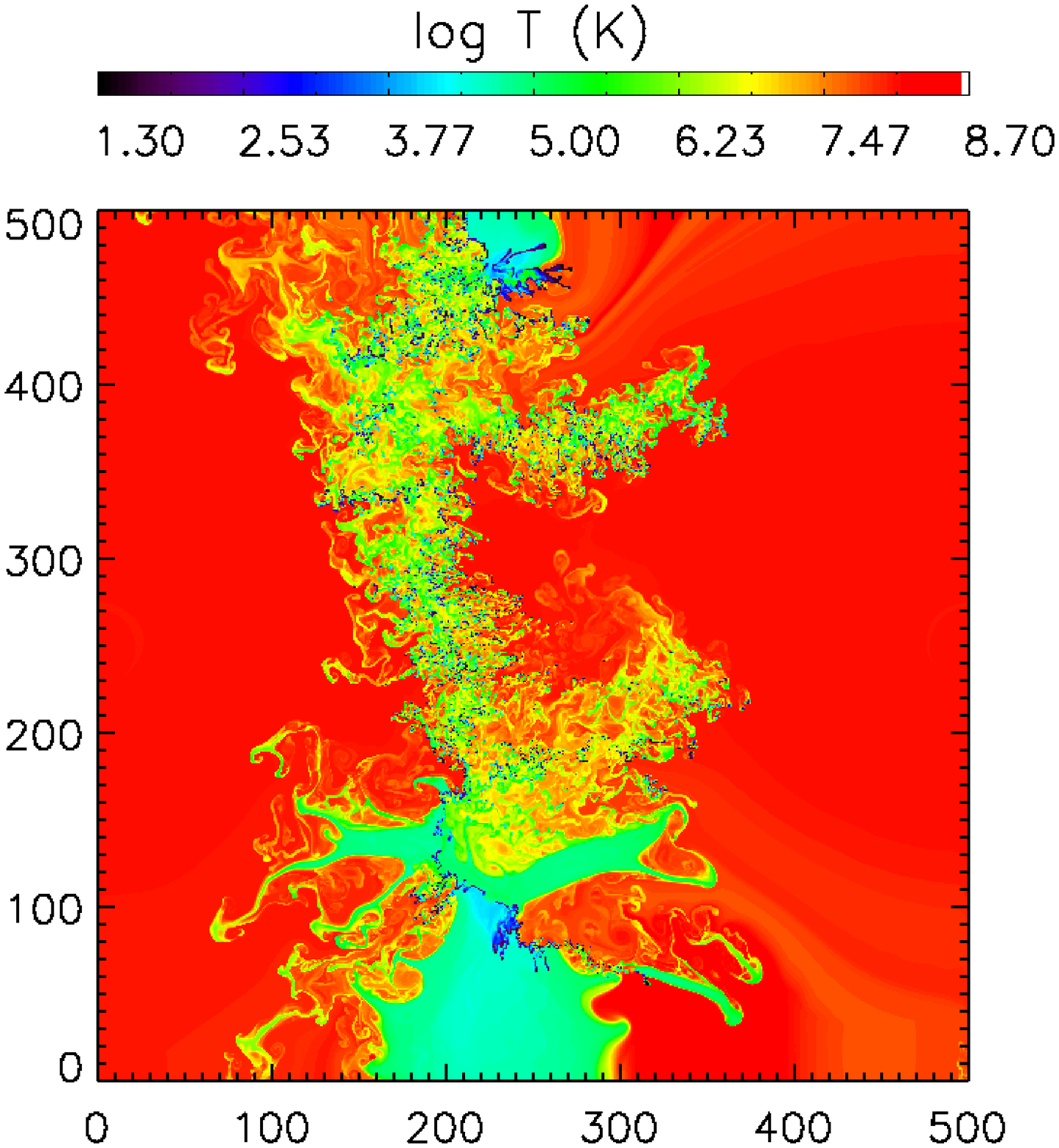}
    \caption{Super-bubble collision snapshots in a turbulent diffuse medium.
Plotted on the top panels is the logarithm of the hydrogen number density in log(cm$^{-3}$)
and on the bottom panels the logarithm of the gas temperature in log(K).
Left: 3 Myrs after star formation, right: 7 Myrs after star formation.}  
   \label{turb}
\end{figure*}

\begin{figure*}
    \includegraphics[width=0.5\linewidth]{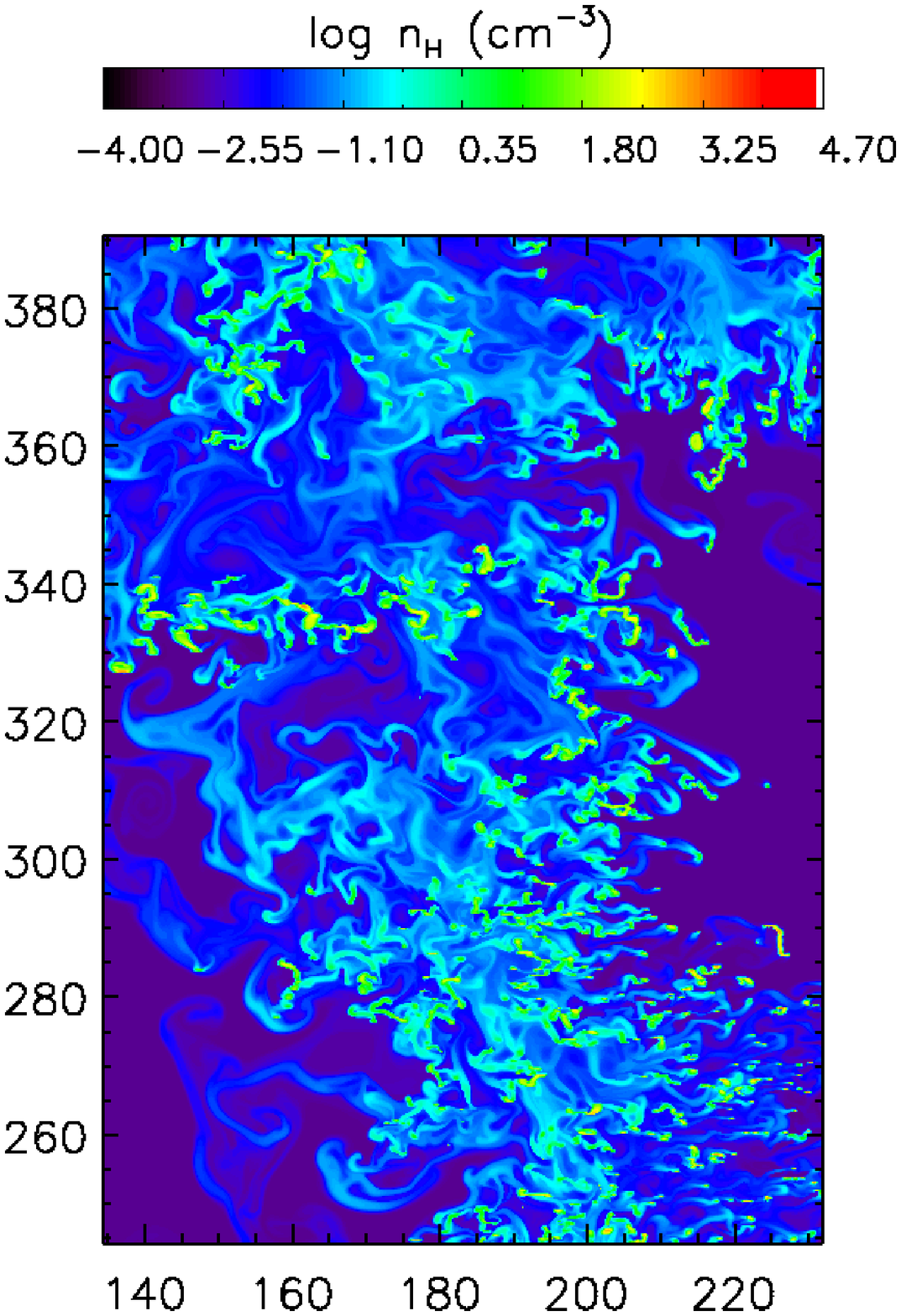} 
    \includegraphics[width=0.5\linewidth]{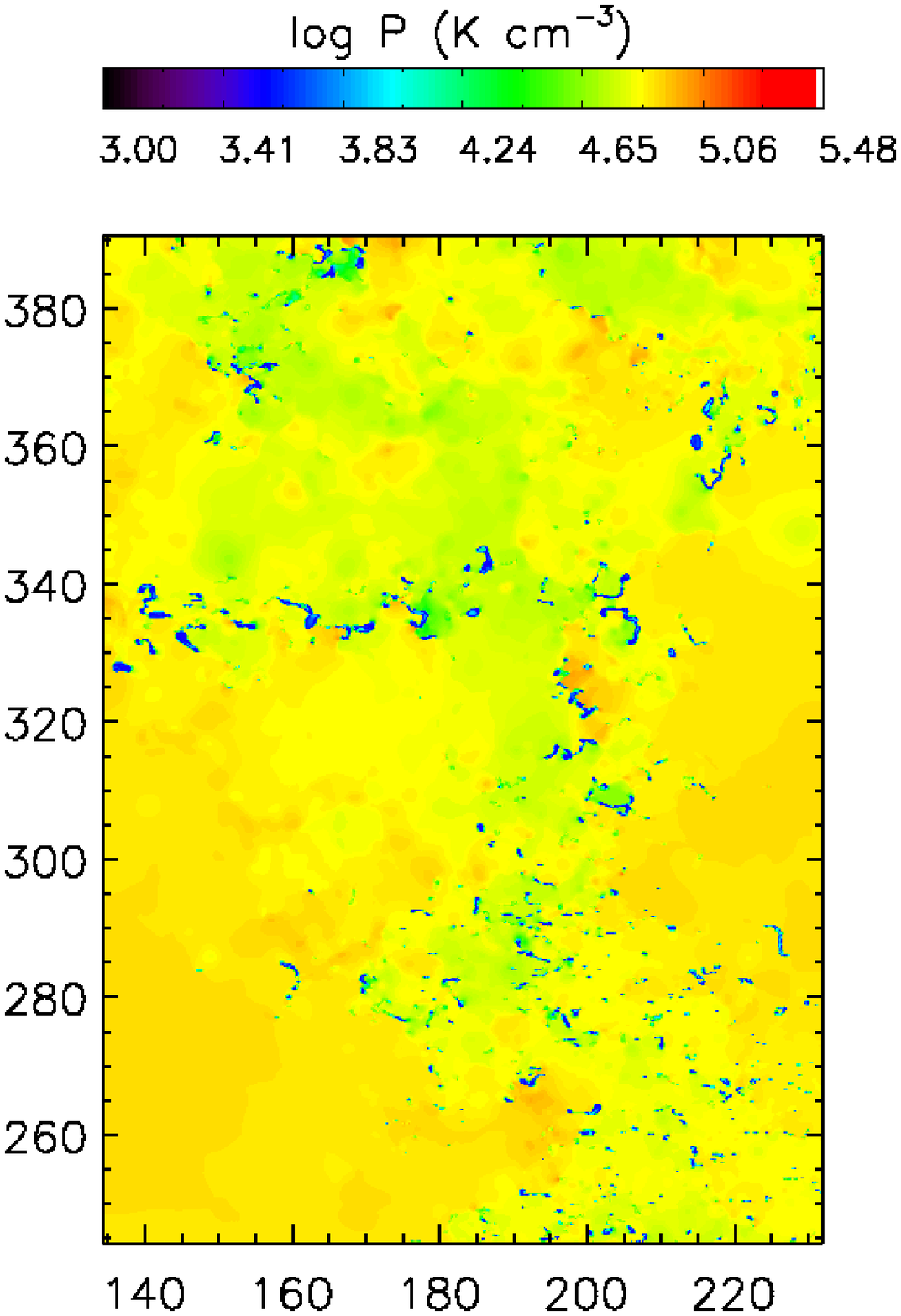}
    \caption{Zoom-in of the last snapshot (7 Myrs after star formation) of the turbulent diffuse medium run.
Plotted on the left is the logarithm of the hydrogen number density in log(cm$^{-3}$)
and on the right the logarithm of the gas pressure in log(K cm$^{-3}$).  The axes coordinates are in parsecs.}
   \label{turb_zoom}
\end{figure*}

As for the uniform medium case presented in the previous section, here as well we observe the formation of cold and dense structures due to the combined action of the NTSI, the Kelvin-Helmholtz and the Thermal Instability.  The main difference in this case is the anisotropy caused by the turbulent density and velocity field to the formation of clumps, as illustrated in figure \ref{turb}. 
The shell on the left-hand side fragments very early on  in some locations, already at less than 1 Myr after star formation, but the shell on the right-hand side fragments much later, at around 3 Myrs after star formation.  Rotating the initial conditions by 180 degrees produces the exactly opposite effect.

This is caused by the difference in the background velocity field and is related to the range in which the Vishniac instability exists \citep{Vishniac_1994}. 
The smallest and largest unstable wavelength both depend on the relative velocity of the shock-bounded slab, in our case the thin expanding shell, and the background medium.  Performing a simple test with a uniform background velocity field indicated that the instability indeed grows first where the relative velocity of the shell with respect to the background medium is largest, in agreement with Vishniac's analysis.

Figure \ref{turb_zoom} shows a zoom-in of the last snapshot, showing a picture very similar to that in figure \ref{noturb_zoom}, where the different gas phases are almost in pressure equilibrium, with clumps located in regions with a pressure almost an order of magnitude lower with respect to their surroundings.

\begin{figure}
  \plotone{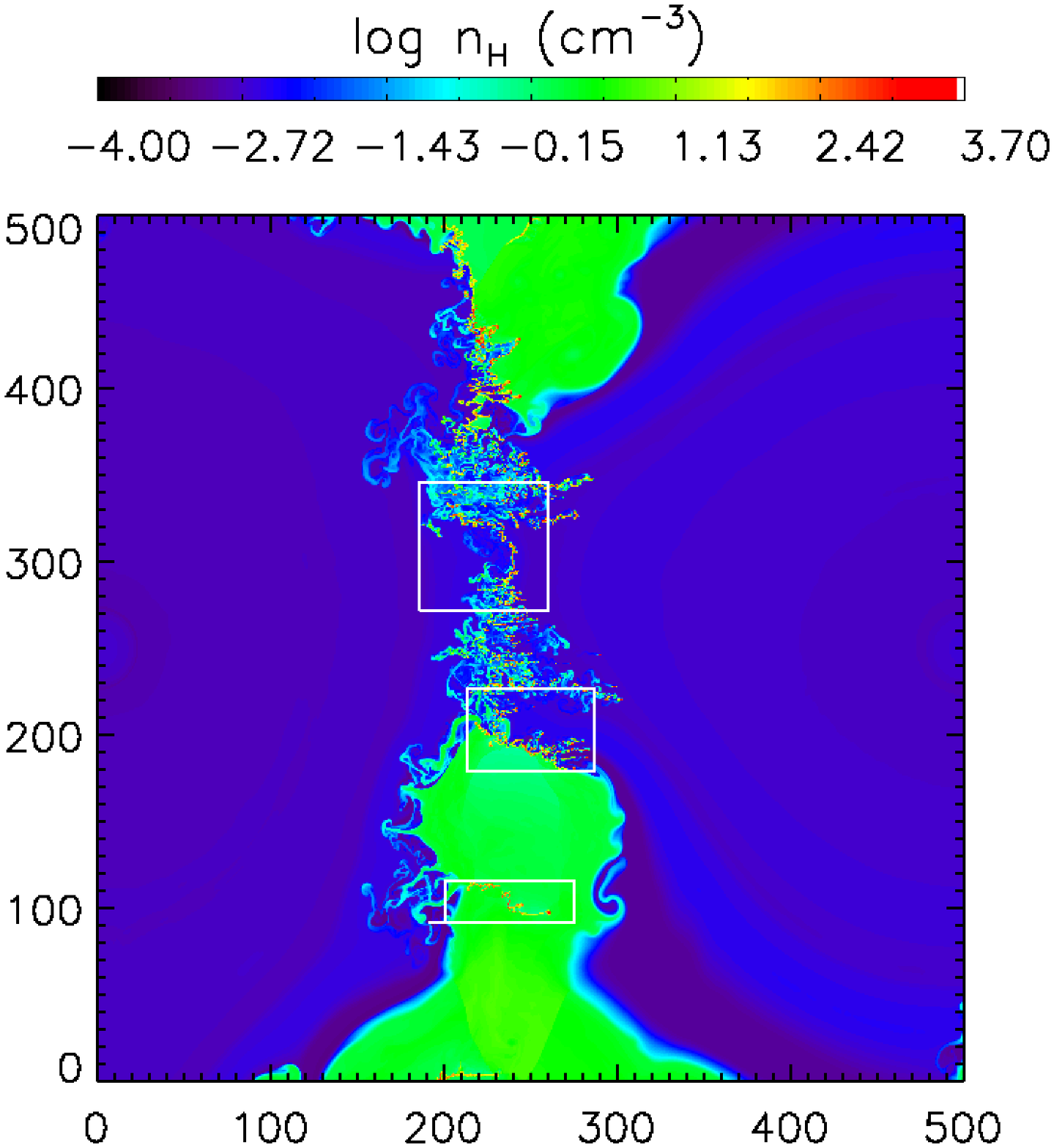}
  \caption{Simulation snapshot at t=5.3 Myrs after star formation.  Plotted here is the logarithm of the hydrogen number density.
      The axes are marked in parsecs.
    The black rectangleds show the positions of the filaments shown in figures \ref{fil1}, \ref{fil2} and the 5.3 Myrs snapshot of figure \ref{fil_time}}
  \label{5myrs_filament}
\end{figure}

\begin{figure}
  \plotone{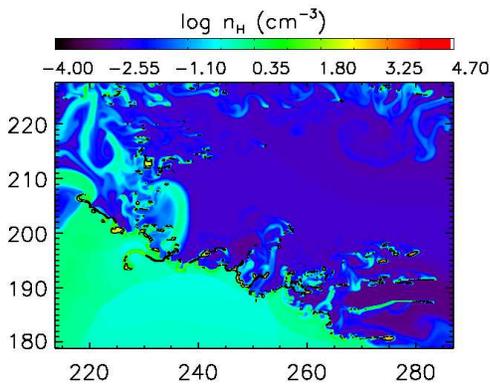}
  \caption{A large filament containing several smaller clumps. 
   Plotted here is the logarithm of the hydrogen number density in log(cm$^{-3}$).  
   The snapshot corresponds to the white box centered at x=250 pc, y=200 pc in the t=5.3 Myrs snapshot shown in figure \ref{5myrs_filament}.
   The axes are marked in parsecs from the axes origin.  The black contour corresponds to $n_H=50cm^{-3}$}
  \label{fil1}
\end{figure}

\begin{figure}
  \plotone{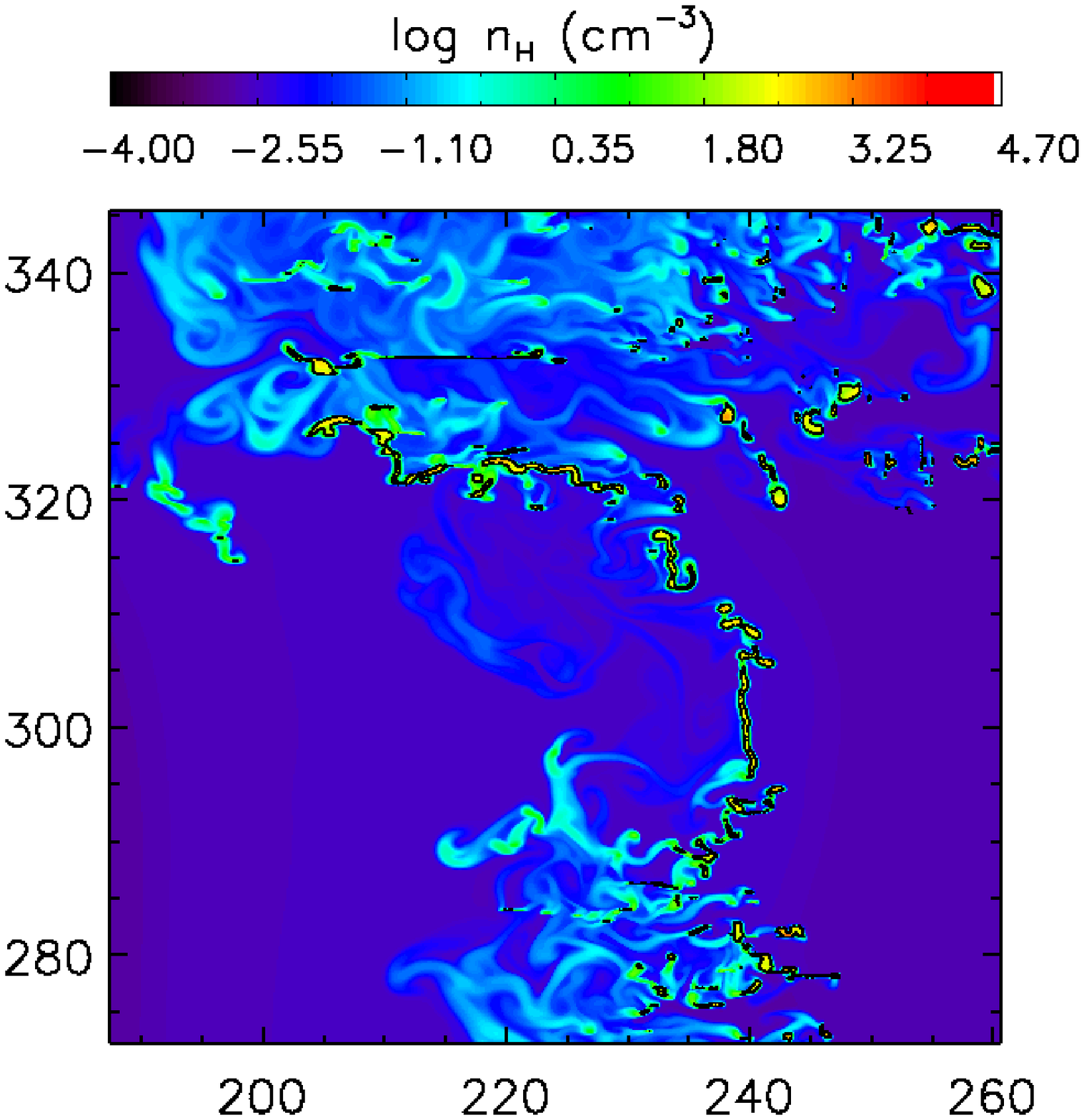}
  \caption{As in figure \ref{fil1}.
    The snapshot corresponds to the white box centered at x=225 pc, y=314 pc in the t=5.3 Myrs snapshot  shown in figure \ref{5myrs_filament}.}
  \label{fil2}
\end{figure}

\begin{figure*}
    \includegraphics[width=0.5\linewidth]{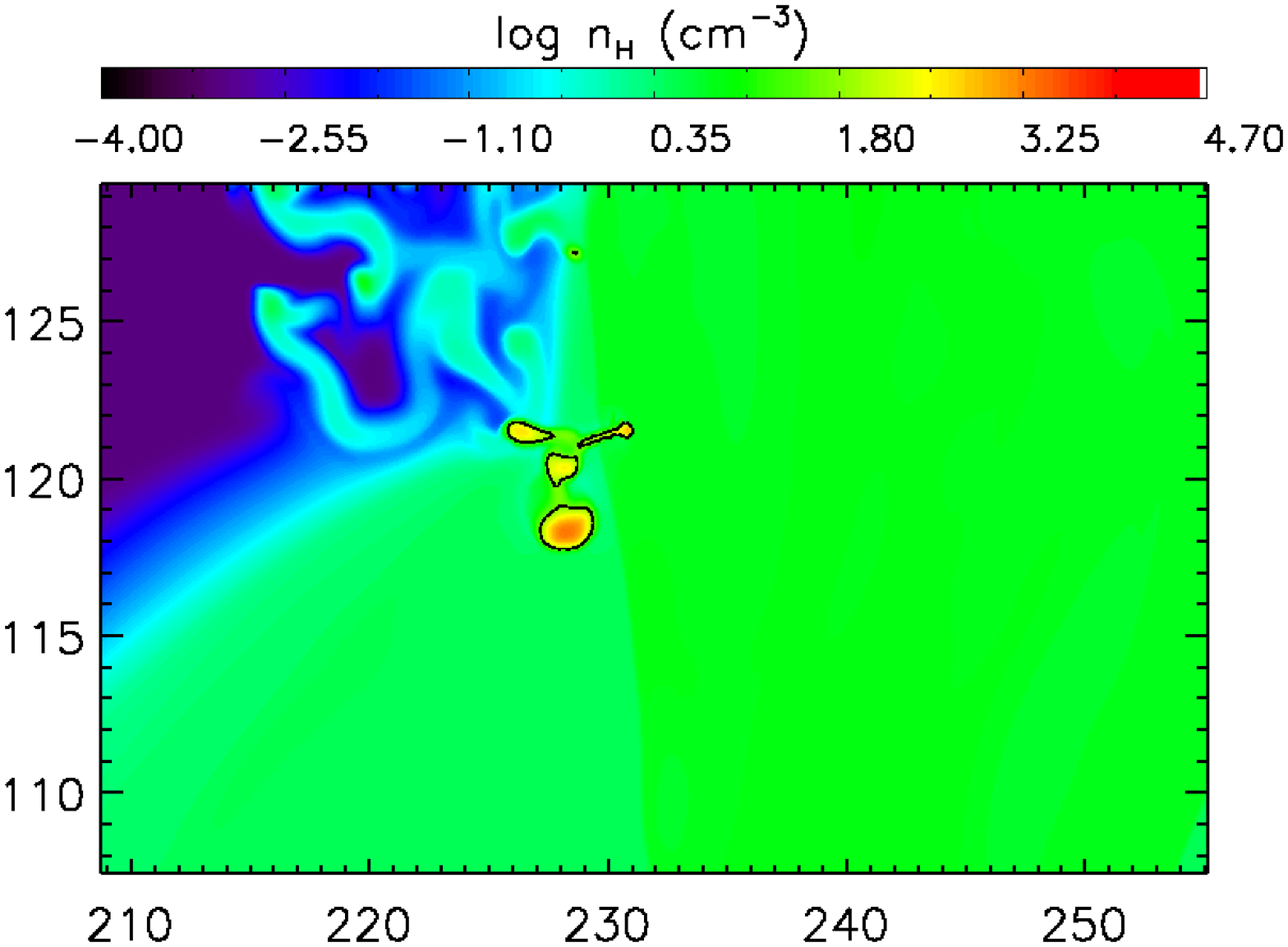} 
    \includegraphics[width=0.5\linewidth]{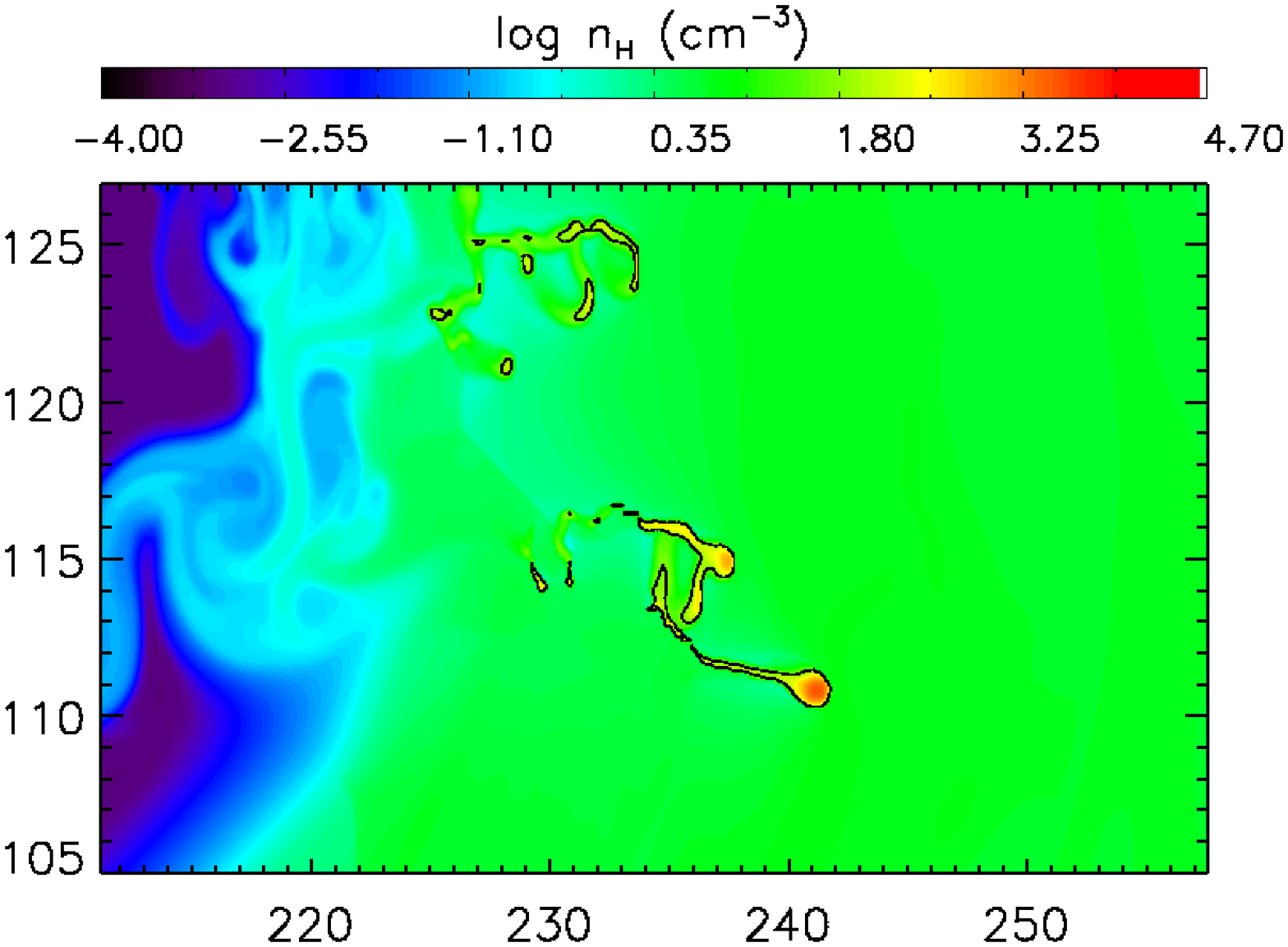} 
    \includegraphics[width=0.5\linewidth]{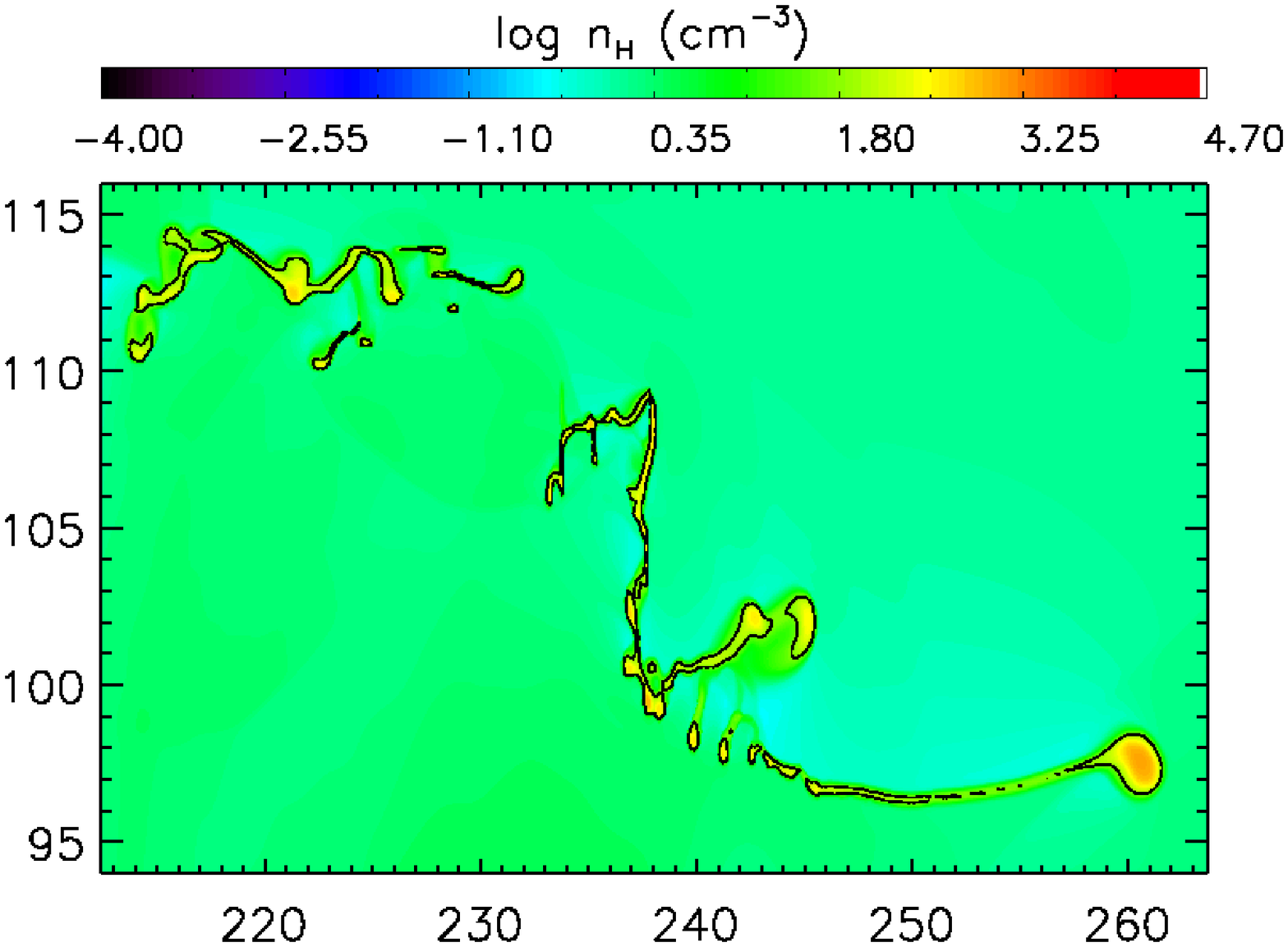} 
    \includegraphics[width=0.5\linewidth]{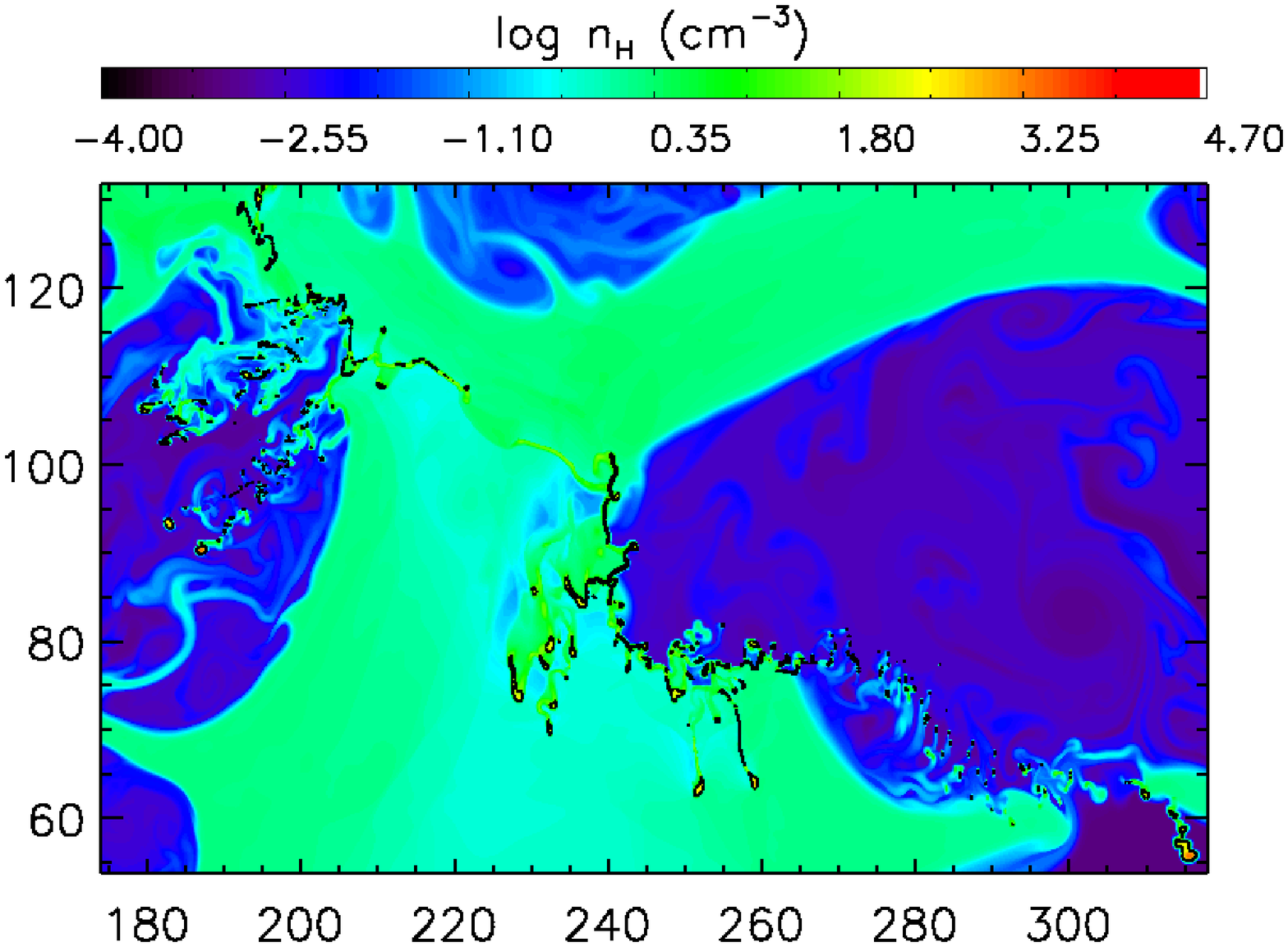}
    \caption{Time evolution of a single filament.  The plots show logarithm of hydrogen number density.
    The black contour shows the level $n_H=50$ cm$^{-3}$.  
    From top left to bottom right: 4.3, 4.6, 5.3 and 7 Myrs after star formation. Note the change in scale between the snapshots.}
   \label{fil_time}
\end{figure*}

After the NTSI starts to grow, the condensed material immediately becomes thermally unstable and we see the formation of cold and dense clumps, as in the case of the uniform medium.  However, due to the background velocity field, as explained above, there are now regions where the NTSI has a faster growth rate with respect to that of a static medium.  This leads to a larger Kelvin-Helmholtz shear, which is the reason why in this simulation longer filaments are formed compared to the uniform background medium simulation.
Some of these structures are illustrated in figures \ref{fil1}, \ref{fil2} and \ref{fil_time}.  These figures are zoomed-in fractions of the full domain, shown in figure \ref{5myrs_filament}. 

Following these filaments from their formation to the end of the simulation, we observe that they increasingly become more elongated and they constantly fragment into smaller clumps.  Figure \ref{fil_time} shows the time evolution of one of the filaments.  Before the shells collide, the structure is a small clump of cold gas.  During the collision, this clump gets caught in a large-scale shear and becomes more and more elongated until, at the end of the simulation, it has reached a total length of about 100 pc.  Note that during the bubble expansion the clump is not located inside the hot bubble, but rather at the edge of the shock.  

The anisotropy in the growth rate of the NTSI leads to the formation of much less clumps in this simulation with respect to the uniform background simulation, as illustrated in figure \ref{number_time}.

\begin{figure}
    \plotone{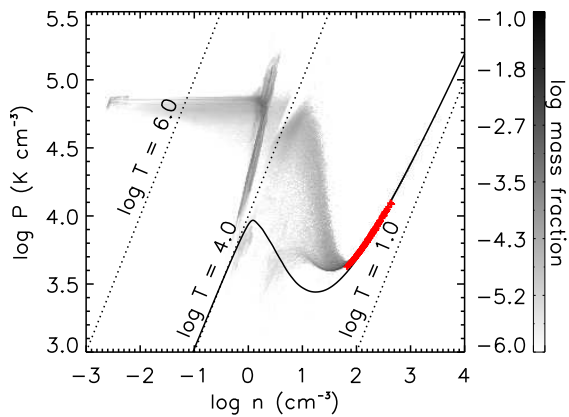}
    \caption{Pressure versus hydrogen density of the fluid in the box for the turbulent medium run.
The plot is from the last snapshot, 7 Myrs after star formation.
The gas has been binned and color-coded according to the mass fraction it represents.
The solid line is the cooling-heating equilibrium curve for the warm and cold gas
and the dashed lines show the locations of three isotherms.
Red crosses are average clump quantities.}
   \label{phases_turb}
\end{figure}

\begin{figure}
   \plotone{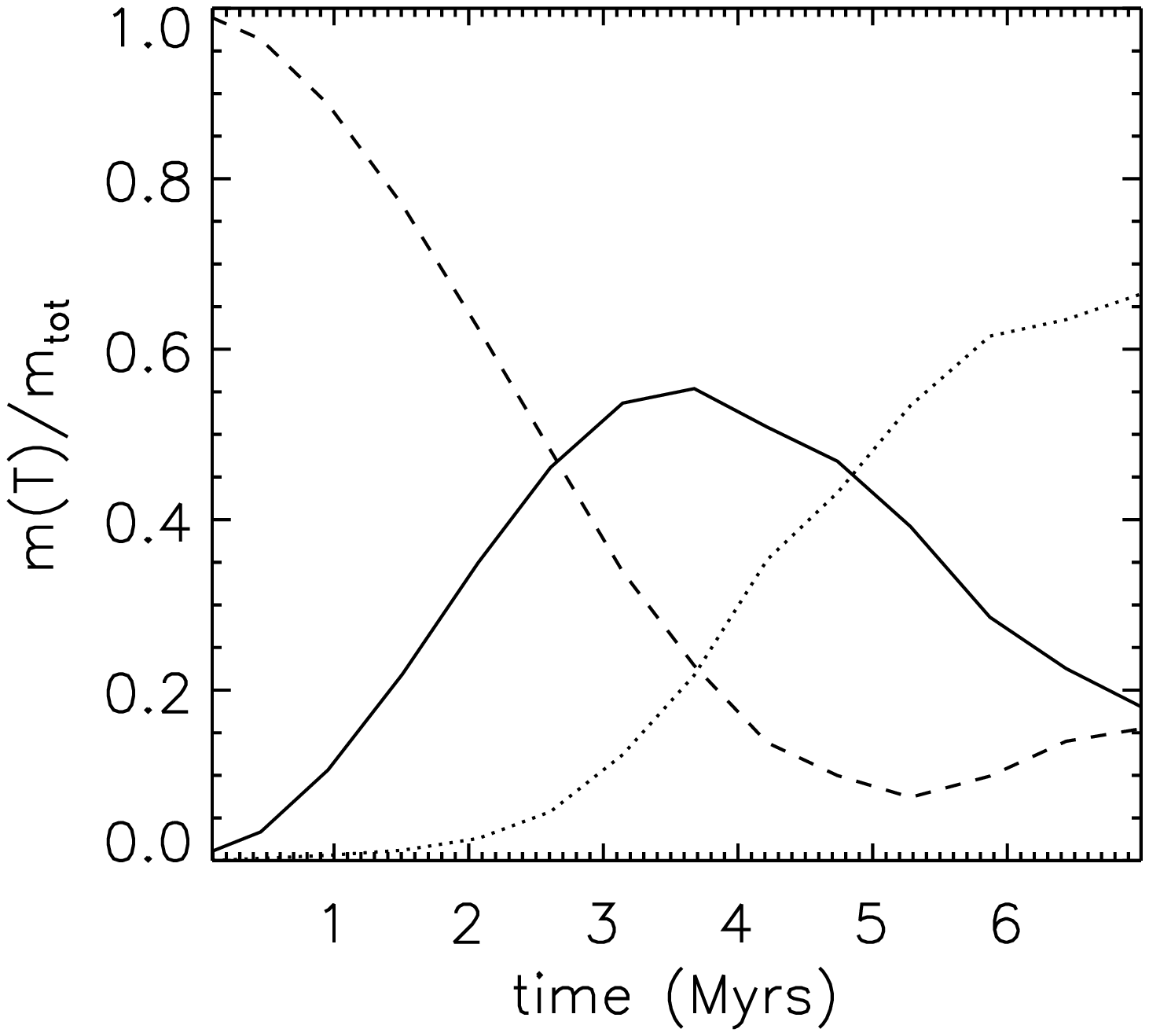}
   \caption{Gas fractions with time for the turbulent medium run.  
   The dashed curve shows the mass fraction in the warm gas phase, the 
   solid curves shows the mass fraction in the hot gas phase and
   the dotted curve the mass fraction in the cold phase.
     (Hot phase: gas with $T\ge25000K$, warm phase: gas with $100K<T<25000K$, cold phase: gas with $T\le100K$)}
   \label{gas_frac_turb}
\end{figure}

Figure \ref{phases_turb} shows the phase diagram for the gas in this simulation, in the same way as figure \ref{phases_noturb}
We note two main differences with respect to the uniform background medium run.  First, that the pressure at low densities (n $<$ 0.1 cm$^{-3}$) is higher than in the uniform medium run.  This effect can be attributed to the additional compression from the turbulence.  
Second, we note that the clumps in this simulation are at not only at lower pressures with respect to the hot medium than the clumps in the uniform background run, but are also at lower absolute pressures than those clumps.  
This can be explained by the formation mechanism of the clumps in these simulations.  Structures created by the Thermal Instability have lower pressures than their surroundings.  When Thermal Instability has stopped acting on them, they move towards pressure equilibrium with the surrounding medium in approximately a sound crossing time, $c_s/L$, where $c_s$ the sound speed and L the typical size of these clumps.    For a typical density of $10^{-21}$ g/cm$^{3}$ and a typical temperature
of 50 K, the sound speed in these clumps is about 0.8 km/sec.  At this velocity sound waves will cross a 2 pc long clump in about 0.6 Myrs. 
In the turbulent diffuse background run, clumps are formed later than in the uniform background run, which means they are less evolved and farther from equilibrium.
According to the above calculations, the time difference between clump formation between the two simulations is about 3 sound crossing times, which is enough for many the clumps in the uniform diffuse background simulation to have reached approximate pressure equilibrium with the hot medium.

Figure \ref{gas_frac_turb} shows the evolution of the mass fractions of the gas in different temperature regimes. 
Unlike the uniform medium run, in this simulation there is a maximum in the hot gas mass fraction at about 3.5 Myrs after star formation and a later minimum
of the warm gas fraction at around 5 Myrs after star formation.  
This can be explained by the delayed formation of cold gas in this simulation.
Until the cold gas is formed, the inserted hot gas just compresses the warm gas, so the hot gas mass fraction increases, while the warm gas mass fraction decreases.
When cold gas starts to form, it quickly dominates the mass, causing the warm and hot mass fractions to drop.


\subsection{Clump properties}\label{clouds}

As shown in figure \ref{number_time}, we form hundreds of clumps in each snapshot of the simulations.  This makes it very difficult to study each of them in detail, especially since they are constantly merging and splitting.  However, we can make some general comments on the morphology of the identified clumps and look into some of their properties.

\begin{figure*}
  \includegraphics[width=0.5\linewidth]{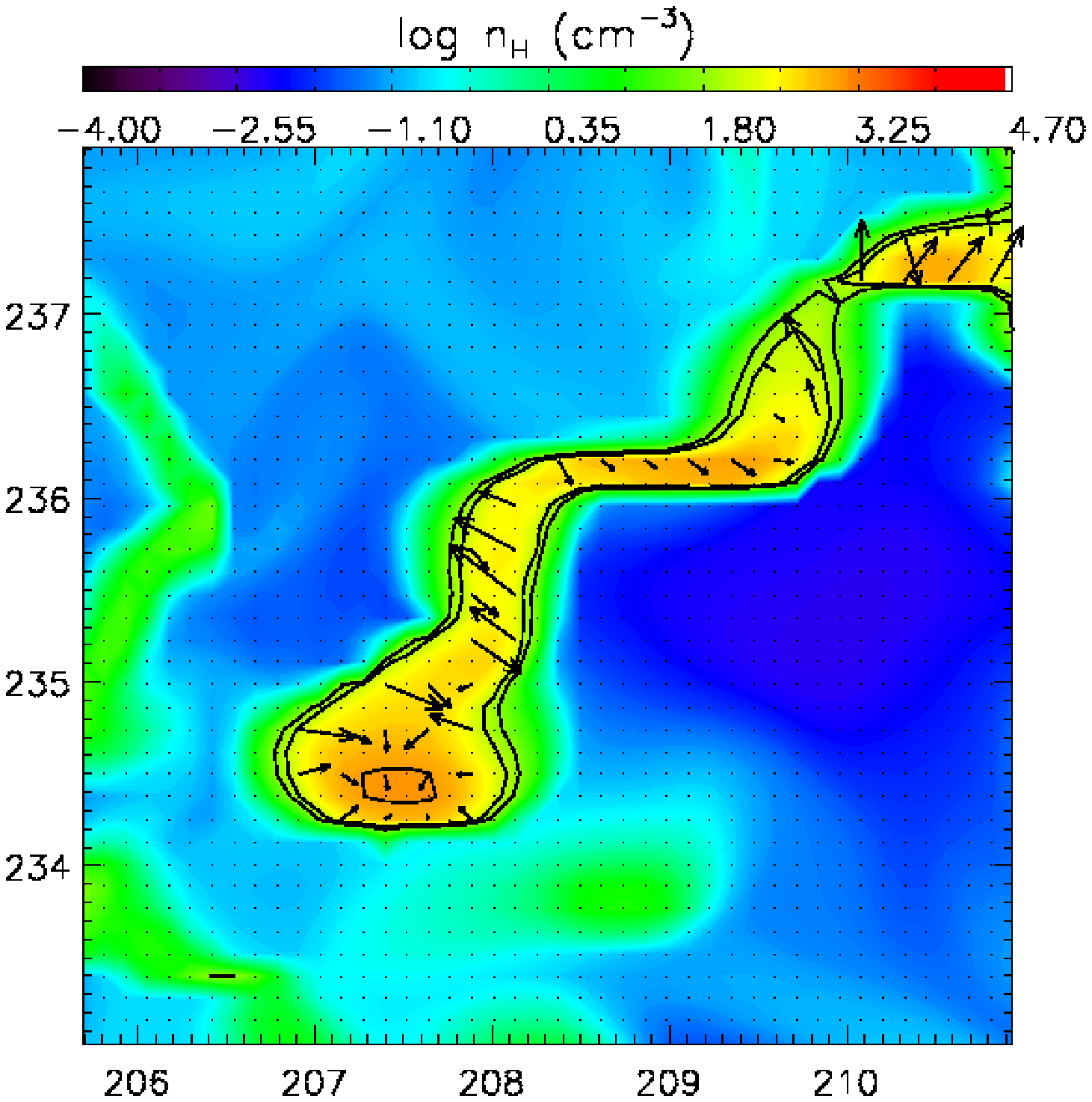} 
  \includegraphics[width=0.5\linewidth]{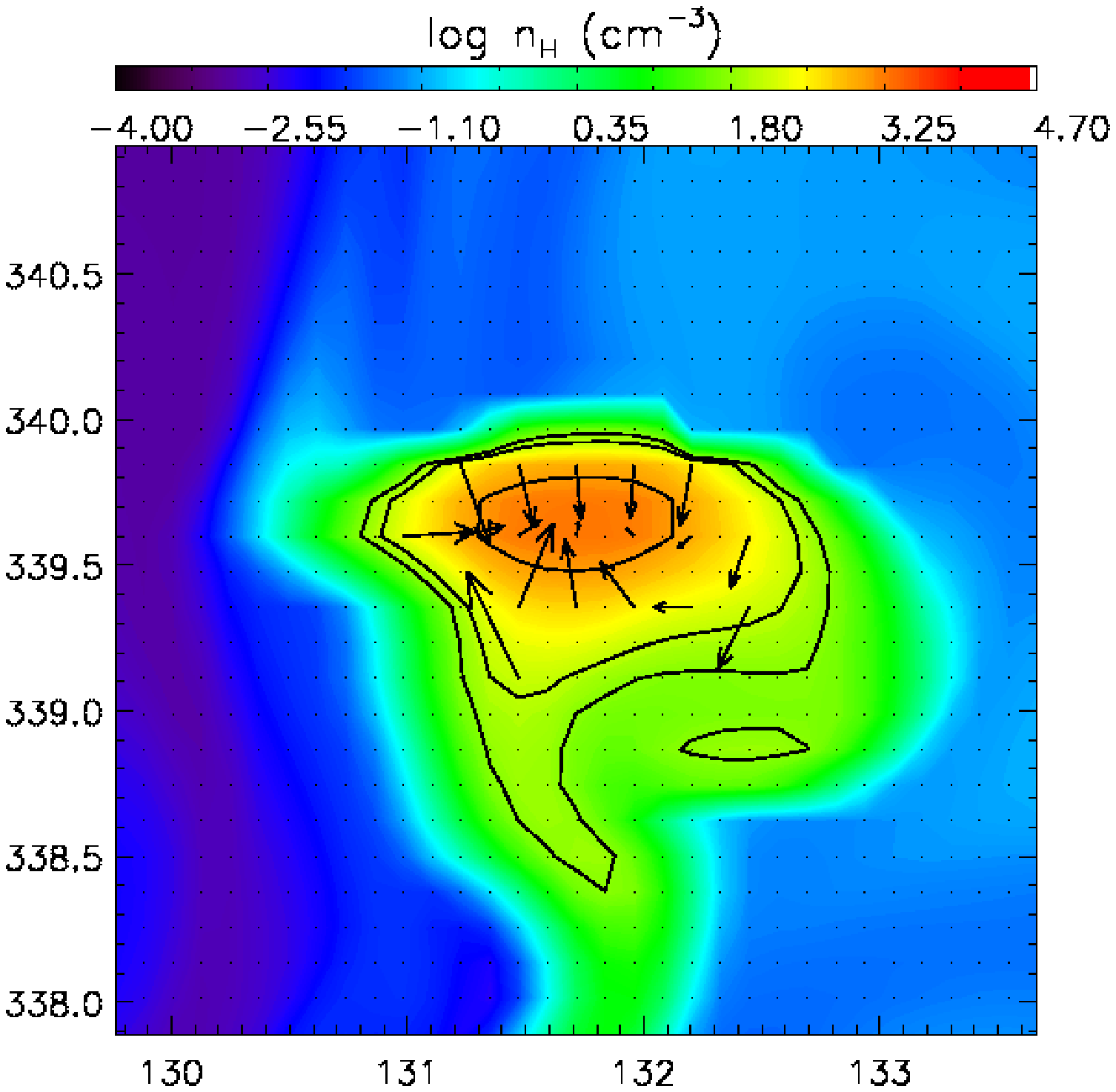} 
  \includegraphics[width=0.5\linewidth]{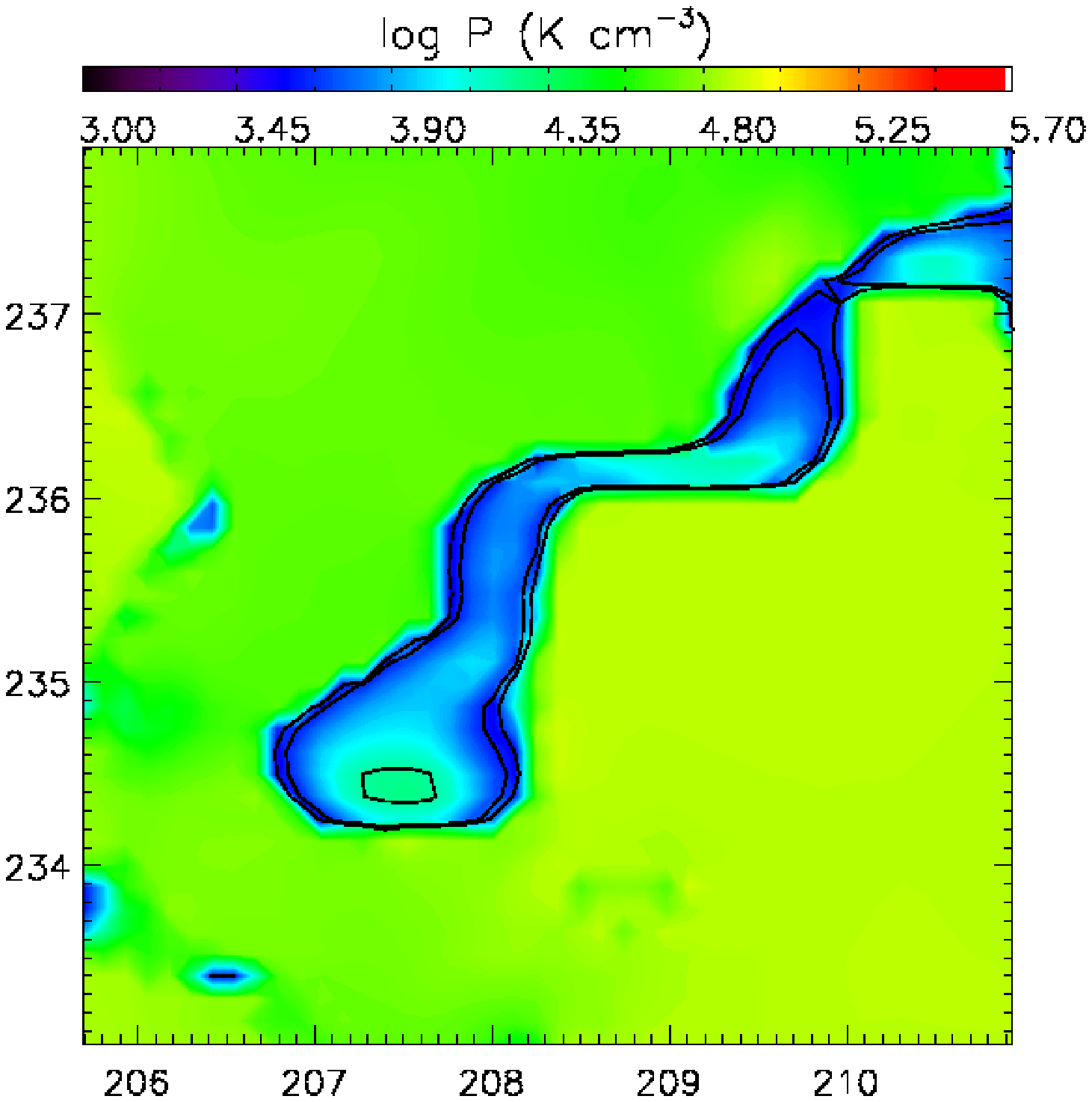} 
  \includegraphics[width=0.5\linewidth]{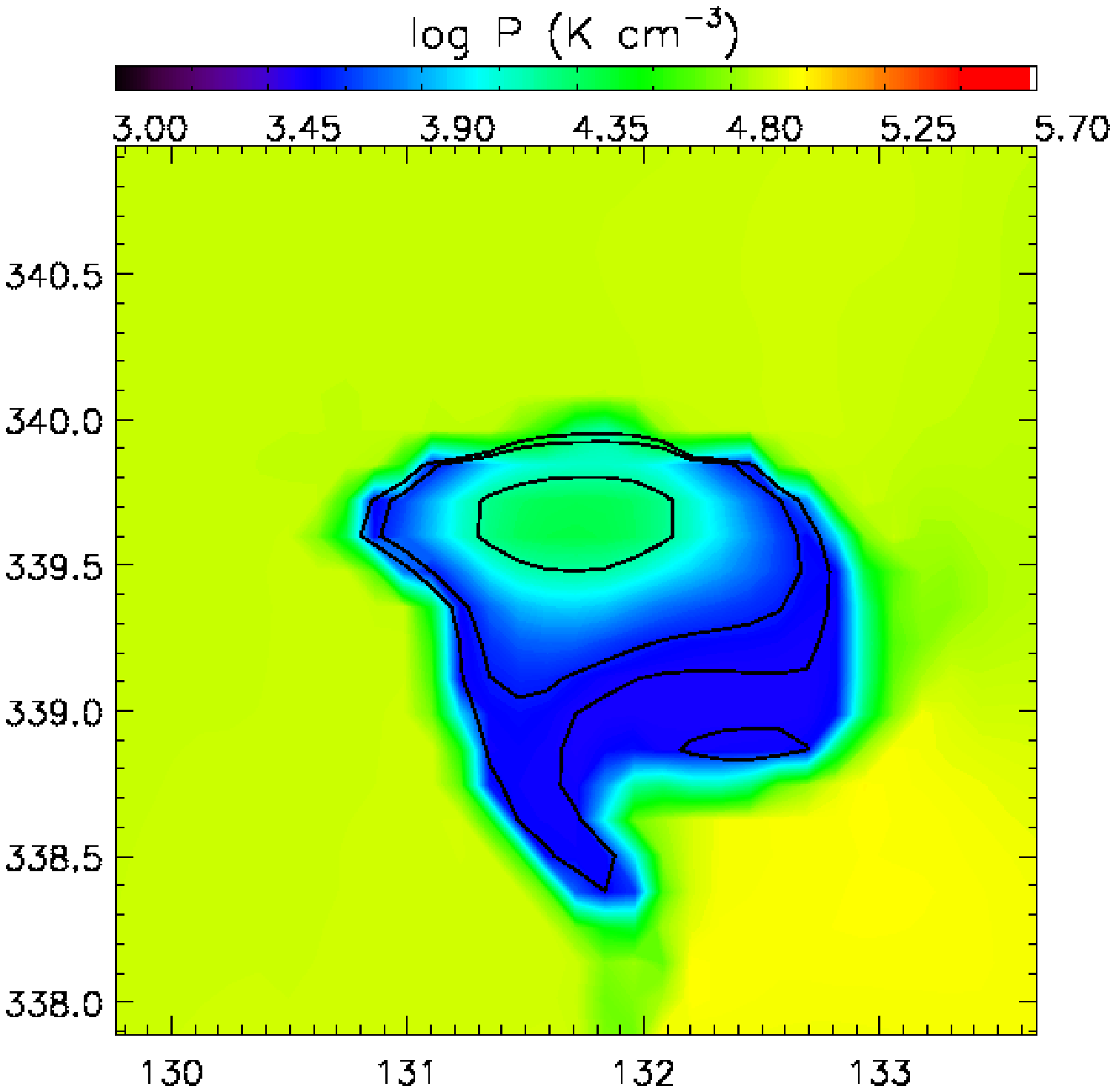}
  \caption{Two examples of condensing clouds.
Top panels show the logarithm of hydrogen number density in log(cm$^{-3}$) and bottom panels show the logarithm of thermal pressure in log(K cm$^{-3}$).
The black arrows show the velocity field with the mean velocity of the central clump subtracted.
Overplotted in black are the contour levels for $n_H$ equal to 50 $cm^{-3}$, 100 $cm^{-3}$ and 1000 $cm^{-3}$. }
  \label{clouds_collapse}
\end{figure*}

\begin{figure*}
  \includegraphics[width=0.5\linewidth]{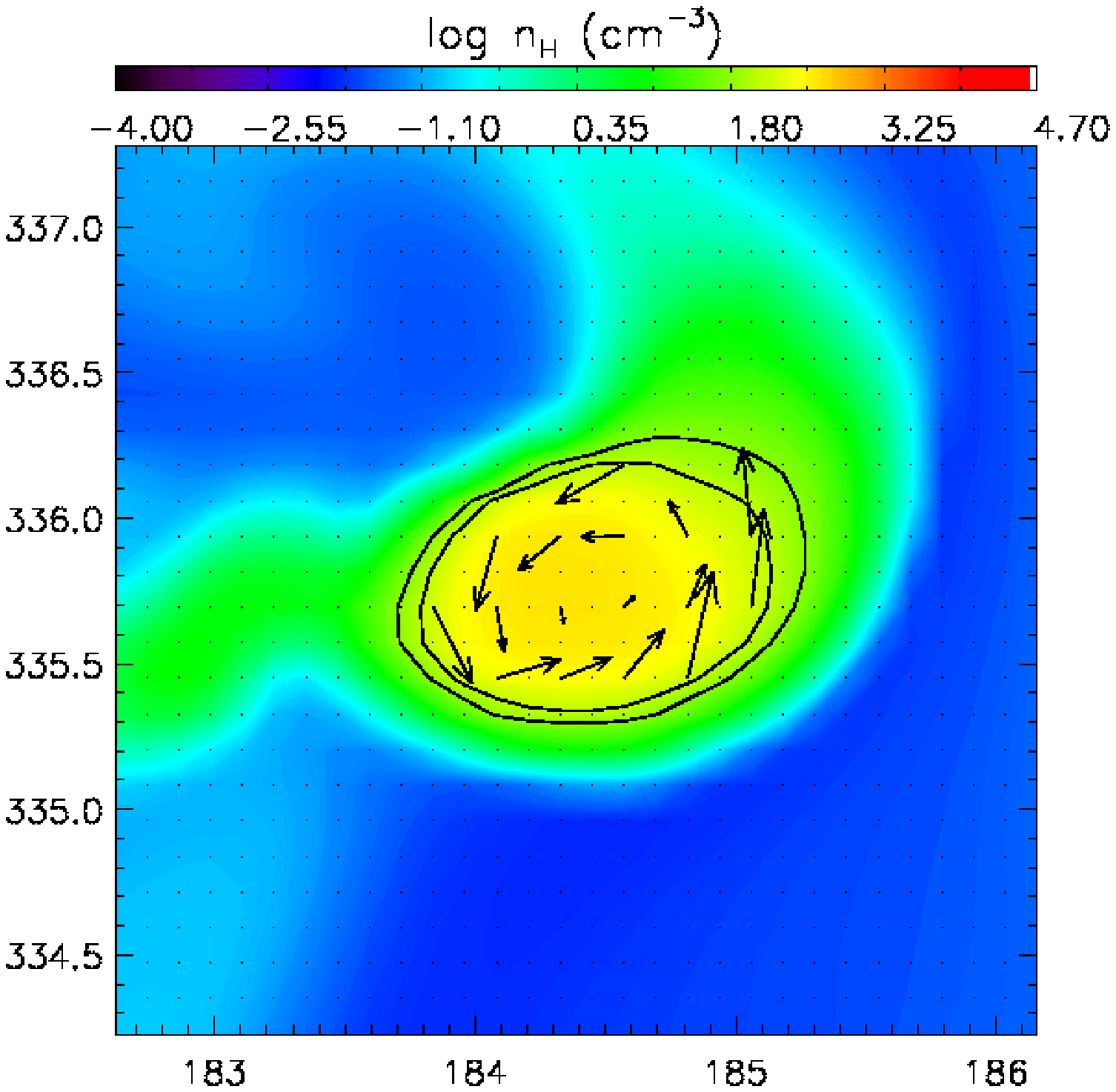} 
  \includegraphics[width=0.5\linewidth]{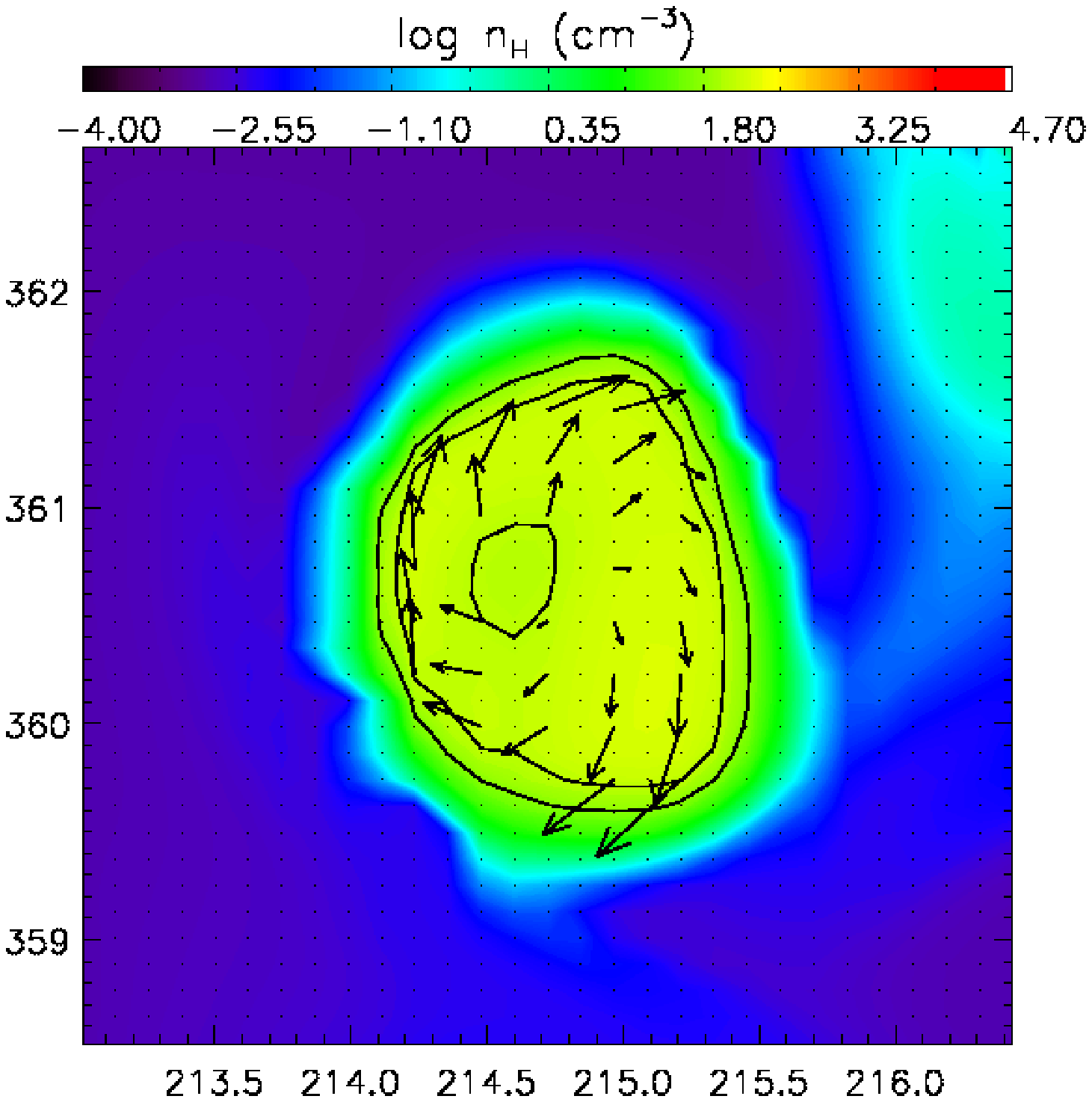} 
  \includegraphics[width=0.5\linewidth]{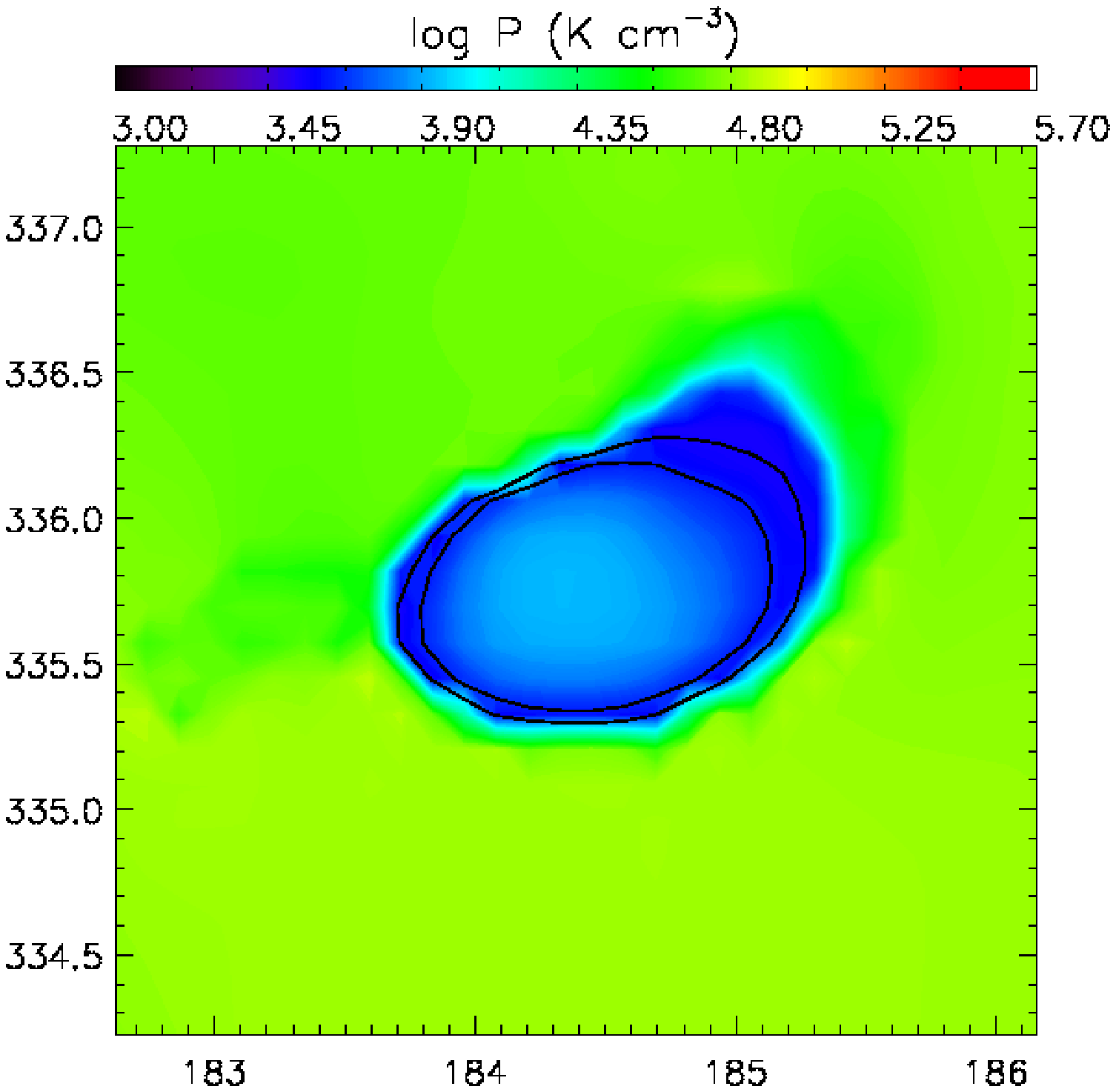} 
  \includegraphics[width=0.5\linewidth]{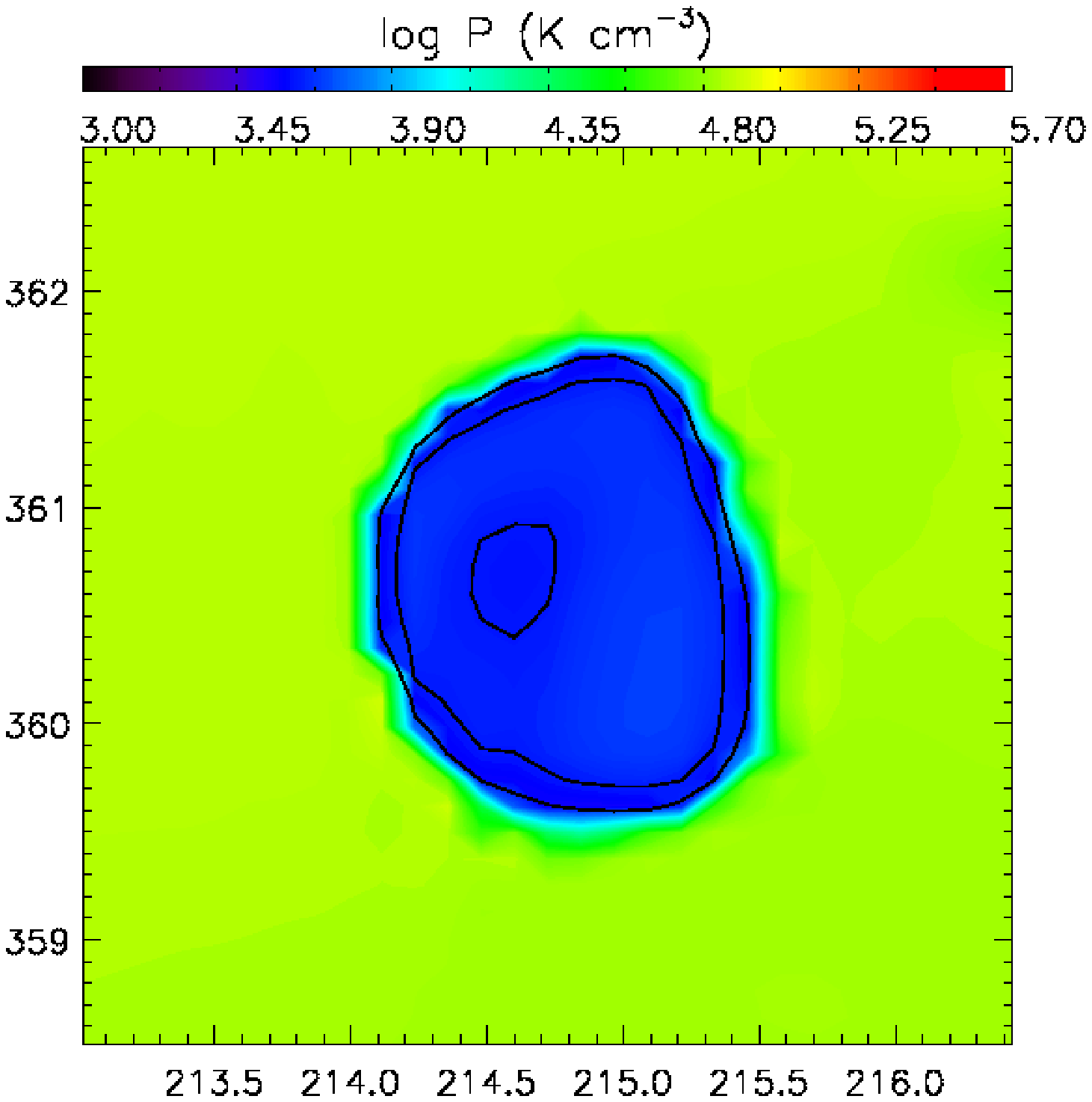}
  \caption{Two examples of rotating clumps.
Top panels show the logarithm of hydrogen number density in log(cm$^{-3}$) and bottom panels show the logarithm of thermal pressure in log(K cm$^{-3}$).
The black arrows show the velocity field with the mean velocity of the central clump subtracted.
Overplotted in black are the contour levels for $n_H$ equal to 50 $cm^{-3}$, 100 $cm^{-3}$ and 1000 $cm^{-3}$. }
\label{clouds_rot}
\end{figure*}

\begin{figure*}
  \includegraphics[width=0.5\linewidth]{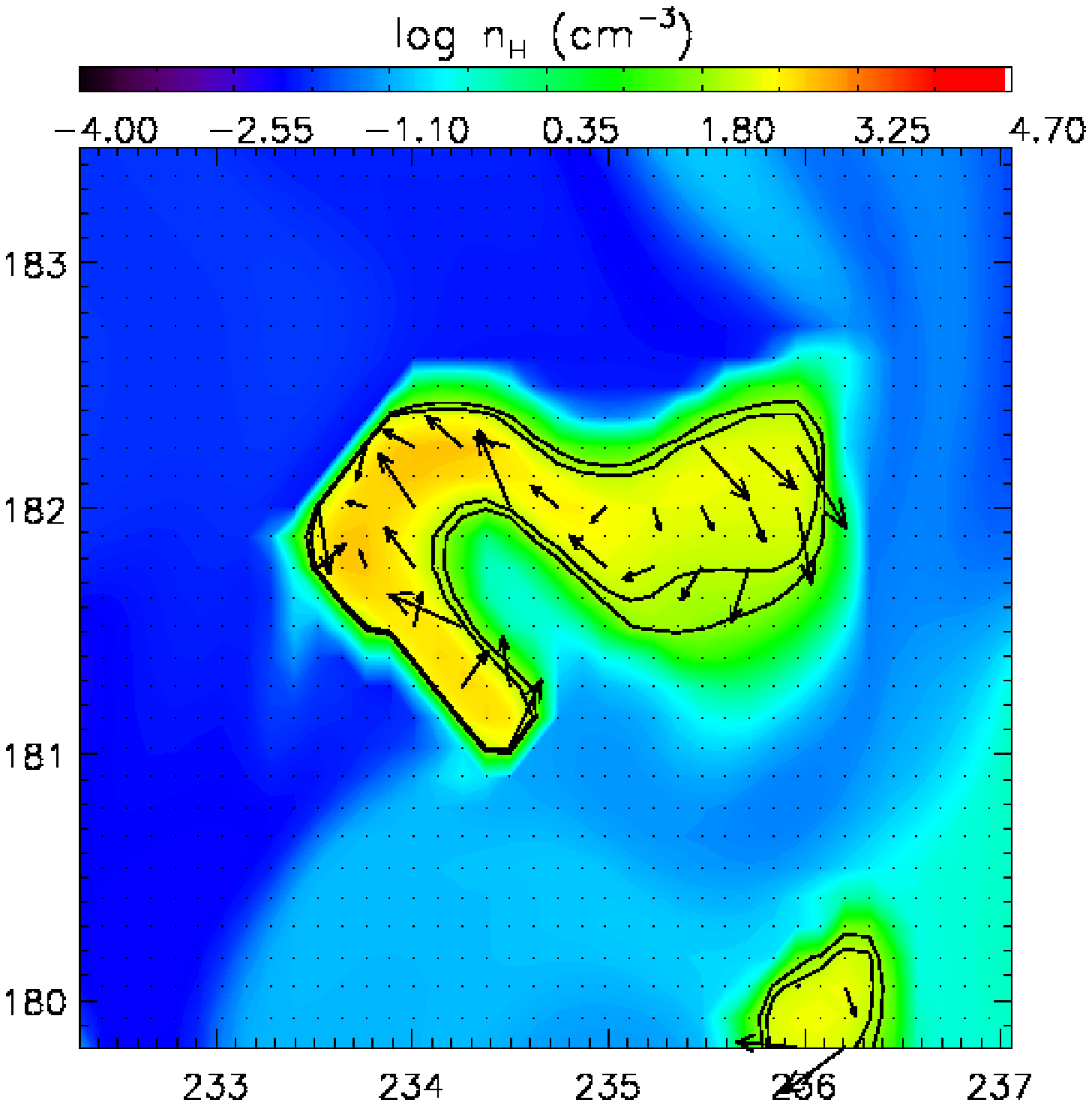} 
  \includegraphics[width=0.5\linewidth]{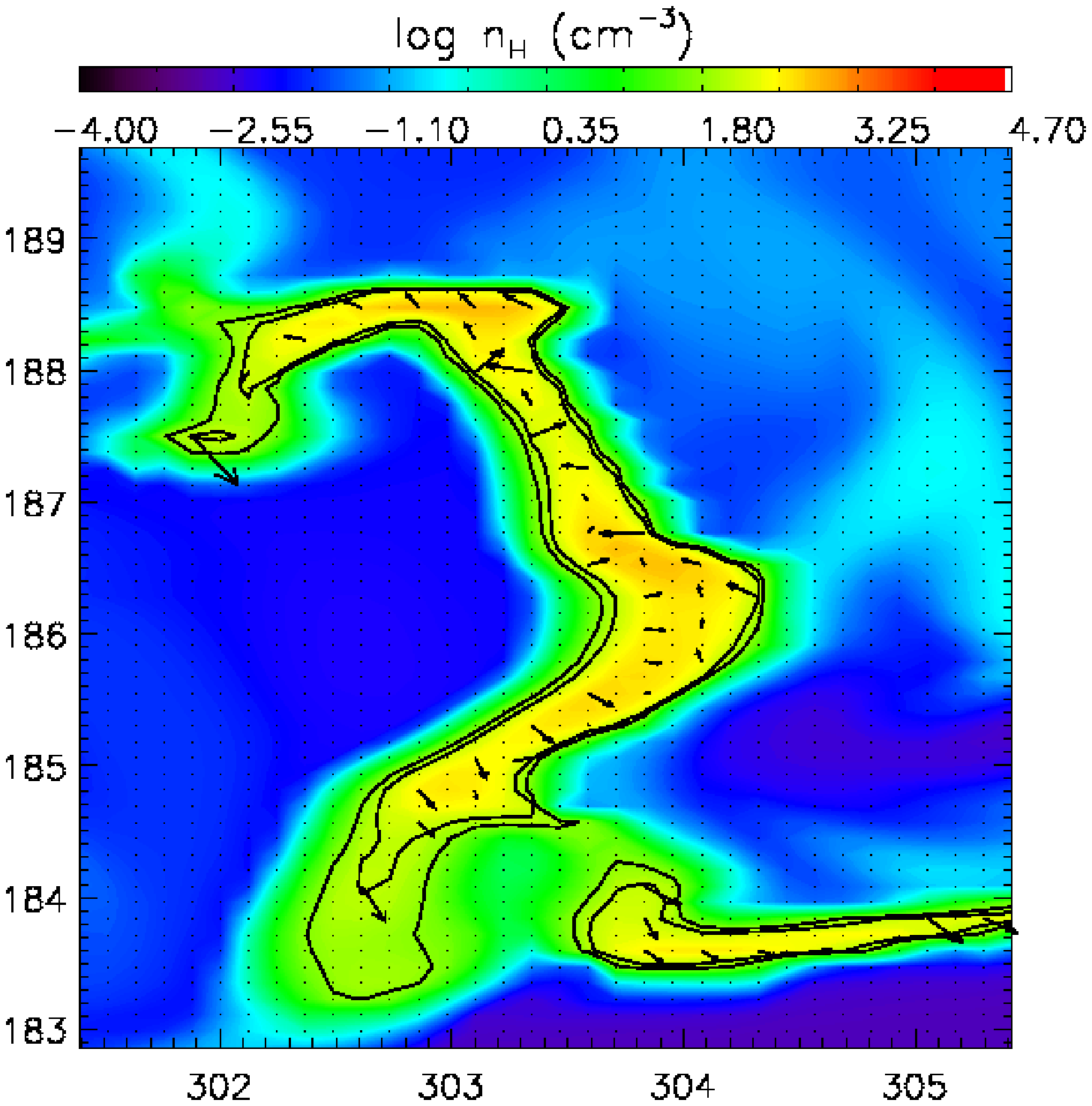} 
  \includegraphics[width=0.5\linewidth]{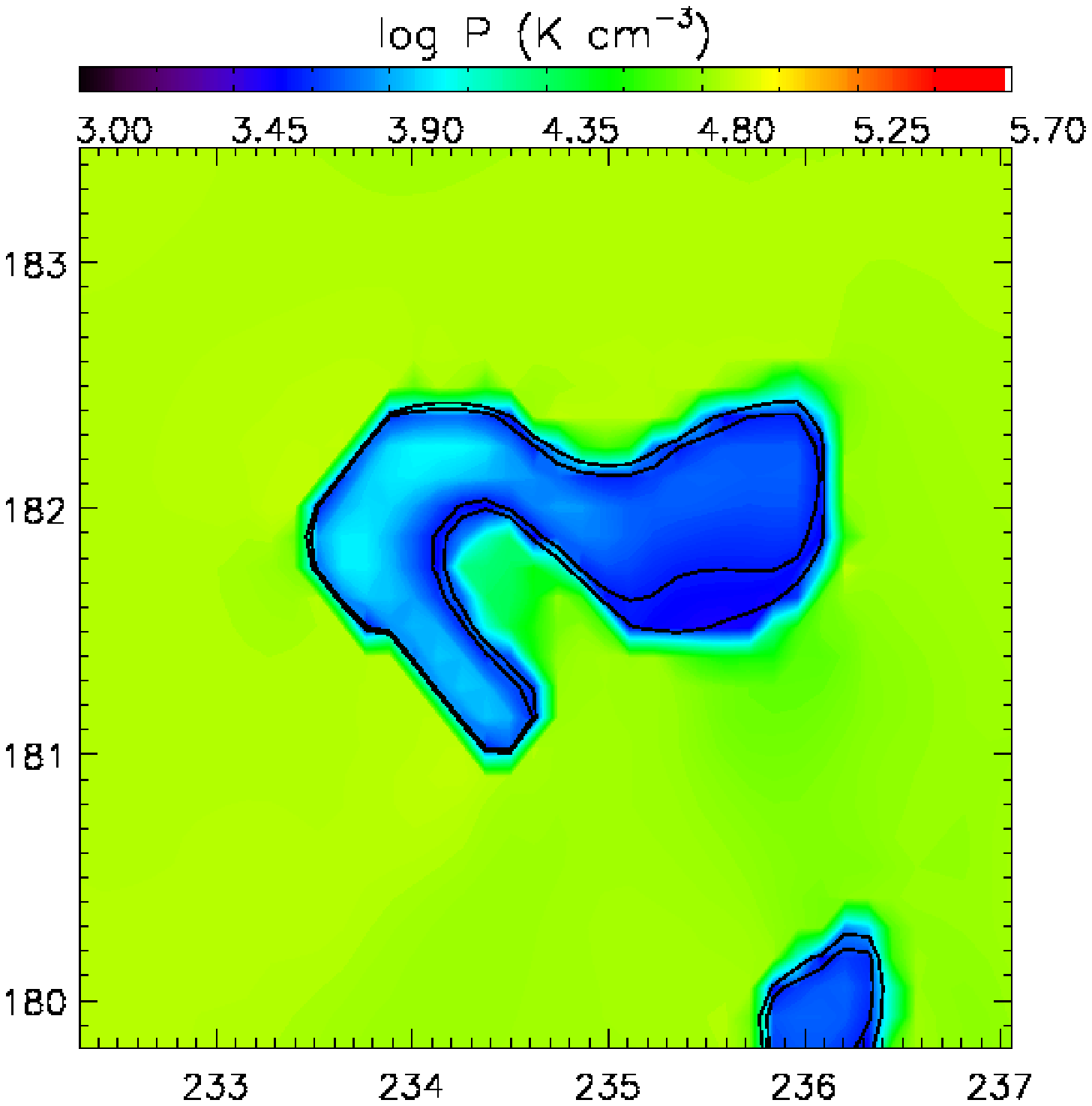} 
  \includegraphics[width=0.5\linewidth]{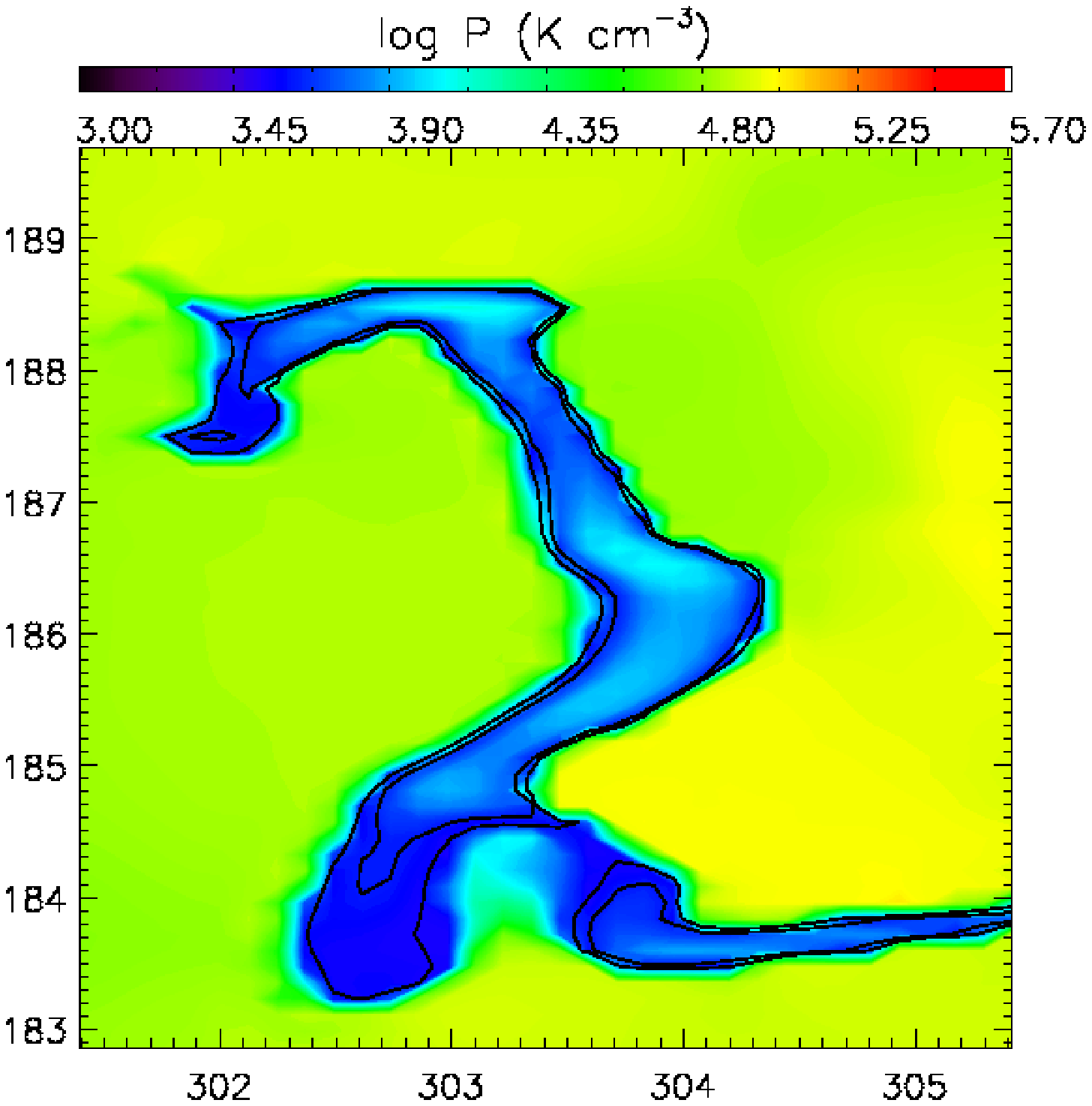}
  \caption{Two examples of clumps hosting random motions.
  Top panels show the logarithm of hydrogen number density in log(cm$^{-3}$) and bottom panels show the logarithm of thermal pressure in log(K cm$^{-3}$).
The black arrows show the velocity field with the mean velocity of the central clump subtracted.
Overplotted in black are the contour levels for $n_H$ equal to 50 $cm^{-3}$, 100 $cm^{-3}$ and 1000 $cm^{-3}$. }
  \label{clouds_random}
\end{figure*}

\begin{figure*}
  \includegraphics[width=0.5\linewidth]{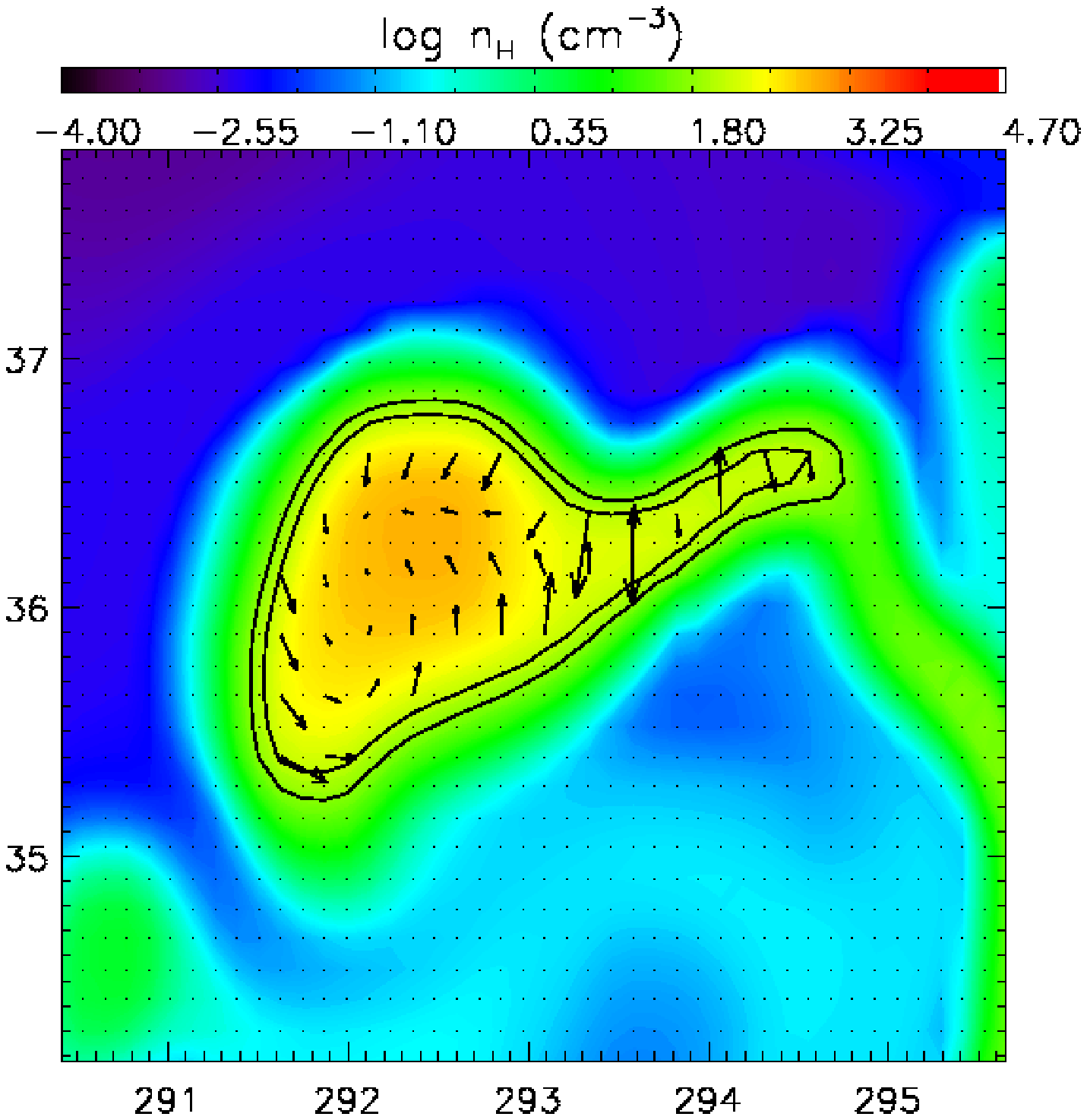} 
  \includegraphics[width=0.5\linewidth]{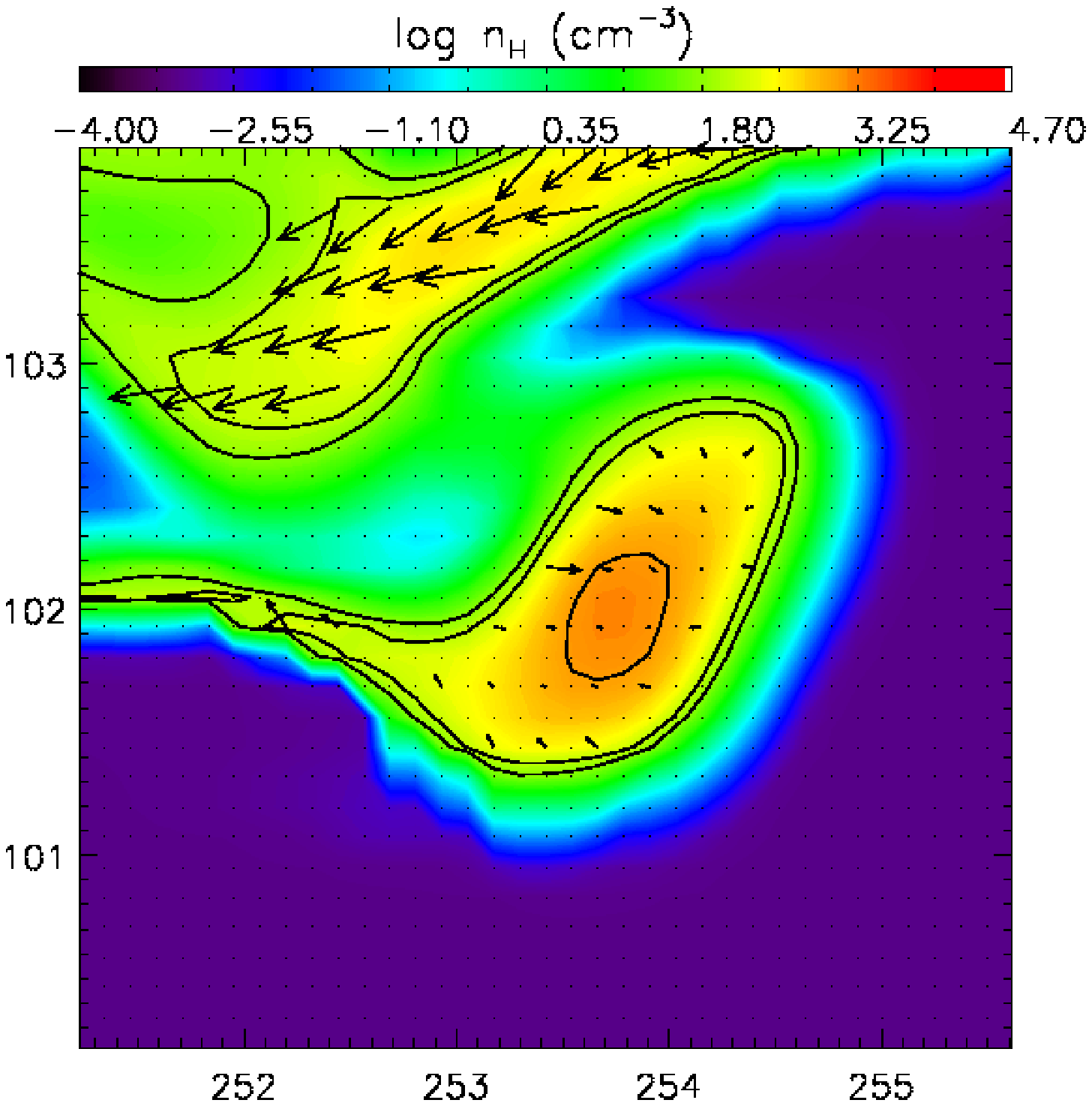} 
  \includegraphics[width=0.5\linewidth]{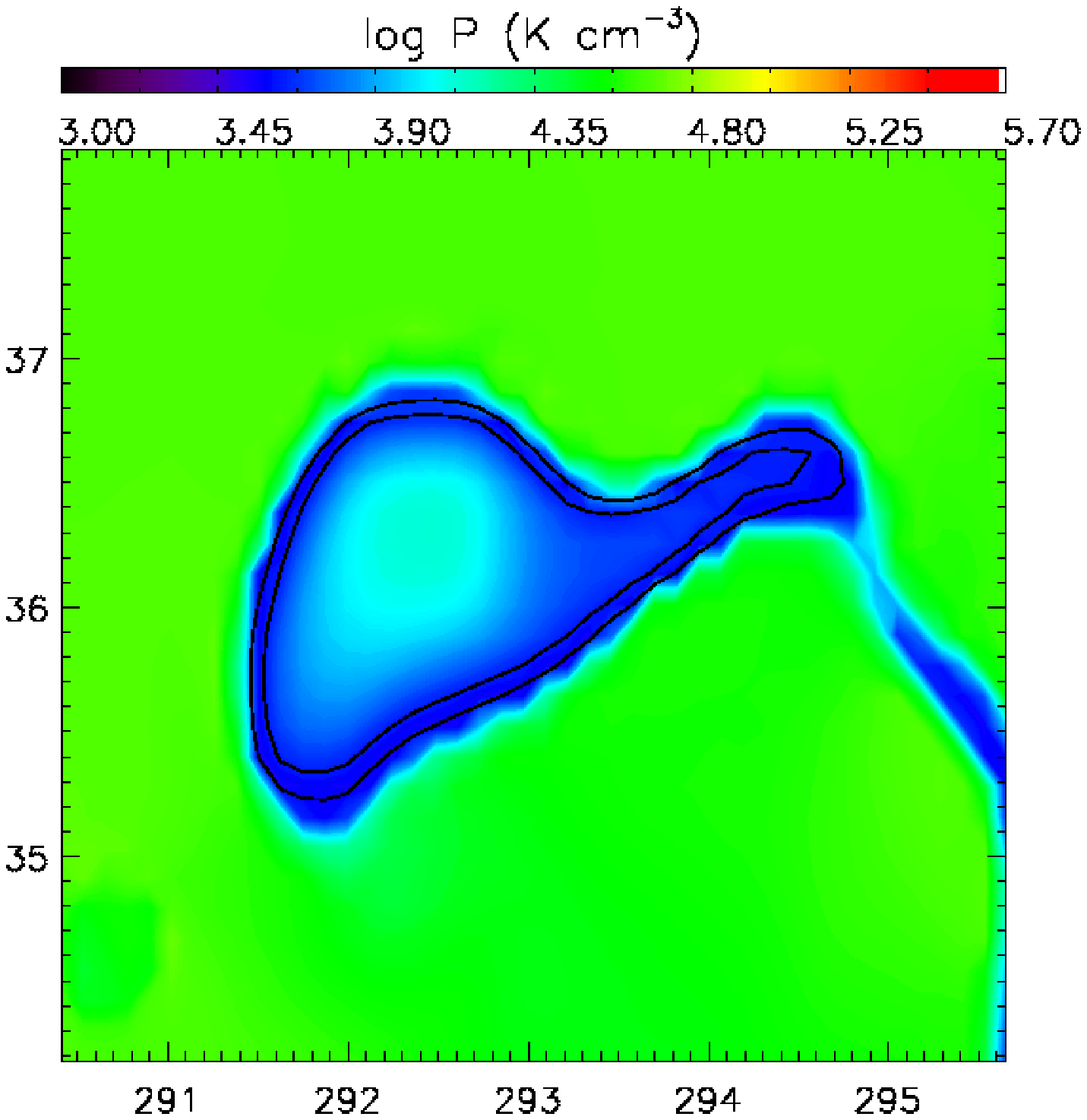} 
  \includegraphics[width=0.5\linewidth]{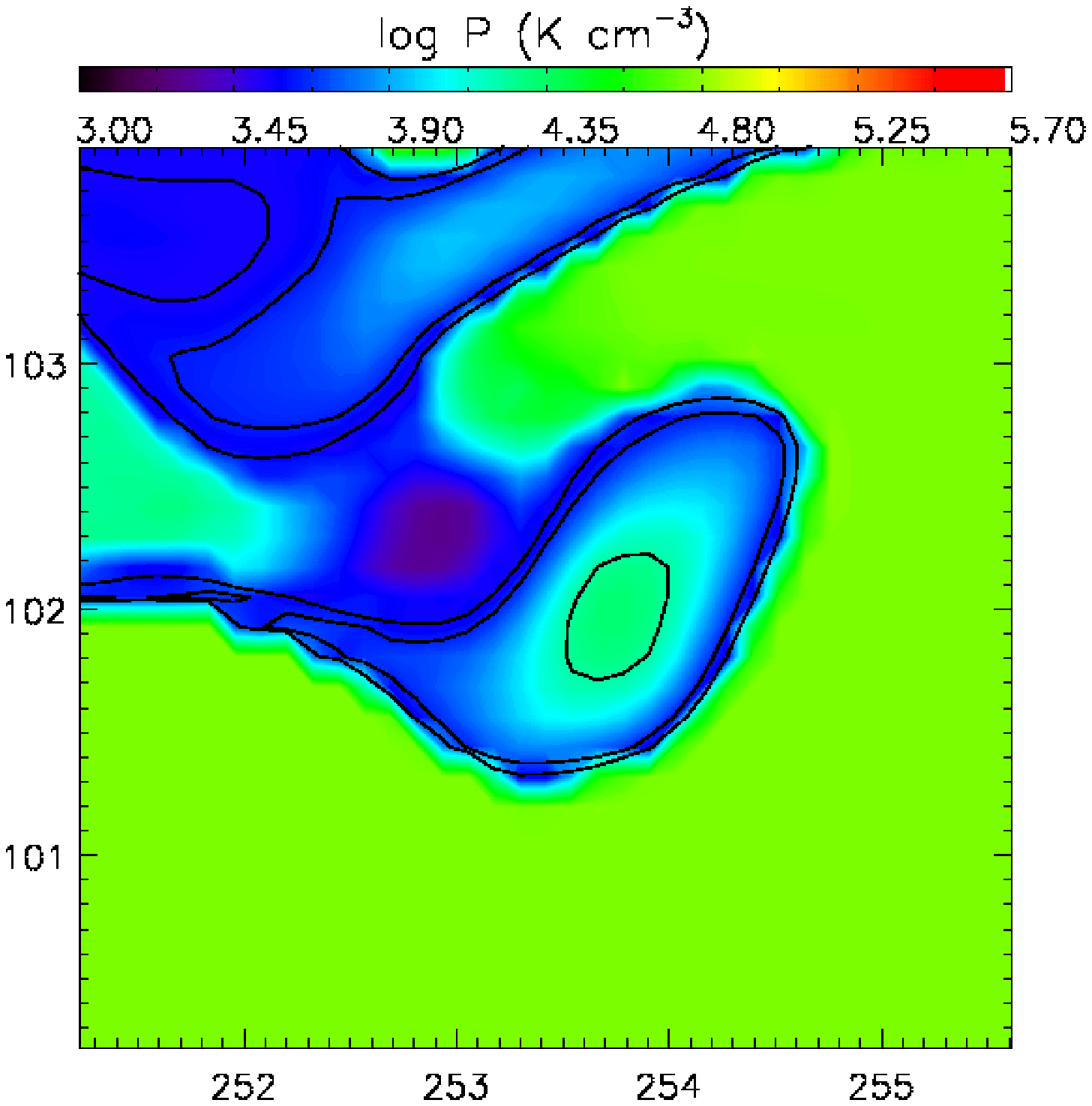}
  \caption{Two examples of clumps with small internal velocities.
Top panels show the logarithm of hydrogen number density in log(cm$^{-3}$) and bottom panels show the logarithm of thermal pressure in log(K cm$^{-3}$).
The black arrows show the velocity field with the mean velocity of the central clump subtracted.
Overplotted in black are the contour levels for $n_H$ equal to 50 $cm^{-3}$, 100 $cm^{-3}$ and 1000 $cm^{-3}$. }
  \label{clouds_low_vel}
\end{figure*}

Clumps are generally at lower pressures with respect to their surrounding gas (-- see figures \ref{clouds_collapse}, \ref{clouds_rot} \ref{clouds_random} and \ref{clouds_low_vel}).  
This means that Thermal Instability is still acting in these regions, causing clumps to condense further.
Two examples of condensing clumps are shown in figure \ref{clouds_collapse}.  Others (about 11\% in the last snapshot of the uniform background run and 16\% in  the turbulent background run) are rotating.  Two examples of rotating structures are shown in figure \ref{clouds_rot}.  Rotation is either combined with compression or it introduces a centrifugal force which makes the clump expand.  Of the rotating cores in the last snapshots of the uniform and the turbulent background runs, about 25\% and about 35\% respectively are at the same time condensing and the rest are expanding.  In very few cases, less than 1\%, the centrifugal force exactly balances the force due to the pressure difference between the interior and the exterior of the clump.

A small fraction of the identified clumps are in pressure equilibrium with their surroundings, at least at their most central parts.  These clumps do not tend to host significant internal motions (-- Figure \ref{clouds_low_vel}).

All clumps are surrounded by a warm corona which is more dilute and at an intermediate pressure between their pressure and the one of the surrounding gas.  This corona usually surrounds more than one clump, indicating that they are parts of larger structures, such as the ones illustrated in figures \ref{fil1}, \ref{fil2} and \ref{fil_time}.

In the following we focus on the velocity dispersions, sizes and possible evolution of the clumps.

\subsubsection{Velocity dispersions}
\begin{figure*}
  \includegraphics[width=0.32\linewidth]{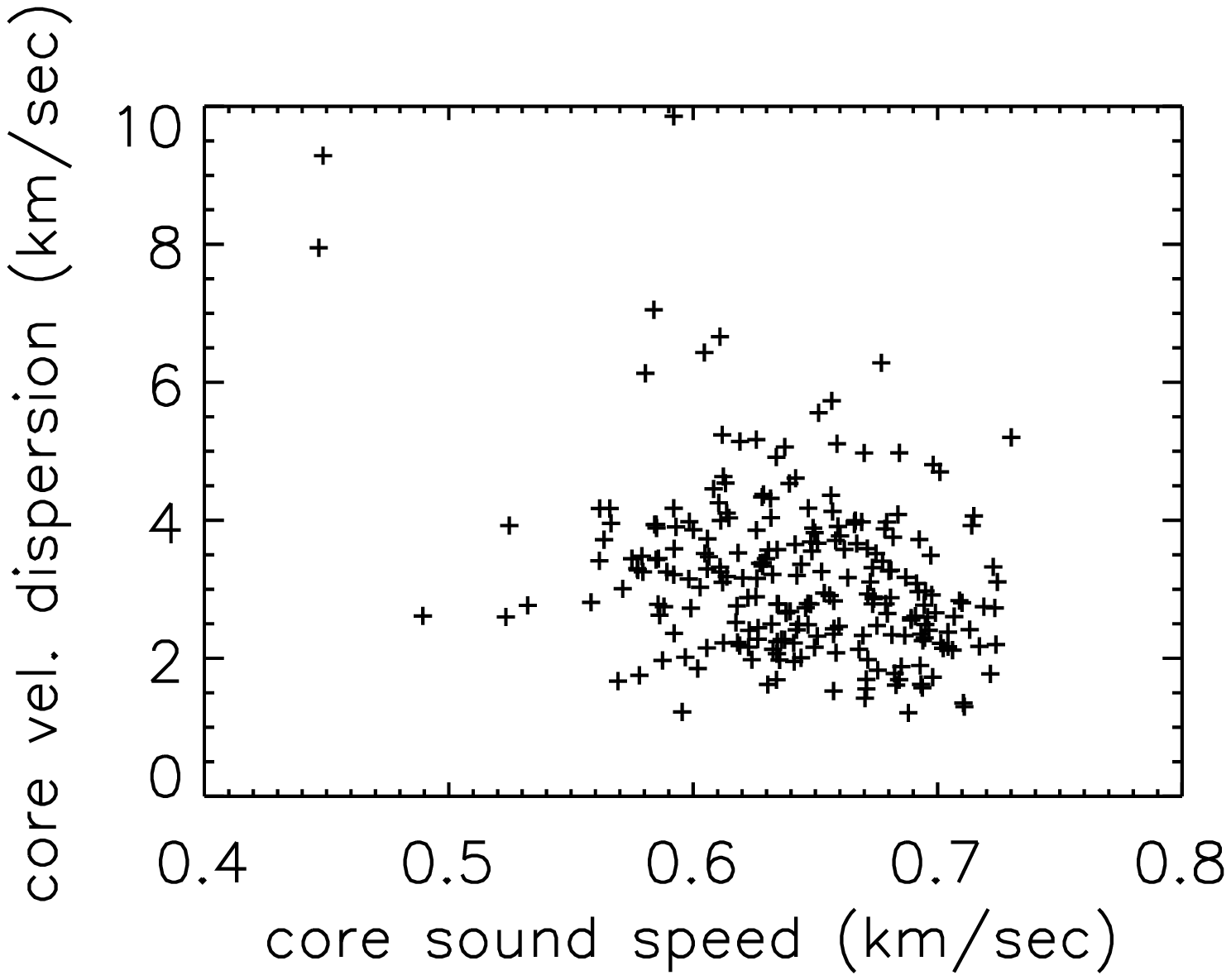} 
  \includegraphics[width=0.32\linewidth]{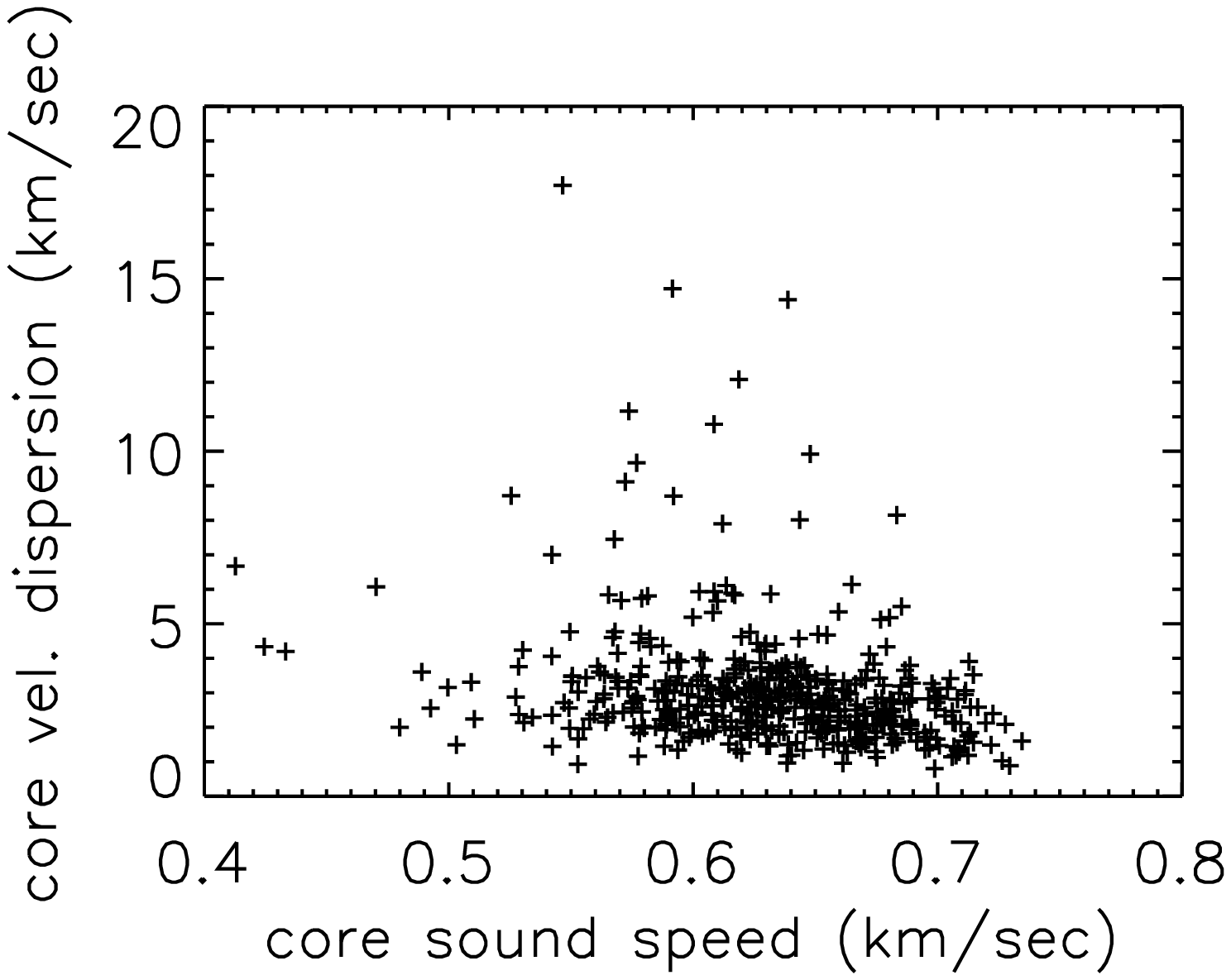}  
  \includegraphics[width=0.32\linewidth]{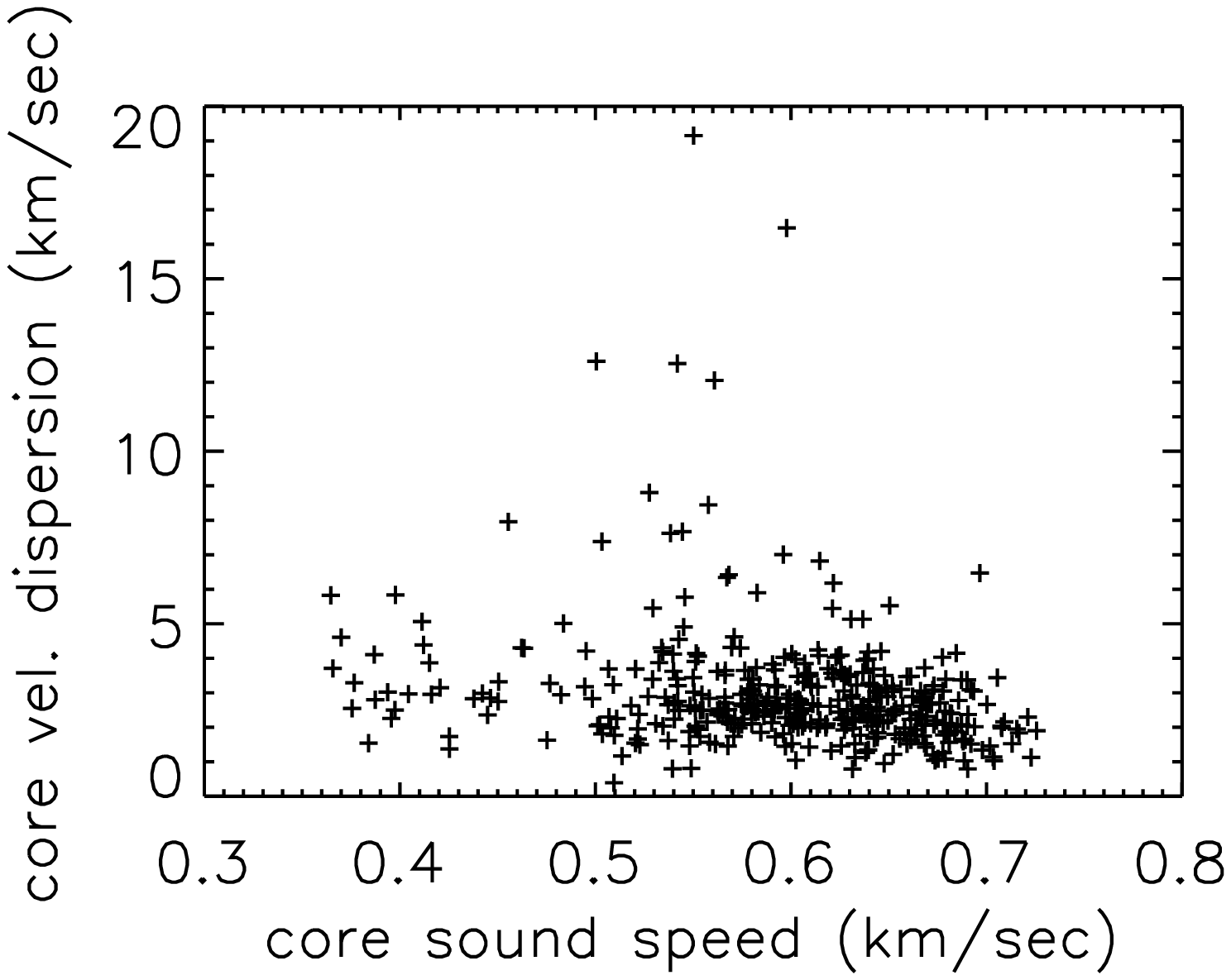} 
  \includegraphics[width=0.32\linewidth]{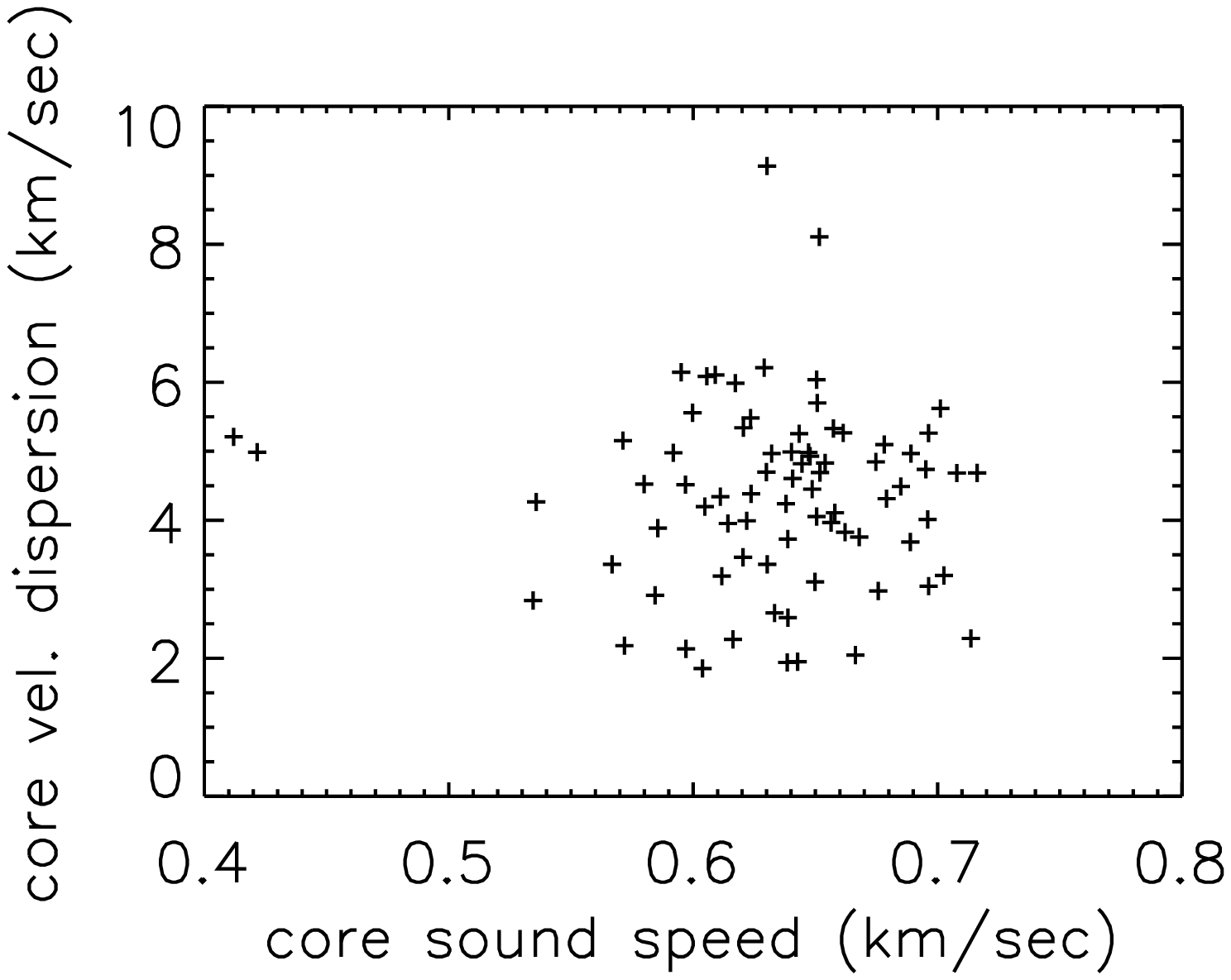}  
  \includegraphics[width=0.32\linewidth]{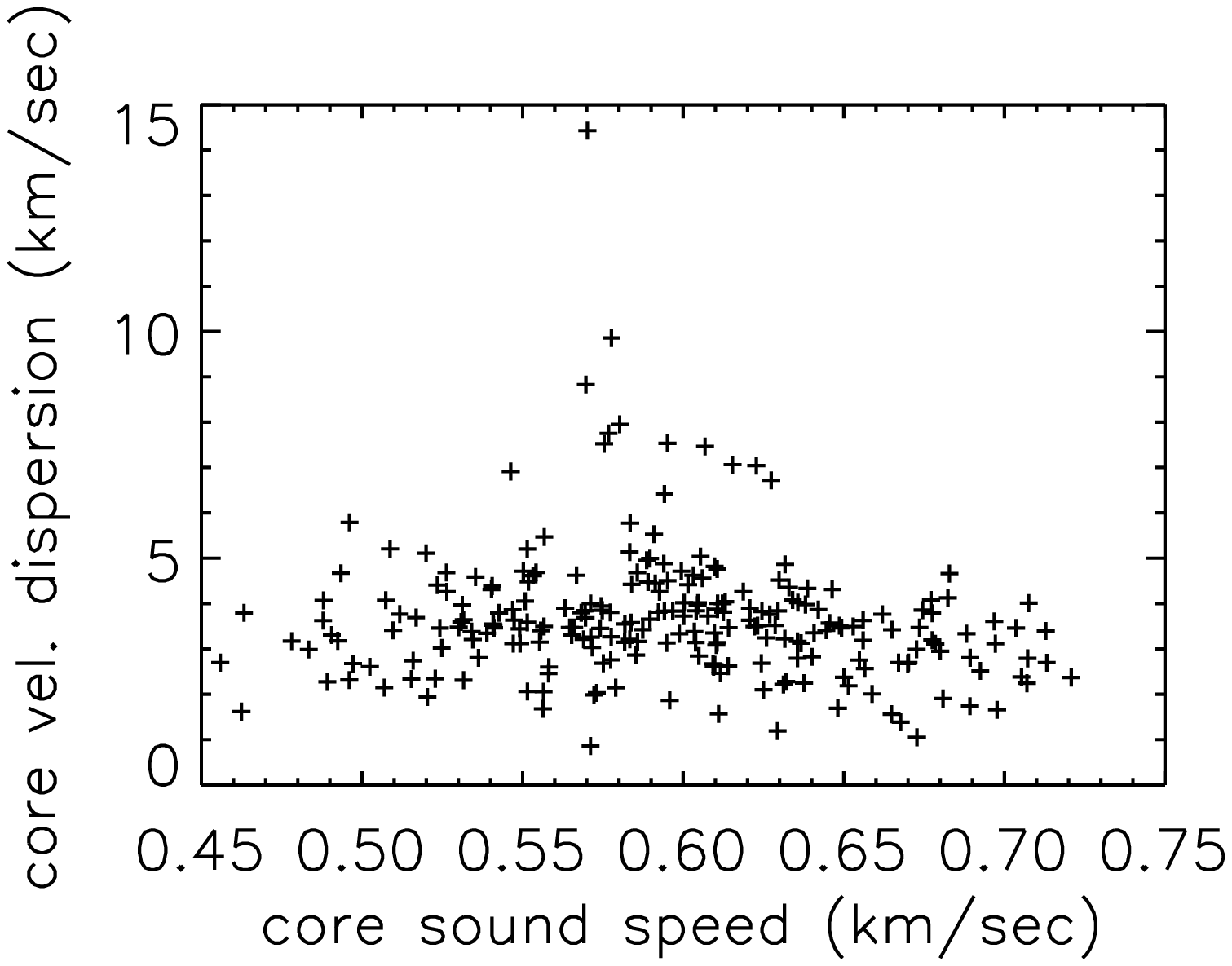} 
  \includegraphics[width=0.32\linewidth]{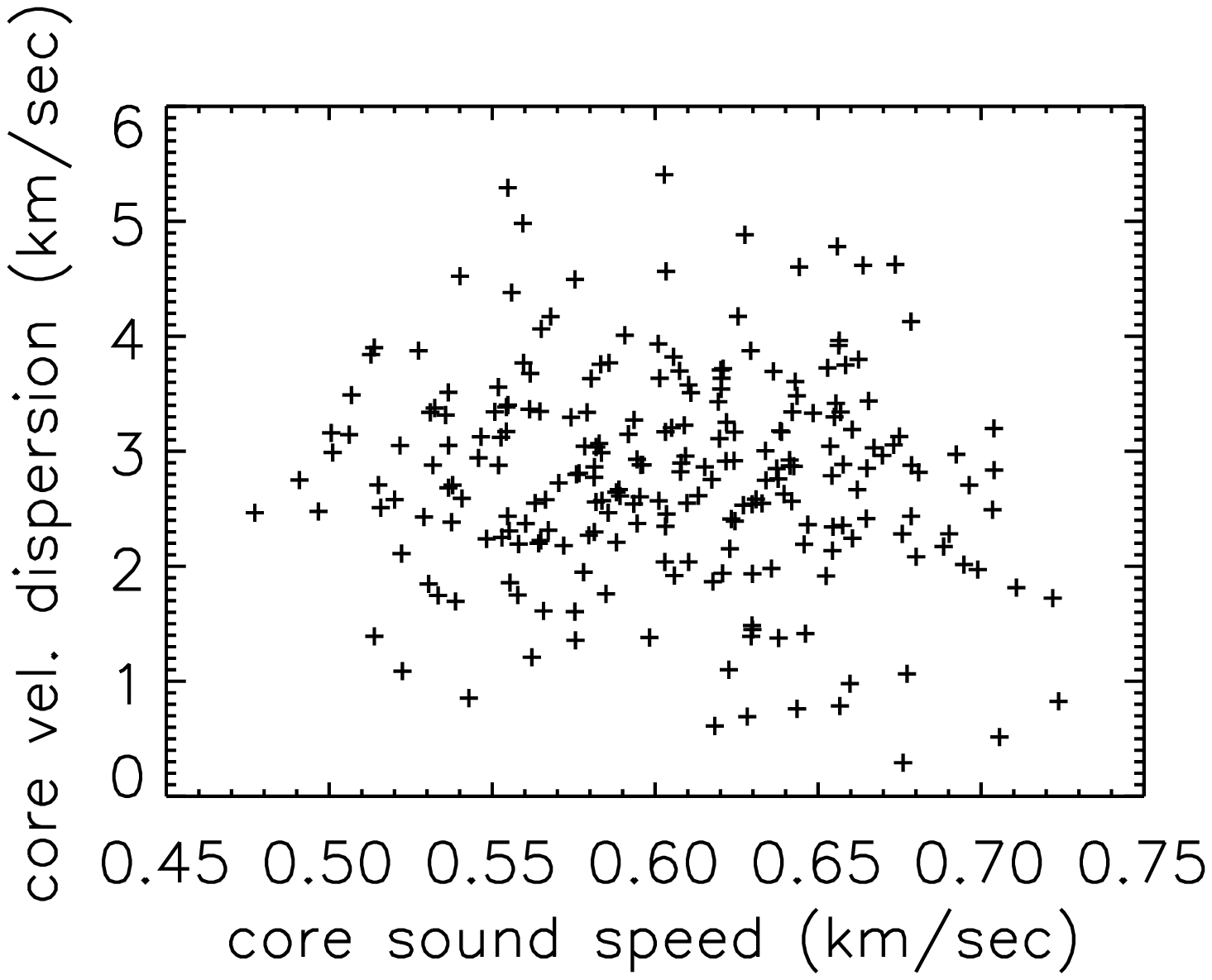}
  \caption{Clump internal velocity dispersion versus their sound speed at different times.  
  From left to right, 3, 5 and 7 Myrs after star formation.
  The top panel corresponds to the uniform background medium run, the bottom panel to
  the turbulent background medium run.}
  \label{vel_dispersions}
\end{figure*}

\begin{figure}
   \plotone{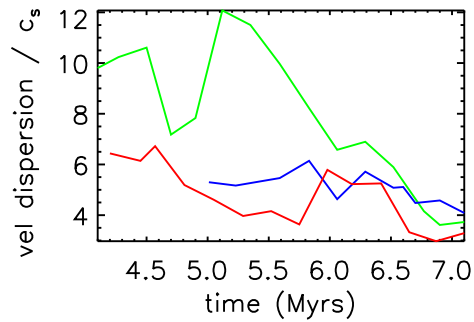}
   \caption{Time evolution of the velocity dispersions over sound speed for three isolated cores.
   The green line corresponds to a condensing clump, the blue line to a rotating clump and the red line to a clump with random motions.
   The data for each clump begin from the snapshot where we can still identify the clump as being the same.}
   \label{vel_dispers_time}
\end{figure}

Figure \ref{vel_dispersions} shows the internal velocity dispersions of the clumps at different times for both simulations. 
The velocity dispersion here is defined as the square root of the variance among all locations which compose the clump.
This means that for many clumps a velocity dispersion may indicate rotation, compression, expansion or random motions.
As random motions here we define any combination of rotation and compression or expansion.
Since the clumps have no significant internal density fluctuations, mass-weighted velocity dispersions are not very different from the ones presented here.

Although many of the formed clumps host supersonic motions in their interior at all times, there is an indication that these motions decrease at later times, as shown in the velocity dispersion plot of the last snapshots.
This effect seems slightly more pronounced for the turbulent background run, where the lagest internal velocity dispersions have disappeared
in the last snapshot.

In the case of compression or expansion, this means that the clumps gradually move to pressure equilibrium with their surrounding gas.  In the case of random motions, it indicates that these motions are inherited by the
turbulent environment that created the clumps but, in absence of any mechanism to sustain them, they die out.
In the case of rotation, though, the situation is a bit more complicated.  Rotation could be an effect of the large-scale Kelvin-Helmholtz shear, it can originate from Thermal Instability accretion or can be a result of structures splitting or merging.  Angular momentum conservation sustains these motions for longer, so we would only expect them to decrease on a viscous timescale.

In order to study if the decrease in internal velocity dispersion is observable for individual clumps, we tracked individual clumps back in previous snapshots 
and plotted their velocity dispersions with time.  Since dense structures are created all the time and clumps merge or split at each snapshot, it is very difficult to construct an algorithm able to automatically identify the same clump in different snapshots.  Instead we identified the clumps by eye according to their positions and translational velocities.  We focus here on three examples.  We selected a rotating clump, a contracting clump  and a clump with random internal motions.  Their velocity dispersions over their corresponding sound speed as a function of time are shown in figure \ref{vel_dispers_time}.
Since clumps evolve almost isothermally, their sound speed does not vary significantly during the interval shown in the plot.
The green line corresponds to a contracting clump, the blue line to a rotating clump and the red line to a clump with random motions.
All three clumps show some decrease in velocity dispersion with time.
Although we are not able to make a complete study at this stage, we observe that the rotating clump practically maintains the same velocity dispersion throughout its existence, showing only a slight decrease, while the condensing clump and the clump with random motions show a decrease in velocity dispersion.  For the condensing clump this decrease is very pronounced. 

\subsubsection{Sizes}

\begin{figure*}
    \includegraphics[width=0.49\linewidth]{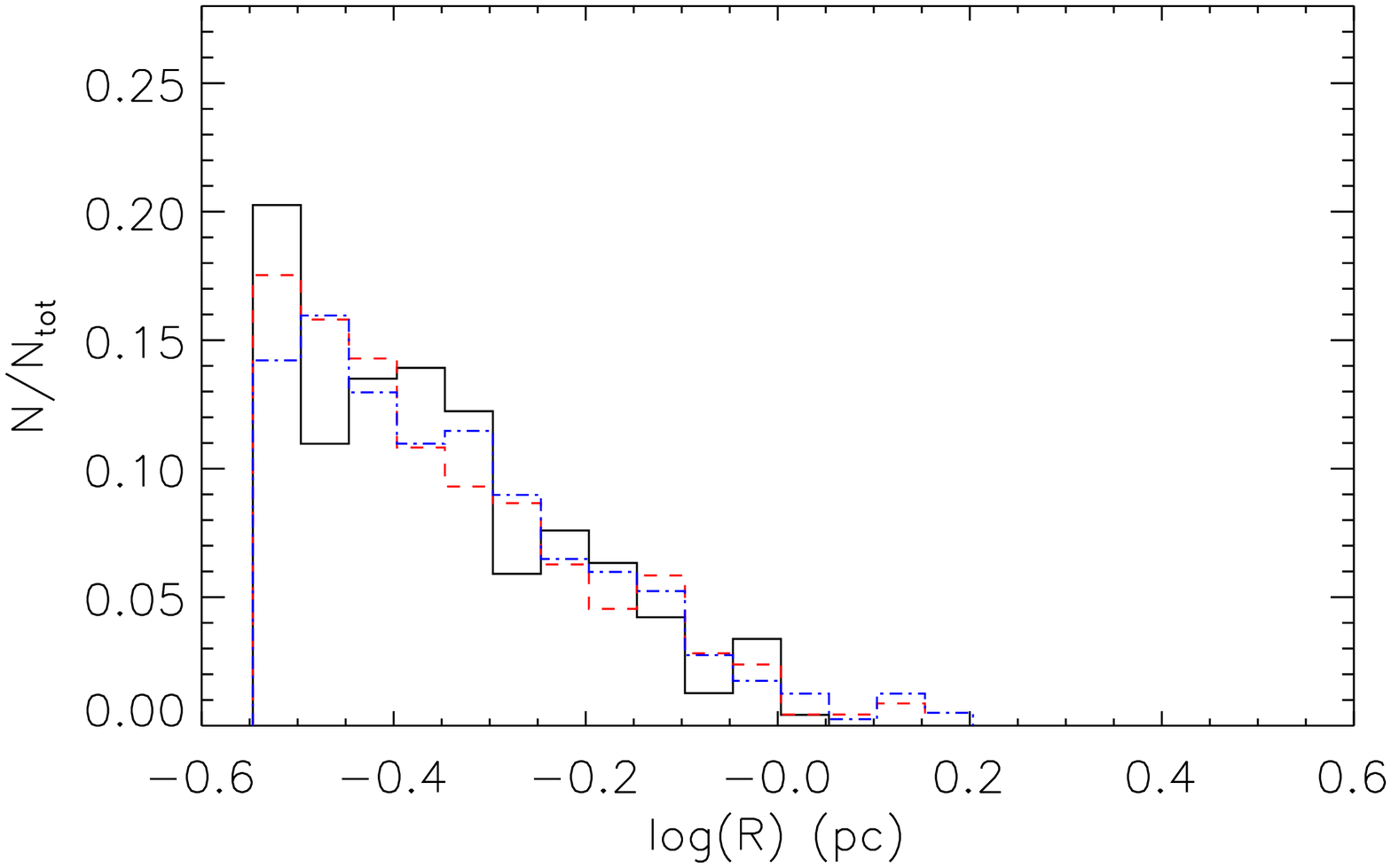} 
    \includegraphics[width=0.49\linewidth]{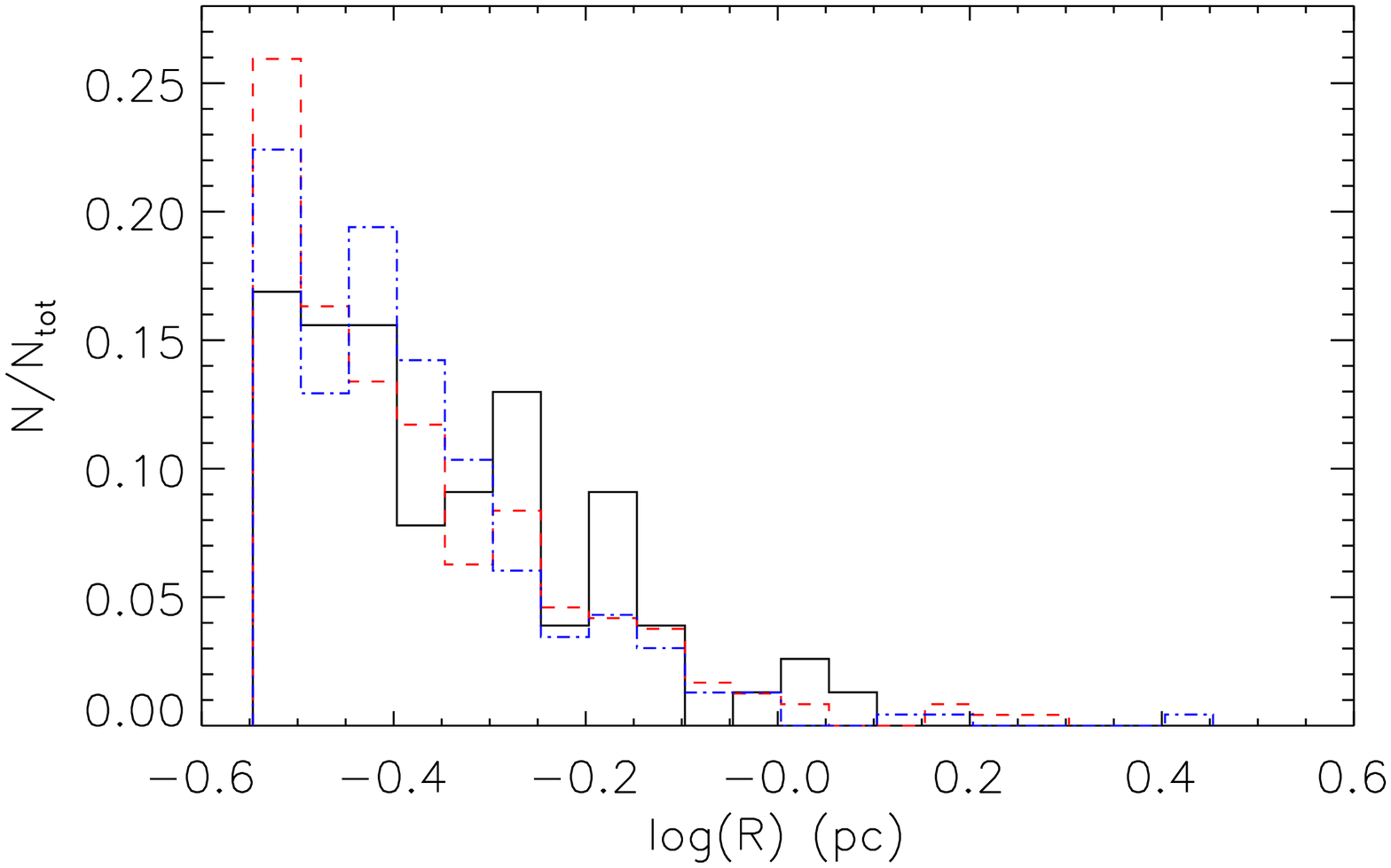}
    \caption{Clump size distributions in parsecs at different times.
 The plot on the left-hand side corresponds to the run in a uniform diffuse medium and the 
 plot on the right to the run in a turbulent diffuse medium.  The solid black, dashed red and dash-dotted blue histograms
 correspond to the 3 Myrs, 5 Myrs and 7 Myrs, respectively.}
  \label{abs_sizes}
\end{figure*}

\begin{figure*}
    \includegraphics[width=0.49\linewidth]{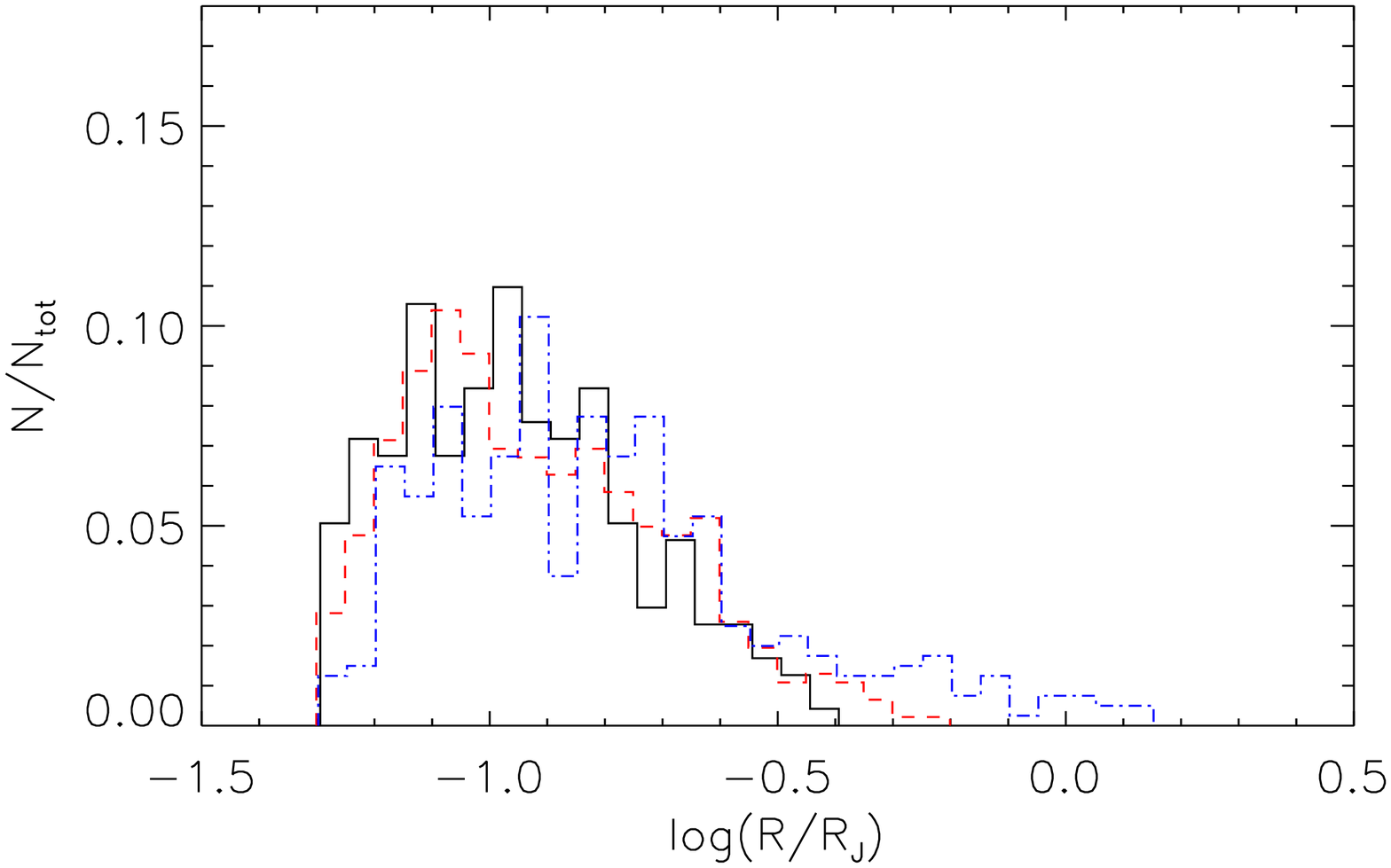} 
    \includegraphics[width=0.49\linewidth]{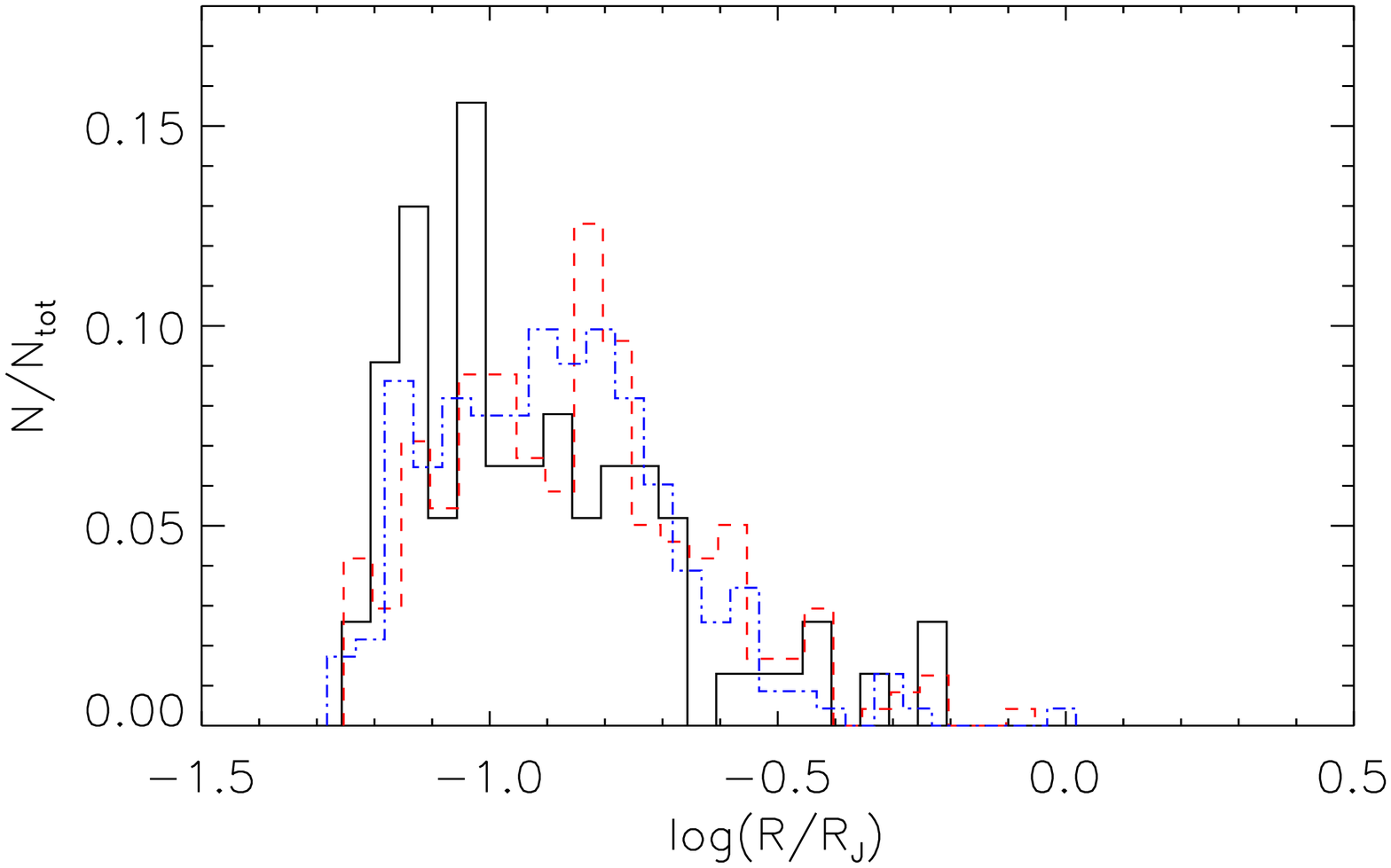}
    \caption{Clump size over clump Jeans length distributions at different times.
     The plot on the left-hand side corresponds to the run in a uniform diffuse medium and the 
 plot on the right to the run in a turbulent diffuse medium. The solid black, dashed red and dash-dotted blue histograms
 correspond to the 3 Myrs, 5 Myrs and 7 Myrs, respectively. }
  \label{sizes}
\end{figure*}

Figure \ref{abs_sizes} shows the size distribution of the clumps at different times for both simulations.  The sizes are calculated as the square root of the area occupied by the clump.  As indicated by the figure, we do not form clumps larger than approximately 3 pc.  However, the maximum clump length can reach about 10 pc.
This means that, although structures may have one very large dimension, they occupy a very small area.

Figure \ref{sizes} shows the distributions of the ratios of clump sizes over their corresponding Jeans length, at different times.
The size distributions of the clumps have a very similar shape and range between the two simulations.
As clumps are fairly uniform in density and temperature, their Jeans length does not vary significantly within each single clump.
Clumps in general seem to be smaller than their corresponding Jeans length but, as time advances, some clumps become large enough to be potentially unstable to gravitational collapse.  Of course, as we have not included gravity in these simulations we cannot know if this would actually be the case.

As mentioned earlier, since the formed structures are very filamentary, they are likely to contain more Jeans lengths along a single dimension.
Note also that the algorithm we use to find clumps favors the identification of the smallest possible structures as separate entities.
As mentioned above however, the clumps we identify are usually parts of larger structures which are dynamically interacting or surrounded by a common warm and more diffuse corona.


\subsubsection{Clump evolution}

Although most of the clumps are in low-pressure regions in the simulation, there are some clumps with approximately the same 
pressure as their surrounding gas.  All of the clumps are surrounded by an intermediate pressure corona, which is also thermally unstable.

This, in combination with the fact that clumps show a tendency of decreasing their internal motions with time and that clouds in pressure equilibrium tend to host smaller internal motions leads us to believe that there might be an evolution from clumps out of equilibrium, with strong internal motions to more quiescent clumps.

\begin{figure}
   \plotone{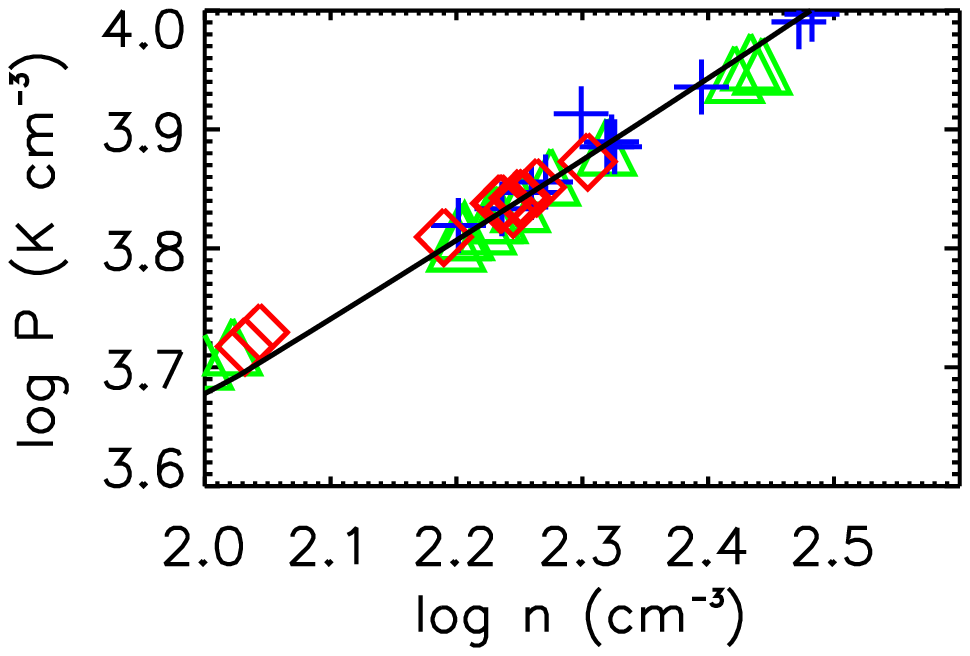}
   \caption{Isolated core tracks on a pressure-density diagram.  
   The green triangles correspond to a condensing clump, the blue crosses to a rotating clump and the red diamonds to a clump with random motions.
   The solid black line is part the cooling-heating equilibrium curve which was shown in figures \ref{phases_noturb} and \ref{phases_turb}.}
   \label{pres_den_time}
\end{figure}

Figure \ref{pres_den_time} shows the tracks of the same three clumps we traced back in time on the pressure-density diagram. 
The dotted line is the cooling-heating equilibrium curve.  The clumps form in an area where cooling dominates, possibly from thermally unstable gas, at the left of the figure.  As time goes by they move on the equilibrium curve and gradually increase their density and pressure, while staying on the curve.  The timescale of this evolution is about 2-3 Myrs, which corresponds to about 4 - 5 clump sound crossing times.


\section{Summary and Discussion}\label{sum}

In this work we have studied the combined effects of the Vishniac, the Kelvin-Helmholtz and the Thermal Instability on the expansion and collision of two superbubbles, created by stellar winds and supernova explosions from young OB associations.

We have presented high-resolution, two-dimensional simulations to show that, even without simulating the effects of gravity, the shells formed by stellar feedback will condense and fragment to form cold and dense filaments.  We have modeled the shell collision in a static, uniform diffuse medium and in a turbulent diffuse medium and found that background turbulence not only introduces an anisotropy in the shell fragmentation, but also helps to produce more filamentary structures.

Our results provide a picture of the ISM similar to \citet{Audit_Hennebelle_2005} and \citet{Hennebelle_Audit_2007}, where the ISM phases are tightly interwoven, with sharp thermal interfaces between them.  Of course, a more detailed comparison is not possible, since that work was done with a much higher resolution, representing a smaller region of the ISM and not including the hot phase.    In our simulations, elongated cold structures sit in warm, thermally unstable coronas, submerged in a hot dilute medium that the stellar winds and supernovae from young OB stars create.  
Clumps are dynamical entities, constantly merging and splitting and in general not in pressure equilibrium with their surrounding gas.
In that sense, the picture we see in our simulations is very different from the classical ISM model proposed by \citet{McKee_Ostriker_1977}, where the clouds are treated as quasi-static spheres.  The structures we find are rather long thin filaments, very similar to the large "blobby sheets" proposed by \citet{Heiles_Troland_2003}.

We identify individual clumps in the simulations by setting density and temperature thresholds and connecting adjacent locations in the simulation domain which exceed these thresholds.  
This method favors the identification of the smallest cold and dense structures.  The clumps we identify tend to be connected in groups within a common warm corona, forming long filaments and are mostly located in areas of lower pressure with respect to their surroundings.  The latter causes many of them to condense to higher densities and smaller volumes, almost isothermally.  Apart from condensing clumps we find rotating clumps and clumps hosting irregular motions.  All these internal motions are reflected in the velocity dispersion as supersonic internal motions.  Approximately 11 to 16\% of the identified clumps are rotating and are interesting candidates for forming protostellar disks.  However, rotation should be studied in three dimensions for more reliable results.  

Internal clump motions tend to decrease with time, as the clumps come closer to thermal pressure equilibrium with their warm surroundings.  However, if gravity and star formation was included in our simulations, we would expect these motions to be sustained for longer times due to gravitational collapse and feedback from the newly formed stars in their interior.

Cold clumps in our simulations form by Thermal Instability condensation of the ambient diffuse medium when it is perturbed by the expanding shells.  This, in combination with the fact that the mass injected by the OB associations is a very small fraction of the total mass in our domain throughout the simulation, shows that the material forming the new clumps could not have been enriched directly by the supernova explosions.  The material ejected from the OB associations will probably enrich the interstellar matter on much larger timescales, after the hot bubbles have mixed with the diffuse medium.

Thermally unstable gas amounts to about 8 to 10\% of the total gas mass in the last snapshots of our simulations, far from the almost 50\% that is commonly detected in observations \citep{Heiles_Troland_2003} or found in simulations of turbulent, thermally bistable flows \citep{Gazol_2001, Gazol_2005}.  This is because our simulations are dominated by what we identify as CNM, that is, gas with temperatures lower than 100K.  Due to our very high resolution, this gas is also very dense, reaching hydrogen number densities of the order of $10^5$ cm$^{-3}$.  This gas would be mostly molecular, so it would not be identified as cold HI in observations.
In simulations with gravity we would not expect to encounter this issue, since gravity would eventually dominate the dynamics of the formed clumps and turn them into stars once they became dense enough, so thermal instability would no longer be responsible for their evolution at these densities.

The simulations presented in this paper are only a first step to modeling triggered molecular cloud formation using physically motivated colliding flow parameters.
We have not attempted a parameter study in this paper, but it will be the object of future work to study the effect of varying the distance between the superbubbles, the number of OB stars creating the superbubbles and the metallicity of the gas.  

In this work we have also not taken into account galactic shear or density stratification.
Assuming a galactic rotation rate of 26 km sec $^{-1}$ kpc $^{-1}$, the relative shear in our computational box would be 13 km sec$^{-1}$.  This velocity would
have a crossing time of approximately 37 Myrs, which is much longer than what our simulations last.
Density stratification with a scale height of H=150pc would cause the superbubbles to expand faster along the vertical direction, 
but since we are interested mostly in what happens at the superbubble collision interface, we can neglect this effect as a first approximation.

Our calculations also ignore magnetic fields and gravity.  
We would expect magnetic fields to play a significant role in the dynamics of the problem, both during the expansion of the superbubbles and also during the more complex collision phase, if they had a preferred orientation, but in two dimensions we would not be able to model all relevant phenomena properly anyway.   
Gravity is essential for studying the evolution of the clumps and for estimating their star forming efficiency.  Combining with the modeling of a third dimension, it would give us useful mass estimates of the formed structures.  Moving to three dimensions and including gravity is work in process and will be presented in a future paper.  

We find the main limitations of our work to be the lack of resolution and bidimensionality.  
As we have pointed out, our simulations do not reach the resolution required to capture all the relevant physical processes on the smallest scales.  We have accounted for small-scale effects by only studying clumps which contain more than 16 grid cells.  Higher numerical resolution will certainly provide more insight on the number of formed clumps and their internal structure.  Although numerical effects do introduce a thermal conduction effect, explicitly modeling thermal conduction would help set the minimum scale for formed clumps and achieve convergence with increasing resolution.  On the other hand, if gravity is simulated it will probably already become important at length scales larger than the Field length.  

The restriction of the presented models to two dimensions is emphasizing the formation of filaments. 
In three dimensions, these structures could have a sheet-like morphology.
Gravity could then be responsible for turning these sheets into filaments \citep{Burkert_Hartmann_2004, Hartmann_Burkert_2007, Heitsch_2008}.
Although combining an increase in resolution with studying the problem in three dimensions is numerically very expensive, it is probably feasible with the use of AMR.


\acknowledgments
The authors would like to thank Rasmus Voss for providing the stellar wind data used in these calculations and Ralf Klessen for useful comments and suggestions.
This research was supported by the DFG cluster of excellence ‘Origin and Structure of the Universe’.


\bibliographystyle{natbib}
\bibliography{colliding_flows}

\label{lastpage}

\end{document}